\RequirePackage{pgf,tikz, tikz-3dplot}

\documentclass[11pt]{article}
\usepackage{graphicx} % Required for inserting images
\usepackage{amsmath, bm} %for bold letters in math mode
\usepackage{amssymb} %for a nice looking empty set
\usepackage[backend=biber, maxbibnames=9, style=numeric]{biblatex} 
\usepackage{amsthm} %for theoremstyle
\usepackage{todonotes} %to add notes
\theoremstyle{definition}
\newtheorem{definition}{Definition}
\newtheorem{theorem}{Theorem}
\newtheorem{lemma}[theorem]{Lemma}
\newtheorem{proposition}[theorem]{Proposition}
\newtheorem{remark}{Remark}
\newtheorem{corollary}[theorem]{Corollary}

\usepackage{multirow}
\usepackage{multicol}
\usepackage{caption}
\usepackage{subcaption}

\usepackage{hyperref}
\usepackage{dirtytalk}
\usepackage{array}

\usetikzlibrary{arrows}
\usetikzlibrary{patterns,intersections}
\usetikzlibrary{math}
\usetikzlibrary{decorations.pathreplacing,angles,quotes}
\usetikzlibrary{decorations.markings}
\usetikzlibrary{positioning}
\usetikzlibrary{through}
\usetikzlibrary{tikzmark}
\usetikzlibrary{angles,quotes}
\usetikzlibrary{calc}
\usetikzlibrary{intersections, shapes}
\usepackage[outline]{contour}% 
\usepackage{xcolor}
\usepackage{soul}

\title{A Generalization of Arrow's Impossibility Theorem Through Combinatorial Topology}

\makeatletter
\renewcommand\@date{{%
  \vspace{-\baselineskip}%
  \large\centering
  \begin{tabular}{@{}c@{}}
    Isaac Lara\textsuperscript{1}, \\
    \normalsize ilara@colmex.mx
  \end{tabular}%
  \quad \quad
  \begin{tabular}{@{}c@{}}
    Sergio Rajsbaum\textsuperscript{2}, \\
    \normalsize rajsbaum@im.unam.mx
  \end{tabular}
  \quad \quad
   \begin{tabular}{@{}c@{}}
    Armajac Raventós-Pujol\textsuperscript{3} \\
    \normalsize 
    aravento@eco.uc3m.es
  \end{tabular}
  \bigskip

  \textsuperscript{1}Centro de Estudios Económicos, El Colegio de México, Ciudad de México, México.\par
  \textsuperscript{2}Instituto de Matemáticas, Universidad Nacional Autónoma de México, Ciudad de México, México.\par
  \textsuperscript{3}Departamento de Economía, Universidad Carlos III de Madrid, Madrid, España.
  \bigskip

  \today
}}
\makeatother

\begin{document}

\maketitle

\begin{abstract}
To the best of our knowledge, a complete characterization of the domains that escape the famous Arrow's impossibility theorem remains an open question. We believe that different ways of proving Arrovian theorems illuminate this problem. This paper presents a new combinatorial topology proof of Arrow's theorem. In PODC 2022, Rajsbaum and Raventós-Pujol proved this theorem using a combinatorial topology approach. This approach uses simplicial complexes to represent the sets of profiles of preferences and that of single preferences. These complexes grow in dimension with the number of alternatives. This makes it difficult to think about the geometry of Arrow's theorem when there are any (finite) number of voters and alternatives.
Rajsbaum and Raventós-Pujol (2022) use their combinatorial topology approach only for the base case of two voters and three alternatives and then proceed by induction to prove the general version. The problem with this strategy is that it is unclear how to study domain restrictions in the general case by focusing on the base case and then using induction. Instead, the present article uses the two-dimensional structure of the high-dimensional simplicial complexes (formally, the $2$-skeleton), yielding a new combinatorial topology proof of this theorem. Moreover, we do not assume the unrestricted domain, but a domain restriction that we call the class of \emph{polarization and diversity over triples}, which includes the unrestricted domain. By doing so, we obtain a new generalization of Arrow's theorem. This shows that the combinatorial topology approach can be used to study domain restrictions in high dimensions through the $2$-skeleton.
\end{abstract}

\section{Introduction}

The famous impossibility theorem by Arrow \cite{Arrow1950,ArrowKennethJoseph1963Scai} says that any social welfare function (SWF) satisfying certain desirable properties is dictatorial. The SWFs for which this theorem applies assume an unrestricted domain, i.e., all preference profiles are allowed as inputs. Restricting the domain can yield both impossibility and possibility results; that is, dictatorship remains the only option or there are others\footnote{This paper mostly focuses on impossibility results. We refer the reader to \cite{BarberaEncyclo,ElkindEtAl2022,gaertner_2001} for surveys on domain restrictions including many possibility results.}. To the best of our knowledge, a complete characterization of which domains escape the impossibility remains as an open question. We believe that proving (im)possibility Arrovian results in different ways is important in the quest for such a characterization. In this paper, we propose a new generalization of Arrow's theorem inspired by a geometric understanding of Arrow's theorem under a combinatorial topology framework. Even if the initial motivation for this generalization was only to exemplify how this geometric understanding helps to find (im)possibility results, this generalization might be interesting on its own as it relates to notions of polarization and diversity over triples of alternatives.

%Some of the most known domains that help escaping the impossibility have been studied extensively; classical examples are the \emph{single-peaked} domains \cite{black1948}, \emph{group-separable} domains \cite{InadaKen-ichi1964ANot,InadaKen-ichi1969TSMD} and \emph{value-restricted} domains \cite{SenAmartyaK.1966APTo}. For extensive reviews on the literature of domain restrictions see \cite{BarberaEncyclo,ElkindEtAl2022,gaertner_2001}.

Arrow's theorem has sparked many alternative proofs, e.g. \cite{ArrowKennethJoseph1963Scai,Baryshnikov,Geanakoplos2005,Hansson,KIRMAN1972267,TANG20091041,NingYu2012}. Key to this paper is an algebraic/combinatorial topology framework that Baryshnikov \cite{Baryshnikov} establishes to prove Arrow's theorem using algebraic topology techniques. Referring to this theorem's traditional combinatorial proof, Baryshnikov \cite[p. 404]{Baryshnikov} says: \say{its proof seems a little irrelevant, as a proof of the fix point theorem without use of algebraic topology, e.g., by means of Sperner’s lemma, that is, in a combinatorial way.} Moreover, in another article \cite[p. 208]{BaryshnikovYuliyM.1997Tads}, he says (referring to the algebraic topology approach): \say{the topology we are exploiting is in fact very geometrical and thus appeals to our intuition much better than dry combinatorial constructions.}

Baryshnikov \cite{Baryshnikov} uses simplicial complexes to represent the set of profiles and preferences, and simplicial maps to represent SWFs. The authors of \cite{RajsbaumR2022preprint-new,RajsbaumArmajacPODC} use Baryshnikov's framework to prove Arrow's theorem's base case, i.e., two voters and three alternatives, with combinatorial topology (they use simple counting and geometric arguments instead of more involved algebraic topology). In particular, \cite{RajsbaumR2022preprint-new} demostrate that Arrow’s theorem can be derived from the index lemma, a result in combinatorial topology that generalizes Sperner’s lemma (which is equivalent to Brouwer's fixed-point theorem). They also give a geometric argument that reveals the intuition behind their index lemma proof, showing that combinatorial topology might be deeply connected with the geometry behind Arrow's theorem. After obtaining the result for the base case, \cite{RajsbaumR2022preprint-new,RajsbaumArmajacPODC} proceed by induction to obtain the general one.

The algebraic/combinatorial topology framework can also be used to study Arrovian (im)possibilities under domain restrictions. The author of \cite{Baryshnikov} points out that the $2$-skeleton of the simplicial complex representing the unrestricted domain provides the required information for his algebraic topology approach to be used to prove a generalization of Arrow's theorem to domains with the \emph{free triple property}, i.e., domains that are unrestricted at the level of any triple of alternatives (see \cite{CAMPBELL200235} for a formal definition). He also suggests a relation between possibility results for some domain restrictions and the homotopy type of the simplicial complex representing the preferences with some deleted simplices of maximal dimension.

%we formalize their geometric argument, without the index lemma, and show that it can be combined with other classical techniques in social choice to yield a non-inductive combinatorial topology proof of a generalization of Arrow’s theorem

%\cite{Baryshnikov} uses simplicial complexes to represent the set of profiles and preferences, and simplicial maps to represent SWFs. Rajsbaum and Raventós-Pujol \cite{RajsbaumArmajacPODC} use Baryshnikov's framework to prove Arrow's theorem's base case, i.e., two voters and three alternatives, with combinatorial topology (they used simple counting and geometric arguments instead of more involved algebraic topology). In particular, \cite{RajsbaumArmajacPODC} use a generalization of the \emph{index lemma} in one of their proofs. Having restricted the combinatorial topology approach to the base case, \cite{RajsbaumArmajacPODC} then proceed by induction to prove Arrow's general case (at least two voters and at least three alternatives). This motivated us to look for a simple and geometrically appealing non-inductive combinatorial topology proof of Arrow's theorem for the general case.

The authors of \cite{RajsbaumR2022preprint-new,RajsbaumArmajacPODC} work with arbitrary domain restrictions, restricted to the base case, by eliminating simplices of maximal dimension from the simplicial complex representing the unrestricted domain. However, \cite{RajsbaumR2022preprint-new,RajsbaumArmajacPODC} do not get into the formal details of how the complexes representing the restricted domains are defined. They also do not prove that the combinatorial topology and the classical Arrovian frameworks are equivalent under domain restrictions (in a sense that we will make precise later). For a proof of this equivalence for the case of the unrestricted domain see \cite{RajsbaumR2022preprint-new}. 

We regard the combinatorial topology approach to domain restrictions as promising because  \cite{RajsbaumR2022preprint-new} use it to provide a schematic (instead of an \textit{ad hoc}) characterization of Arrow's theorem for the base case in the context of a broad class of domains. Also, \cite{baryshnikov2024topological} recently applied the algebraic topology approach to prove another very important result in social choice and mechanism design: the Gibbard-Satterthwaite theorem \cite{Gibbard,SATTERTHWAITE1975187}. Therefore, one could expect that a combinatorial topology proof could also be possible. Additionally, combinatorial topology has been very useful in distributed computing \cite{HerlihyFeichtnerKozlovRajsbaum14}, and, as \cite{RajsbaumArmajacPODC} point out, there are interesting analogies between this field and social choice. 
 
Finally, the characterization by \cite{RajsbaumR2022preprint-new} that we just mentioned is based on deleting certain triangles (formally: $2$-simplices), called \textit{critical profiles} by these authors, that live in a torus that is part of the simplicial complex representing the set of all profiles. When one deals with a higher number of alternatives, the dimension of the simplicial complexes grows (the number of voters does not affect the dimension), but one can still look at their $2$-skeletons (intuitively: the two-dimensional level of the whole structure). In the $2$-skeleton, there are toruses isomorphic to the one in the base case. In this paper, we use certain triangles on these toruses that generalize the critical profiles and the connection between the toruses to construct and prove our generalization of Arrow's theorem. By doing this, we show how to use the combinatorial topology approach when facing any (finite) number of alternatives and voters, as opposed to \cite{RajsbaumR2022preprint-new,RajsbaumArmajacPODC}, who only use this approach for the base case.

\subsection{Our Contributions}
 We present a generalization of Arrow's theorem to a class of preferences domains, called the class of \emph{polarization and diversity over triples}, denoted $\mathcal{D}^{\text{PT}}\cap\mathcal{D}^{\text{DT}}$, i.e., if a domain belongs to this class it does not escape Arrow's theorem (in particular, the unrestricted domain belongs to $\mathcal{D}^{\text{PT}}\cap\mathcal{D}^{\text{DT}}$). We provide a combinatorial topology representation of the domains in this class. An arbitrary domain $D$ in $\mathcal{D}^{\text{PT}}\cap \mathcal{D}^{\text{DT}}$ is such that, for any partition of society and every triple of alternatives, we require certain profiles exhibiting the following property: the two coalitions (subsets of voters) given by the partition in question disagree over how to rank the alternatives in two pairs of alternatives of the triple and agree on how to rank the alternatives in the remaining pair. Corresponding to the notion of diversity, when there are at least three voters, domains in $\mathcal{D}^{\text{PT}}\cap\mathcal{D}^{\text{DT}}$ contain a profile that violates the classical property of value-restriction introduced by \cite{SenAmartyaK.1966APTo}. Furthermore, we show that any domain having as a subset any domain in $\mathcal{D}^{\text{PT}}\cap\mathcal{D}^{\text{DT}}$ also fails to escape Arrow's theorem. Following \cite{FISHBURNP.C1997Sadw}, a domain $D$ is \emph{super-Arrovian} if it does not escape Arrow's theorem and satisfies that for every domain $D'$ such that $D\subseteq D'$, $D'$ does not escape Arrow's theorem too. So, $\mathcal{D}^{\text{PT}}\cap\mathcal{D}^{\text{DT}}$ is a class of super-Arrovian domains.

We formalize the combinatorial topology framework used by \cite{RajsbaumR2022preprint-new,RajsbaumArmajacPODC} to deal with domain restrictions. We prove that this yields a combinatorial topology framework equivalent to the classical Arrovian one in a sense captured by Theorem \ref{thrm:equivalence} in Section \ref{sec:transition_comb_top}.

In the context of at least two voters and at least three alternatives, we provide a combinatorial topology proof of our generalization of Arrow's theorem. This proof involves a formalization and generalization (in particular, to any number of voters) of the geometric argument employed by \cite{RajsbaumR2022preprint-new,RajsbaumArmajacPODC} to prove Arrow's theorem, we will be more specific later. Also, our proof combines the combinatorial topology framework with a classical approach in the Arrovian literature: using \emph{(almost-)decisive} coalitions\footnote{A coalition $G$ is \emph{almost-decisive} if whenever every voter in $G$ prefers an alternative $x$ over an alternative $y$ and every voter in $G^c$ (the complement of $G$ w.r.t. the society) prefers $y$ over $x$, then the social preference ranks $x$ over $y$. A coalition is \emph{decisive} if whenever every voter in $G$ prefers $x$ over $y$, the social preference ranks $x$ over $y$ no matter the preferences of the voters in $G^c$.} together with ultrafilters from set theory.

\subsection{Additional Related Work}

As we said before, we do not know of a complete characterization of the domains that escape Arrow's impossibility. However, this problem is solved for \emph{common}~\cite{KALAI1977457} and \emph{Cartesian domains}~\cite{EssentialBlair}\footnote{Following \cite{LEBRETON2011191}, a domain $D$ is \emph{Cartesian} if it can be written as the Cartesian product of sets of admissible preferences, one for each voter; it is \emph{common} if every voter has the same set of admissible preferences.}. On the side of impossibility results, \cite{KalaiMullerSatterthwaite} generalize Arrow's theorem in the context of common domains by showing that if a domain is \emph{saturating} (see \cite{KalaiMullerSatterthwaite} for the definition), then it cannot escape Arrow's theorem. Moreover, \cite{CAMPBELL200235} show that if a domain satisfies the \emph{chain property} (see \cite{CAMPBELL200235} for the definition), then it cannot escape Arrow's theorem. Both the saturating condition and the chain property imply the existence of at least one free triple, while the class $\mathcal{D}^{\text{PT}}\cap\mathcal{D}^{\text{DT}}$ does not. 

For the case of weak orders, \cite{kelly1994free} generalizes the free triple property to a condition called \emph{k-flexible triple}, $k$-FT, (see \cite{kelly1994free} for the definition). As \cite{kelly1994free} says, the free triple property is equivalent to $13$-FT assumption (the lesser the $k$, the weaker is the assumption). However, as \cite{kelly1994free} points out, most proofs of Arrow's theorem (by the time \cite{kelly1994free} was published) only require the $4$-FT assumption. Moreover, \cite{kelly1994free} argues that $3$-FT is enough, but $2$-FT is not. Adapting the $k$-FT definition to strict orders, it would be easy to show that $k$-FT is stronger than $\mathcal{D}^{\text{PT}}\cap\mathcal{D}^{\text{DT}}$ for $k\ge 4$ and that these two conditions are logically unrelated for $n\ge 4$ voters if $1\le k\le 3$. Saturating domains, the chain property, $k$-FT, and $\mathcal{D}^{\text{PT}}\cap\mathcal{D}^{\text{DT}}$ are all properties defined at the level of triples of alternatives.

Our proof that the domains in $\mathcal{D}^{\text{PT}}\cap\mathcal{D}^{\text{DT}}$ cannot escape Arrow's impossibility shares ideas of well-known proofs of Arrow theorem based on (almost-)decisive coalitions and ultrafilters \cite{CAMPBELL200235,Hansson,KIRMAN1972267}. As noted by \cite[p. 49]{CAMPBELL200235}, the ultrafilter approach in turn contains the same elements as Arrow's proof in \cite{ArrowKennethJoseph1963Scai} arranged in a different manner. One main idea in \cite{CAMPBELL200235,Hansson,KIRMAN1972267} is to fix an arbitrary SWF $F$ satisfying the assumptions of interest and show that the set of (almost-)decisive coalitions w.r.t. $F$ forms an ultrafilters w.r.t. the set of all voters. Then, the existence of a dictator can be deduced. However, the proofs of \cite{CAMPBELL200235,Hansson,KIRMAN1972267} use the classical Arrovian framework, as opposed to a combinatorial topology one. Also, the proofs of  \cite{Hansson,KIRMAN1972267} assume the unrestricted domain, the proof of \cite{CAMPBELL200235} domains satisfying the chain property, and the proof of our generalization of Arrow's theorem assumes domains in $\mathcal{D}^{\text{PT}}\cap\mathcal{D}^{\text{DT}}$. Our proof of Lemma \ref{lem:contagioLemma} in Section \ref{sec:generalization} is similar in spirit to the \say{local approach} started by \cite{KalaiMullerSatterthwaite} (see \cite{LEBRETON2011191} for more on this approach).

The authors of \cite{DASGUPTAS1999Otso,FISHBURNP.C1997Sadw} work on finding super-Arrovian domains of minimal cardinality. In particular, \cite{FISHBURNP.C1997Sadw} characterizes super-Arrovian domains in the context of domains that do not escape Arrow's theorem. However, we did not use this characterization to show that the domains in $\mathcal{D}^{\text{PT}}\cap\mathcal{D}^{\text{DT}}$ are super-Arrovian. Also, some of the domains that appear in \cite{DASGUPTAS1999Otso,FISHBURNP.C1997Sadw} are very similar in spirit to the domains in $\mathcal{D}^{\text{PT}}\cap\mathcal{D}^{\text{DT}}$ or belong to this class. We will be more specific about this in Section \ref{sec:generalization}.

Lastly, as we said before, our motivation to use the word \say{diversity} in the name of $\mathcal{D}^{\text{DT}}\cap \mathcal{D}^{\text{PT}}$ comes from fact that, when there are at least three voters, its definition implies the violation of the value-restriction condition. The concept of diversity has also been explicitly used in works dealing with the \emph{single-profile model} (i.e. domains with only one profile), e.g. \cite{FeldmanSerrano2008}, instead of the more traditional multiple-profiles model that we use.

%Our motivation to use the word \say{diversity} in the name of the class $\mathcal{D}^{\text{PT}}\cap\mathcal{D}^{\text{DT}}$ comes from the fact that, when there are at least three voters, its definition implies the violation of the value-restriction condition. The concept of diversity has also been explicitly used in works dealing with the \emph{single-profile model} (i.e., domains with only one profile), e.g. \cite{FeldmanSerrano2008}, instead of the more traditional multiple-profiles model that we use. Value-restriced preferences play an important role in social choice: they are extended into a probabilistic framework \cite{Regenwetter1998} and into the \emph{net value-restricted} preferences \cite{Feld1986,Gaertner1978}. Moreover, Regenwetter et al. \cite{Regenwetter2003} show that value-restricted preferences in an electoral framework do not fit the empirical data, but net value-restricted preferences do. Hence, it is reasonable to suppose a violation of value-restriction. 

\section{Preliminaries}

\subsection{Preference Domains}\label{sec:pref-dom}

Let $X$ be a set of alternatives and $\{1,\dots,n\}$ a set of voters, also denoted $N$. A \emph{coalition} is a subset of $N$. Throughout, we assume that $|X|\ge 3$ and $n \ge 2$. Let $Y\subseteq X$. We denote by $W(Y)$ the set of all strict total orders on $Y$. If $(x,y)$ is an order pair of alternatives in $Y$, we denote it as $xy$. We might denote $xy\in P$ as $xPy$. If $P\in W(Y)$, we can write $P$ as $x_1x_2\dots x_{|Y|}$, where $x_iPx_j$ if and only if  $i<j$. A \emph{preference profile} (or just \emph{profile}) $\vec{P}$ on $Y$ is an $n$-tuple of preferences $(P_1,\dots,P_n)$, where $P_i$ is interpreted as the stated preference ranking of voter $i$ over the alternatives in $Y$. We read $xP_iy$ as \say{voter $i$ prefers alternative $x$ over alternative $y$}. Let $W(Y)^n$ be the set of all preference profiles on $Y$. A \emph{preference domain} $D$ (or just \emph{domain} if no confusion can arise) is a non-empty subset of $W(X)^n$.

Let $P\in W(X)$. The \emph{restriction of $P$ to $Y$}, denoted $P|_Y$, is a strict total order on $Y$ defined in the following way: for all $x,y\in Y$, we have that $xP|_Yy$ iff $xPy$. If $\vec{P}\in W(X)^n$, the
\emph{restriction of $\vec{P}$ to $Y$}, denoted $\vec{P}|_Y$ is the profile $(P_1|_Y,\dots,P_n|_Y)\in W(Y)^n$. As in \cite{CAMPBELL200235}, we denote by $D|_Y$ the set $\{\vec{P}|_Y\colon \vec{P}\in D\}$. If $\vec{P}$ belongs to $D|_Y$ we say that $\vec{P}$ is a \emph{subprofile of} $D$. If $\vec{P}\in W(Y)^n$ and $\vec{P}'\in D$, we say that $\vec{P}$ is a \emph{subprofile of} $\vec{P}'$ if $\vec{P}=\vec{P}'|_Y$, in which case $\vec{P}$ is a subprofile of $D$ (w.r.t. $Y$).

\subsection{Social Welfare Functions}\label{sec:SWFs}

Let $D$ be a preference domain. A \emph{social welfare function} (\emph{SWF} or \emph{SWFs} for plural) is a function of the form $F\colon D\to W(X)$. In words, a SWF is a function that assigns to each profile of preferences in $D$ a strict total order, which is commonly referred as the \emph{social preference}.

Now we define some desirable properties that we would like SWFs to have.
Let $x,y\in X$ and $\vec{P}, \vec{P}' \in D$. A SWF $F$ satisfies:
\begin{itemize}
\item \emph{unanimity} if we have the following: if for all $i\in N$ we have that $xP_iy$, then $x F(\vec{P}) y$. 
\item \emph{independence of irrelevant alternatives (IIA)} if $\vec{P}|_{\{x,y\}}=\vec{P}'|_{\{x,y\}}$ implies $F(\vec{P})|_{\{x,y\}}=F(\vec{P}')|_{\{x,y\}}$. 
\item \emph{non-dictatorship} if there is not a voter $i\in N$ such that $x F(\vec{P}) y$ whenever $x P_i y$ . Such a voter is called a \emph{dictator}. So, a SWF satisfies non-dictatorship if there is no dictator. If there is a dictator for $F$, then $F$ is said to be \emph{dictatorial}.
\end{itemize}

We might refer to SWFs satisfying IIA as \emph{IIA-SWFs}. The following is a strict total orders version of Arrow's theorem, which is a common in the social choice literature:

\begin{theorem}[Arrow's impossibility theorem] \label{arrow_thm}
If $|X|\ge 3$, every unanimous IIA-SWF with domain $W(X)^n$ must be dictatorial. 
\end{theorem}

As we mentioned in the introduction, one way of escaping Arrow's theorem is by allowing for SWFs defined on preference domains other than the unrestricted one, $W(X)^n$. We now want to give names to the domains that escape Arrow's theorem and those that do not.
Let $D$ be a domain and remember that $|X|\ge 3$. Following  \cite{LEBRETON2011191}, we say that $D$ is an \emph{Arrow-inconsistent} domain if we have that any unanimous IIA-SWF defined on $D$ must be a dictatorship. Also following 
\cite{LEBRETON2011191}, we say that $D$ is an \emph{Arrow-consistent} domain if there exists a non-dictatorial unanimous IIA-SWF defined on $D$. We next introduce the notion of value-restricted preferences proposed by \cite{SenAmartyaK.1966APTo}, as it is relevant for our results.

\begin{definition} 
A profile $\vec{P}$ on $X$ is \emph{value-restricted} over $Y\subseteq X$ if for every triple of distinct alternatives $\alpha,\beta,\gamma\in Y$, at least one of the three alternatives is never placed as the most-preferred, the middle-preferred or the least-preferred in the individual rankings of $\{\alpha,\beta,\gamma\}$ induced by $\vec{P}$.
\end{definition}

\subsection{Ultrafilters}\label{sec:ultrafilters}

As we said in the introduction, some of the proofs of Arrow's theorem mix decisive or almost-decisive coalitions with ultrafilters. Since we plan to provide a proof that does this, we introduce the concept of ultrafilter.

\begin{definition} \label{def:ultrafilter}
    An \emph{ultrafilter} is a non-empty collection $\mathcal{U}$ of subsets of a set $A$ that satisfies three conditions:
\begin{enumerate}
    \item The empty set, $\varnothing$, does not belong to $\mathcal{U}$.
    \item If $B\subseteq A$, then $B\in \mathcal{U}$ or $B^c\in \mathcal{U}$.
    \item If $B, B'\in \mathcal{U}$, then $B\cap B'\in \mathcal{U}$.
\end{enumerate}
\end{definition}

The following result has been used in proofs that use ultrafilters to prove Arrow's theorem or some generalization of it, like those of \cite{KIRMAN1972267} and \cite{CAMPBELL200235}, respectively. We will also use it to prove some of our results.

\begin{theorem}\label{dictator_in_ultra}
    If $\mathcal{U}$ is an ultrafilter of a finite set $A$, then there is some $a\in A$ such that $\mathcal{U}=\{B\subseteq A\colon a\in B\}$ 
\end{theorem}

The proof can be found in \cite{CAMPBELL200235}. For more references on ultrafilters see \cite{bourbaki1995general}.

\subsection{Simplicial Complexes and Maps}

For the definitions in this section, we mostly follow \cite{HerlihyFeichtnerKozlovRajsbaum14} with some slight adaptations.

A collection $K$ of finite and non-empty subsets of a set $V$ is an (\emph{abstract}) \emph{simplicial complex} if the following condition is satisfied: if $s\in K$ and $t$ is a non-empty subset of $s$, then $t\in S$. 

Let $K$ be a simplicial complex w.r.t. a set $V$. A \emph{vertex} is an element of $V$. The set of all vertices of $K$, i.e., $V$, can also be denoted $V(K)$.
A \emph{simplex} is an element of $K$. The \emph{dimension of a simplex} $s$, $\text{dim}(s)$, is the number $|s|-1$. A $k$\emph{-simplex} is a simplex of dimension $k$. A simplex $t$ is a \emph{face} of $s$ if $t\subseteq s$. A simplex $s$ in $K$ is a \emph{facet} if it is maximal w.r.t. inclusion, i.e., if there is no simplex $t$ of $K$ such that $s$ is strictly contained in $t$. The \emph{dimension} of $K$, $\text{dim}(K)$, is the maximum dimension among the dimensions of all its facets. The simplicial complex $K$ is \emph{pure} if all its facets are of the same dimension. 

A simplicial complex $C$ is a \emph{subcomplex} of $K$ if every simplex of $C$ is a simplex of $K$. Let $l$ be a non-negative integer. The $l$\emph{-skeleton} of $K$, $\text{skel}^l(K)$, is the set of simplices of $K$ with dimension at most $l$. It is not hard to see that the $\text{skel}^l(K)$ is a simplicial complex. 

If $K$ and $C$ are simplicial complexes with sets of vertices $V(K)$ and $V(C)$, a \emph{vertex map} is a function of the form $\mu\colon V(K)\to V(C)$. In words: a vertex map is a function that assigns to each vertex of $K$ a vertex of $C$. A vertex map is called a simplicial map if it maps simplices to simplies. Formally, a vertex map $\mu\colon V(K)\to V(C)$ is a \emph{simplicial map} if for all simplices $s$ of $K$, we have that $\mu(s)$ is a simplex of $C$. If $\mu\colon V(K)\to V(C)$ is a simplicial map, we will always abuse notation and denote it $\mu\colon K\to C$. A simplicial map $\mu\colon K\to C$ is \emph{rigid} if for each simplex $s\in K$ it holds that $|s|=|\mu(s)|$. Informally, simplicial map is rigid if it preserves the cardinality of simplices.     

If $K$ is a simplicial complex, a \emph{$m$-labeling} (also \emph{labeling}) is a function of the form $l\colon V(K)\to A$, where $A$ is a set of cardinality $m$. An \emph{$m$-coloring} (also \emph{coloring}), denoted $\chi$, is a m-labeling such that if $u$ and $v$ are two different vertices in some simplex $t$ of $K$, then $\chi (u)\neq \chi (v)$. A \emph{chromatic simplicial complex} is a simplicial complex $K$ together with a coloring $\chi$. If $K$ and $C$ are two chromatic simplicial complexes with $m$-colorings $\chi_K$ and $\chi_C$, respectively, then a simplicial map $\phi\colon K \to C$  is \emph{chromatic} if for every $v\in V(K)$, we have that $\chi_K(v)=\chi_C(\phi(v))$. %Informally, a simplicial map is chromatic if it preserve colors.

\subsection{$W(X)$ as a Simplicial Complex}

 The author of \cite{Baryshnikov} uses two simplicial complexes, denoted $N_{W(X)}$ and $N_{W(X)^n}$, to represent the set of all preferences, $W(X)$, and the unrestricted domain, $W(X)^n$, respectively. To do so, he establishes a bijection between the set of all facets of $N_{W(X)}$ and $W(X)$ and another bijection between the set of all facets of $N_{W(X)^n}$ and $W(X)^n$. In this paper, we follow \cite{Baryshnikov} in representing $W(X)$ as $N_{W(X)}$ (and hence, we present $N_{W(X)}$ as part of this preliminaries' section), but work within a framework that allows us to represent any domain $D$ (not only the unrestricted domain) as a simplicial complex, that we denote $N_D$, and any IIA-SWF defined on $D$ with a chromatic simplicial map from $N_D$ to $N_{W(X)}$. 

We introduce some notation needed to define the simplicial complex $N_{W(X)}$. Let $\sigma\in \{+,-\}$. We define $-\sigma\in \{+,-\}$ as follows: $-\sigma=+$ iff $\sigma=-$.
Let $\alpha, \beta \in X$ and  $U_{\alpha\beta}^{\sigma}$ be the set $\{P\in W(X)\colon\, \alpha P \beta \: \text{iff} \: \sigma=+ \}$. For example, if $X=\{x,y,z\}$, $U_{xy}^+=\{xyz,xzy,zxy\}$, we have $U_{zy}^+=\{zyx, zxy, xzy\}$, $U_{zx}^+=\{zxy, zyx, yzx\}$ and notice that $U_{xy}^+\cap U_{zy}^+\cap U_{zx}^+=\{zxy\}$ and $U_{xy}^+\cap U_{yz}^+\cap U_{zx}^+=\varnothing$. Again for general $|X|\ge 3$, it is easy to see that $U_{\alpha\beta}^+=U_{\beta\alpha}^-$.

\begin{definition}
Let $N_{W(X)}$ be the simplicial complex defined as follows: 
\begin{itemize}
\item its set of vertices, denoted $V(N_{W(X)})$, is 
\begin{align*}
        \left\{U_{\alpha\beta}^{\sigma}:\sigma\in \{+,-\} \text{ and } \alpha, \beta \in X, \alpha \neq \beta \right\}
\end{align*}
\item a non-empty subset $S\subseteq V(N_{W(X)})$, where $S=\{v_1,\dots,v_k\}$, is a $(k-1)$-simplex of $N_{W(X)}$ iff $\bigcap_{i=1}^k v_i \neq \varnothing$.
\end{itemize}
\end{definition}

\begin{figure}[t]
        \centering
        \resizebox{0.42\textwidth}{!}{\begin{tikzpicture}[scale=1, every node/.style={scale=1}]
\begin{scope}[shift={(2.9,-1)}
        , scale=0.5, every node/.style={scale=0.6}, rotate=0
                ]

    \tikzmath{\r = 1.1; }

    \node (A) at  (-90:\r) {};
    \filldraw[black] (A) circle (1.5pt) node[above, xshift=-0.035cm, yshift=0.22cm] {$U_{xy}^+$};
    
    \node (B) at (-210:\r) {};
    \filldraw[black] (B) circle (1.5pt) node[below right, xshift=0.05cm, yshift=0.05cm] {$U_{yz}^+$};
    
    \node (C) at (-330:\r) {};
   \filldraw[black] (C) circle (1.5pt) node[below left, xshift=-0.1cm, yshift=0.01cm] {$U_{zx}^+$};
    
    \node (A2) at  (90:4) {};
    \filldraw[black] (A2) circle (1.5pt) node[above] {$U_{xy}^-$};
    
    \node (B2) at (330:4) {};
    \filldraw[black] (B2) circle (1.5pt) node[below] {$U_{yz}^-$};
    
    \node (C2) at (210:4) {};
    \filldraw[black] (C2) circle (1.5pt) node[below] {$U_{zx}^-$};

    \draw [thick] (A) -- (B) -- (C) -- (A);
    \draw [thick] (A2) -- (B2) -- (C2) -- (A2);
    \draw [thick] (B2) -- (A) -- (C2);
    \draw [thick] (A2) -- (C) -- (B2);    
    \draw [thick] (A2) -- (B) -- (C2);

    \draw [fill=red!10, fill opacity = 0.7] (-90:\r) -- (-210:\r) -- (210:4) -- cycle;
    \draw [fill=red!10, fill opacity = 0.7] (-90:\r) -- (330:4) -- (-330:\r) -- cycle;
    \draw [fill=red!10, fill opacity = 0.7] (-90:\r) -- (210:4) -- (330:4) -- cycle;
    \draw [fill=red!10, fill opacity = 0.7] (-330:\r) -- (330:4) -- (90:4) -- cycle;
    \draw [fill=red!10, fill opacity = 0.7] (90:4) -- (-330:\r) -- (-210:\r) -- cycle;
    \draw [fill=red!10, fill opacity = 0.7] (90:4) -- (-210:\r) -- (210:4) -- cycle;
    
    \node[align=center] at (-90:1.7) {$x z y$};
    
    \node[align=center] at (215:1.7) {$x y z$};

    \node[align=center, rotate=60, yshift=-0.25cm] at (-205:1.88) {$y x z$};
    
    \node[align=center] at (90:1.7) {$y z x$};
    
    \node[align=center] at (325:1.7) {$z x y$};
    
    \node[align=center, rotate=-60, yshift=-0.25cm] at (-335:1.88) {$z y x$};

\end{scope}
\end{tikzpicture}}

        \caption{The simplicial complex $N_{W(\{x,y,z\})}$. Adapted from Figure 4 in \cite{RajsbaumR2022preprint-new}, which in turn is adapted from Figure 1 in \cite{RajsbaumArmajacPODC}.} 
        \label{fig:N_Wxyz}
\end{figure}
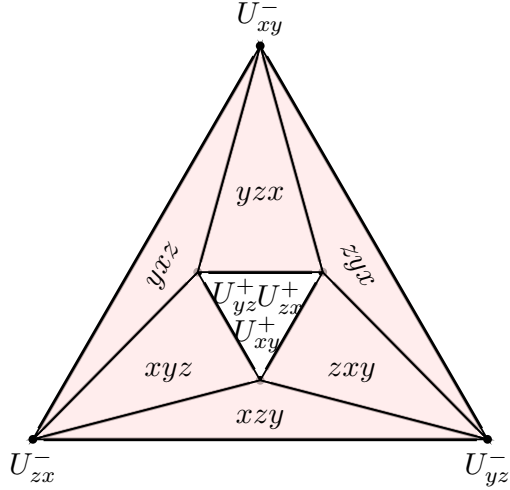

Checking that $N_{W(X)}$ is in fact a simplicial complex is easy. If $X=\{x,y,z\}$, a depiction of $N_{W(X)}$ is shown in Figure \ref{fig:N_Wxyz}. In this figure, the triangle ($2$-simplex) $\{U_{xy}^-, U_{yz}^+, U_{zx}^-\}$ represents the strict total order $yxz$. Notice that this triangle shares an edge ($1$-simplex) with the $yzx$ triangle since $yxz$ and $yzx$ coincide in two pairwise comparisons of alternatives, i.e., they coincide on how they rank $x$ relative to $y$ and $y$ relative to $z$, however they differ in how they rank $x$ relative to $z$. 

In Appendix \ref{sec:appendix}, we prove the result by \cite{Baryshnikov} that there is in fact a bijection from $W(X)$ to the facets of $N_W(X)$.

\section{Domain Restrictions in the Combinatorial Topology Framework}\label{sec:transition_comb_top}

In this section, we define $N_D$ and the chromatic simplicial maps that represent IIA-SWFs on $D$. These objects are straightforward generalizations of $N_{W(X)^n}$ and the chromatic simplicial maps from $N_{W(X)^n}$ to $N_{W(X)}$, and were already used in \cite{RajsbaumR2022preprint-new,RajsbaumArmajacPODC} for the base case and without specifying the formal details of their definition. In Appendix \ref{sec:appendix}, we prove that there are bijections between the subprofiles of $D$ and the simplices of $N_D$ and that there is a bijection from IIA-SWFs on $D$ and chromatic simplicial maps from $N_D$ to $N_{W(X)}$. Using this last bijection, we prove the equivalence between the classical and the combinatorial topology versions of the Arrovian framework in Theorem \ref{thrm:equivalence}.

\subsection{$D$ as a Simplicial Complex}

We want to represent each profile (or even better, any subprofile) in $D$ with a simplex of the simplicial complex $N_{D}$. To illustrate how this representation works, suppose $X=\{w,x,y,z\}$ and $D$ is a domain that has $(xyz, xzy)$ as a subprofile. It is not hard to see that the profile $(xyz, xzy)$ is a subprofile of at most $16$ profiles in $D$ (one of them $(wxyz, xwzy)$ if this profile is in $D$). We represent $(xyz, xzy)$ with a triangle (a $2$-simplex) which is a face of at most $16$ different tetrahedra, one for each of the $16$ profiles possibly in $D$ having $(xyz, xzy)$ as a subprofile. 
%This way of representing subprofiles and simplices defines bijections between collections of these objects, to see these bijections and the proofs that they work, see Appendix \ref{sec:appendix}.

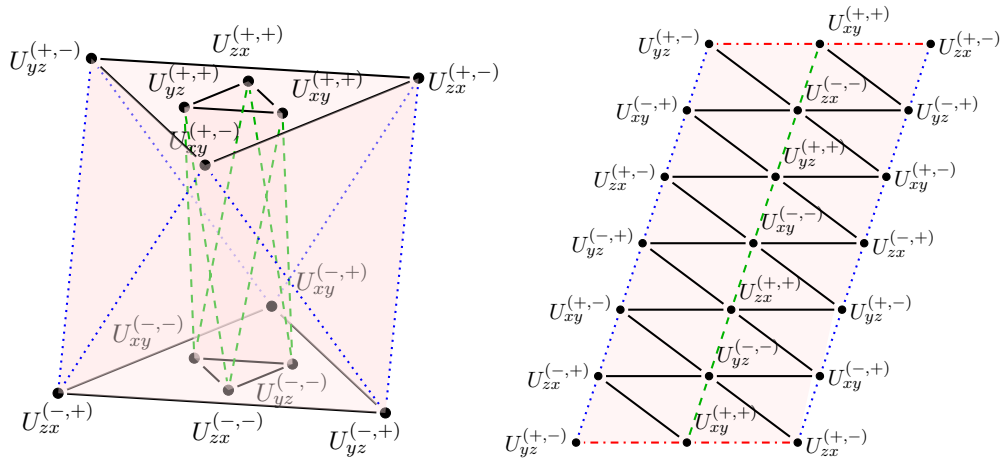
\begin{figure}[ht]
    \centering
\resizebox{0.8\textwidth}{!}{
    \begin{subfigure}{0.4\textwidth}
\tdplotsetmaincoords{70}{110}
\pgfdeclarelayer{background}
\pgfdeclarelayer{foreground}
\pgfsetlayers{background, main, foreground}

\tikzmath{\x = 0.6; \y =2; } 

\begin{tikzpicture}[tdplot_main_coords, scale=1.3, every node/.style={scale=0.9}]

    \tdplotsetrotatedcoords{10}{0}{60}

    \node[tdplot_rotated_coords] (A) at  (\x,0,3) {};
    \filldraw[black] (A) circle (1.5pt) node[above right] {$U_{xy}^{(+,+)}$};

    \node[tdplot_rotated_coords] (B) at ({\x*cos(120)},{\x*sin(120)},3) {};
    \filldraw[black] (B) circle (1.5pt) node[above, yshift=0.25cm] {$U_{zx}^{(+,+)}$};

    \node[tdplot_rotated_coords] (C) at ({\x*cos(-120)},{\x*sin(-120)},3) {};
    \filldraw[black] (C) circle (1.5pt) node[above] {$U_{yz}^{(+,+)}$};

    \node[tdplot_rotated_coords] (A2) at  (-\x,0,0) {};
    \filldraw[black] (A2) circle (1.5pt) node[above left] {$U_{xy}^{(-,-)}$};
    
    \node[tdplot_rotated_coords] (B2) at ({\x*cos(300)},{\x*sin(300)},0) {};
    \filldraw[black] (B2) circle (1.5pt) node[below, yshift=-0.2cm] {$U_{zx}^{(-,-)}$};
    
    \node[tdplot_rotated_coords] (C2) at ({\x*cos(60)},{\x*sin(60)},0) {};
    \filldraw[black] (C2) circle (1.5pt) node[below] {$U_{yz}^{(-,-)}$};

%\begin{pgfonlayer}{background}
    \draw[fill=red!10, fill opacity = 0.4, tdplot_rotated_coords, draw= none] (\x,0,3) -- ({\x*cos(300)},{\x*sin(300)},0) -- ({\x*cos(-120)},{\x*sin(-120)},3) -- (\x,0,3);
    \draw[fill=red!10, fill opacity = 0.4, tdplot_rotated_coords, draw= none] ({\x*cos(120)},{\x*sin(120)},3) -- ({\x*cos(60)},{\x*sin(60)},0) -- (\x,0,3) -- ({\x*cos(120)},{\x*sin(120)},3);
    \draw[fill=red!10, fill opacity = 0.4, tdplot_rotated_coords, draw= none] ({\x*cos(-120)},{\x*sin(-120)},3) -- (-\x,0,0) -- ({\x*cos(120)},{\x*sin(120)},3) -- ({\x*cos(-120)},{\x*sin(-120)},3);
    \draw[fill=red!10, fill opacity = 0.4, tdplot_rotated_coords, draw= none] (-\x,0,0) -- ({\x*cos(120)},{\x*sin(120)},3) -- ({\x*cos(60)},{\x*sin(60)},0) -- (-\x,0,0);
    \draw[fill=red!10, fill opacity = 0.4, tdplot_rotated_coords, draw= none] ({\x*cos(300)},{\x*sin(300)},0) -- ({\x*cos(-120)},{\x*sin(-120)},3) -- (-\x,0,0) -- ({\x*cos(300)},{\x*sin(300)},0);
    \draw[fill=red!10, fill opacity = 0.4, tdplot_rotated_coords, draw= none] ({\x*cos(60)},{\x*sin(60)},0) -- (\x,0,3) -- ({\x*cos(300)},{\x*sin(300)},0) -- ({\x*cos(60)},{\x*sin(60)},0);
%\end{pgfonlayer}

    \draw[thick] (A) -- (B) -- (C) -- (A);
    \draw[thick] (A2) -- (B2) -- (C2) -- (A2);

    \draw[thick, black!30!green, dashed] (B) -- (A2) -- (C) -- (B2) -- (A) -- (C2) -- (B);

%%%%%%%%%%%%%%%%%%%%%%%%%%%%%

    \tdplotsetrotatedcoords{10}{0}{0}

    \node[tdplot_rotated_coords] (Ax) at  (\y,0,3) {};
    \filldraw[black] (Ax) circle (1.5pt) node[above] {$U_{xy}^{(+,-)}$};

    \node[tdplot_rotated_coords] (Bx) at ({\y*cos(120)},{\y*sin(120)},3) {};
    \filldraw[black] (Bx) circle (1.5pt) node[right] {$U_{zx}^{(+,-)}$};

    \node[tdplot_rotated_coords] (Cx) at ({\y*cos(-120)},{\y*sin(-120)},3) {};
    \filldraw[black] (Cx) circle (1.5pt) node[left] {$U_{yz}^{(+,-)}$};

    \node[tdplot_rotated_coords] (Ax2) at  (-\y,0,0) {};
    \filldraw[black] (Ax2) circle (1.5pt) node[above left, xshift=1.6cm] {$U_{xy}^{(-,+)}$};
    
    \node[tdplot_rotated_coords] (Bx2) at ({\y*cos(300)},{\y*sin(300)},0) {};
    \filldraw[black] (Bx2) circle (1.5pt) node[below] {$U_{zx}^{(-,+)}$};
    
    \node[tdplot_rotated_coords] (Cx2) at ({\y*cos(60)},{\y*sin(60)},0) {};
    \filldraw[black] (Cx2) circle (1.5pt) node[below, xshift=-0.3cm] {$U_{yz}^{(-,+)}$};

\begin{pgfonlayer}{foreground}    
    \draw[thick] (Ax) -- (Bx) -- (Cx) -- (Ax);
\end{pgfonlayer}
\begin{pgfonlayer}{background}
    \draw[thick] (Cx2) -- (Ax2) -- (Bx2) -- (Cx2);
\end{pgfonlayer}

\begin{pgfonlayer}{foreground}
    \draw[fill=red!10, fill opacity = 0.4, tdplot_rotated_coords, draw= none] (\y,0,3) -- ({\y*cos(300)},{\y*sin(300)},0) -- ({\y*cos(-120)},{\y*sin(-120)},3) -- (\y,0,3);
    \draw[fill=red!10, fill opacity = 0.4, tdplot_rotated_coords, draw= none] ({\y*cos(120)},{\y*sin(120)},3) -- ({\y*cos(60)},{\y*sin(60)},0) -- (\y,0,3) -- ({\y*cos(120)},{\y*sin(120)},3);
    \draw[fill=red!10, fill opacity = 0.4, tdplot_rotated_coords, draw= none] ({\y*cos(60)},{\y*sin(60)},0) -- (\y,0,3) -- ({\y*cos(300)},{\y*sin(300)},0) -- ({\y*cos(60)},{\y*sin(60)},0);
\end{pgfonlayer}
    
\begin{pgfonlayer}{background}
    \draw[fill=red!10, fill opacity = 0.4, tdplot_rotated_coords, draw= none] ({\y*cos(-120)},{\y*sin(-120)},3) -- (-\y,0,0) -- ({\y*cos(120)},{\y*sin(120)},3) -- ({\y*cos(-120)},{\y*sin(-120)},3);
    \draw[fill=red!10, fill opacity = 0.4, tdplot_rotated_coords, draw= none] (-\y,0,0) -- ({\y*cos(120)},{\y*sin(120)},3) -- ({\y*cos(60)},{\y*sin(60)},0) -- (-\y,0,0);
    \draw[fill=red!10, fill opacity = 0.4, tdplot_rotated_coords, draw= none] ({\y*cos(300)},{\y*sin(300)},0) -- ({\y*cos(-120)},{\y*sin(-120)},3) -- (-\y,0,0) -- ({\y*cos(300)},{\y*sin(300)},0);
\end{pgfonlayer}

\begin{pgfonlayer}{foreground}
    \draw[thick, blue, dotted] (Bx) -- (Cx2) -- (Ax) -- (Bx2) -- (Cx);
\end{pgfonlayer}
\begin{pgfonlayer}{background}
    \draw[thick, blue, dotted] (Bx) -- (Ax2) -- (Cx);
\end{pgfonlayer}

\end{tikzpicture}
    \end{subfigure}
    \begin{subfigure}{0.4\textwidth}
    \tikzmath{\a = 1; \b =1.5; }
\pgfdeclarelayer{background}
\pgfsetlayers{background,main}

\begin{tikzpicture}[scale=1, every node/.style={scale=1}]

\begin{scope}[shift={(0,0)}
        , scale=0.95, every node/.style={scale=0.78}, rotate=71.56505118
                ]
    
    \node (A) at (\a,0) {};
    \filldraw[black] (A) circle (1.5pt) node[right] {$U_{zx}^{(+,-)}$};
    \node (B) at (2*\a,0) {};
    \filldraw[black] (B) circle (1.5pt) node[right] {$U_{xy}^{(-,+)}$};
    \node (C) at (3*\a,0) {};
    \filldraw[black] (3*\a,0) circle (1.5pt) node[right] {$U_{yz}^{(+,-)}$};
    \node (D) at (4*\a,0) {};
    \filldraw[black] (D) circle (1.5pt) node[right] {$U_{zx}^{(-,+)}$};
    \node (E) at (5*\a,0) {};
    \filldraw[black] (E) circle (1.5pt) node[right] {$U_{xy}^{(+,-)}$};
    \node (F) at (6*\a,0) {};
    \filldraw[black] (F) circle (1.5pt) node[right] {$U_{yz}^{(-,+)}$};
    \node (G) at (7*\a,0) {};
    \filldraw[black] (G) circle (1.5pt) node[right] {$U_{zx}^{(+,-)}$};

    \node (A1) at (0.5*\a,\b) {};
    \filldraw[black] (A1) circle (1.5pt) node[above right] {$U_{xy}^{(+,+)}$};
    \node (B1) at (1.5*\a,\b) {};
    \filldraw[black] (B1) circle (1.5pt) node[above right] {$U_{yz}^{(-,-)}$};
    \node (C1) at (2.5*\a,\b) {};
    \filldraw[black] (C1) circle (1.5pt) node[above right] {$U_{zx}^{(+,+)}$};
    \node (D1) at (3.5*\a,\b) {};
    \filldraw[black] (D1) circle (1.5pt) node[above right] {$U_{xy}^{(-,-)}$};
    \node (E1) at (4.5*\a,\b) {};
    \filldraw[black] (E1) circle (1.5pt) node[above right] {$U_{yz}^{(+,+)}$};
    \node (F1) at (5.5*\a,\b) {};
    \filldraw[black] (F1) circle (1.5pt) node[above right] {$U_{zx}^{(-,-)}$};
    \node (G1) at (6.5*\a,\b) {};
    \filldraw[black] (G1) circle (1.5pt) node[above right] {$U_{xy}^{(+,+)}$};
    
    \node (A2) at (0,2*\b) {};
    \filldraw[black] (A2) circle (1.5pt) node[left] {$U_{yz}^{(+,-)}$};
    \node (B2) at (1*\a,2*\b) {};
    \filldraw[black] (B2) circle (1.5pt) node[left] {$U_{zx}^{(-,+)}$};
    \node (C2) at (2*\a,2*\b) {};
    \filldraw[black] (C2) circle (1.5pt) node[left] {$U_{xy}^{(+,-)}$};
    \node (D2) at (3*\a,2*\b) {};
    \filldraw[black] (D2) circle (1.5pt) node[left] {$U_{yz}^{(-,+)}$};
    \node (E2) at (4*\a,2*\b) {};
    \filldraw[black] (E2) circle (1.5pt) node[left] {$U_{zx}^{(+,-)}$};
    \node (F2) at (5*\a,2*\b) {};
    \filldraw[black] (F2) circle (1.5pt) node[left] {$U_{xy}^{(-,+)}$};
    \node (G2) at (6*\a,2*\b) {};
    \filldraw[black] (G2) circle (1.5pt) node[left] {$U_{yz}^{(+,-)}$};
    
    \draw[thick] (A) -- (B1) -- (B) -- (C1) -- (C) -- (D1) -- (D) -- (E1) -- (E) -- (F1) -- (F) -- (G1);
    \draw[thick] (A1) -- (B2) -- (B1) -- (C2) -- (C1) -- (D2) -- (D1) -- (E2) -- (E1) -- (F2) -- (F1) -- (G2);
    
\begin{pgfonlayer}{background}    
    \draw[fill=red!10, fill opacity = 0.4, draw= none] (\a,0) -- (1.5*\a,\b) -- (B) -- (\a,0);
    \draw[fill=red!10, fill opacity = 0.4, draw= none] (2*\a,0) -- (2.5*\a,\b) -- (C) -- (2*\a,0);
    \draw[fill=red!10, fill opacity = 0.4, draw= none] (C) -- (3.5*\a,\b) -- (4*\a,0) -- (C);
    \draw[fill=red!10, fill opacity = 0.4, draw= none] (4*\a,0) -- (4.5*\a,\b) -- (5*\a,0) -- (4*\a,0);
    \draw[fill=red!10, fill opacity = 0.4, draw= none] (5*\a,0) -- (5.5*\a,\b) -- (6*\a,0) -- (5*\a,0);
    \draw[fill=red!10, fill opacity = 0.4, draw= none] (6*\a,0) -- (6.5*\a,\b) -- (7*\a,0) -- (6*\a,0);
    
    \draw[fill=red!10, fill opacity = 0.4, draw= none] (0.5*\a,\b) -- (\a,0) -- (1.5*\a,\b) -- (0.5*\a,\b);
    \draw[fill=red!10, fill opacity = 0.4, draw= none] (1.5*\a,\b) -- (2*\a,0) -- (2.5*\a,\b) -- (1.5*\a,\b);
    \draw[fill=red!10, fill opacity = 0.4, draw= none] (2.5*\a,\b) -- (3*\a,0) -- (3.5*\a,\b) -- (2.5*\a,\b);
    \draw[fill=red!10, fill opacity = 0.4, draw= none] (3.5*\a,\b) -- (4*\a,0) -- (4.5*\a,\b) -- (3.5*\a,\b);
    \draw[fill=red!10, fill opacity = 0.4, draw= none] (4.5*\a,\b) -- (5*\a,0) -- (5.5*\a,\b) -- (4.5*\a,\b);
    \draw[fill=red!10, fill opacity = 0.4, draw= none] (5.5*\a,\b) -- (6*\a,0) -- (6.5*\a,\b) -- (5.5*\a,\b);
    
    \draw[fill=red!10, fill opacity = 0.4, draw= none] (0.5*\a,\b) -- (1.5*\a,\b) -- (1*\a,2*\b) -- (0.5*\a,\b);
    \draw[fill=red!10, fill opacity = 0.4, draw= none] (1.5*\a,\b) -- (2.5*\a,\b) -- (2*\a,2*\b) -- (1.5*\a,\b);
    \draw[fill=red!10, fill opacity = 0.4, draw= none] (2.5*\a,\b) -- (3.5*\a,\b) -- (3*\a,2*\b) -- (2.5*\a,\b);
    \draw[fill=red!10, fill opacity = 0.4, draw= none] (3.5*\a,\b) -- (4.5*\a,\b) -- (4*\a,2*\b) -- (3.5*\a,\b);
    \draw[fill=red!10, fill opacity = 0.4, draw= none] (4.5*\a,\b) -- (5.5*\a,\b) -- (5*\a,2*\b) -- (4.5*\a,\b);
    \draw[fill=red!10, fill opacity = 0.4, draw= none] (5.5*\a,\b) -- (6.5*\a,\b) -- (6*\a,2*\b) -- (5.5*\a,\b);
    
    \draw[fill=red!10, fill opacity = 0.4, draw= none] (0,2*\b) -- (0.5*\a,\b) -- (1*\a,2*\b) -- (0,2*\b);
    \draw[fill=red!10, fill opacity = 0.4, draw= none] (1*\a,2*\b) -- (1.5*\a,\b) -- (2*\a,2*\b) -- (1*\a,2*\b);
    \draw[fill=red!10, fill opacity = 0.4, draw= none] (2*\a,2*\b) -- (2.5*\a,\b) -- (3*\a,2*\b) -- (2*\a,2*\b);
    \draw[fill=red!10, fill opacity = 0.4, draw= none] (3*\a,2*\b) -- (3.5*\a,\b) -- (4*\a,2*\b) -- (3*\a,2*\b);
    \draw[fill=red!10, fill opacity = 0.4, draw= none] (4*\a,2*\b) -- (4.5*\a,\b) -- (5*\a,2*\b) -- (4*\a,2*\b);
    \draw[fill=red!10, fill opacity = 0.4, draw= none] (5*\a,2*\b) -- (5.5*\a,\b) -- (6*\a,2*\b) -- (5*\a,2*\b);
\end{pgfonlayer}
    
    \draw[thick, black!30!green, dashed] (A1) -- (B1) -- (C1) -- (D1) -- (E1) -- (F1) -- (G1);
    
    \draw[thick, blue, dotted] (A) -- (B) -- (C) -- (D) -- (E) -- (F) -- (G);
    \draw[thick, blue, dotted] (A2) -- (B2) -- (C2) -- (D2) -- (E2) -- (F2) -- (G2);
    
    \draw[thick, red, dashdotted] (A) -- (A1) -- (A2);
    \draw[thick, red, dashdotted] (G) -- (G1) -- (G2);

\end{scope}
\end{tikzpicture}
    \end{subfigure}}
    \caption{The simplicial complex $N_{W(\{x,y,z\})^2}$. It consists of two cylinders on the left joined together by the torus on the right by identifying vertices according to the patterns of the edges. Adapted from Figure 6 in \cite{RajsbaumR2022preprint-new}, which in turn is adapted from Figure 3 in \cite{RajsbaumArmajacPODC}.} 
    \label{fig:NWn}
\end{figure}

Before presenting the formal definition of $N_D$, the reader may want to check how this simplicial complex looks if $X=\{x,y,z\}$, $n=2$ and $D=W(X)^n$. This is depicted in Figure \ref{fig:NWn}. The complex $N_{W(\{x,y,z\})^2}$ has a triangle representing each of the $(3!)^2=36$ profiles in $W(\{x,y,z\})^2$. It consists of two cylinders and a torus glued to the cylinders in a certain way (see the figure's description). One of these cylinders consists of all the profiles of complete agreement (w.r.t. the pairwise comparisons of the alternatives), and the other one of those of complete disagreement. The torus consists of all the profiles that has some degree of disagreement (but not complete). In particular, the triangles on the torus that have a blue-dotted edge are referred by \cite{RajsbaumR2022preprint-new} as \emph{critical profiles}. They play a key role in their characterization of Arrow's theorem for a broad class of domains. These triangles are characterized by the fact that voter $1$ and $2$ differ on how they rank the alternatives in two pairs of alternatives and coincide on how they rank the alternatives in the remaining pair. In Section \ref{sec:generalization}, we will generalize critical profiles to the case of any finite number of voters.

Before proceeding with the formal definition of $N_D$, we introduce some notation (similar, but not exactly analogous, to the one introduced for $N_{W(X)}$). Let $\vec{\sigma}\in \{+,-\}^n$, i.e., $\vec{\sigma}$ is an $n$-tuple whose components are $+$ or $-$ signs. For example, $\vec{\sigma}=(-,+,-,-)$. The $i$-component of $\vec{\sigma}$ is denoted $\vec{\sigma}_i$. We define $-\vec{\sigma}$ as follows: for all $i\in N$, we have that $(-\vec{\sigma})_i=+$ iff $\vec{\sigma}_i=-$. For example if $\vec{\sigma}=(+,-,+)$, then $-\vec{\sigma}=(-,+,-)$.

If $D$ is a preference domain, let $L$ denote the following set of labels:
\begin{align*}
        \left\{U_{\alpha\beta}^{\vec{\sigma}}:\vec{\sigma}\in \{+,-\} \text{ and } \alpha, \beta \in X, \alpha \neq \beta \right\}
        %\bigcup_{\substack{
        %\vec{\sigma}\in \{+,-\}^n \\
        %\alpha, \beta \in X, \alpha \neq \beta }}\{U_{\alpha\beta}^{\vec{\sigma}}\}.
    \end{align*}
For every label $U_{\alpha\beta}^{\vec{\sigma}}\in L$, let $s_D(U_{\alpha\beta}^{\vec{\sigma}})$ denote the set 
\begin{align*}
    \{\vec{P}\in D\colon \text{for all} \: i\in N, \alpha P_i \beta \: \text{iff} \: \vec{\sigma}_i = +\}
\end{align*}

If $D=\{(xwyz,xzyw, xzyw), (yxwz, zwyx, zwxy), (wyxz, wyxz, wxzy)\}$, consider the following examples: \\ $s_D(U_{xy}^{(-,-,+)})=\{(yxwz,zwyx,zwxy), (wyxz, wyxz, wxzy)\}$, $s_D(U_{zx}^{(+,-,-)})=\varnothing$, $s_D(U_{yz}^{(+,-,-)})=\{(xwyz,xzyw, xzyw),(yxwz, zwyx, zwxy)\}$, \\and observe that $s_D(U_{xy}^{(-,-,+)})\cap  s_D(U_{yz}^{(+,-,-)})=\{(yxwz, zwyx, zwxy)\}$.

Notice that for every $\vec{\sigma}$ and every $\alpha,\beta\in X$, $\alpha\neq \beta$, we have $s_D(U_{\beta\alpha}^{-\vec{\sigma}})=s_D(U_{\alpha\beta}^{\vec{\sigma}})$. For our purposes, this fact allows us to treat the element $U_{\alpha\beta}^{\vec{\sigma}}$ of $L$ and the element $U_{\beta\alpha}^{-\vec{\sigma}}$ of $L$ as if they were the same element and write $U_{\alpha\beta}^{\vec{\sigma}}=U_{\beta\alpha}^{-\vec{\sigma}}$. For a formal justification of this, see Appendix \ref{sec:appendix}.

\begin{definition}
    Let $N_D$ denote the simplicial complex defined as follows: 
\begin{itemize}
    \item its set of vertices, denoted $V(N_D)$, is $\{u\in L\colon s_D(u) \neq \varnothing \}$.
    \item a non-empty subset $S\subseteq V(N_D)$, where $S=\{v_1,\dots,v_k\}$, is a $(k-1)$-simplex of $N_D$ iff $\bigcap_{i=1}^k s_D(v_i) \neq \varnothing$.
    \end{itemize}
\end{definition}

As with $N_{W(X)}$, checking that $N_D$ is in fact a simplicial complex is easy\footnote{The construction of $N_D$ is a generalization of the way $N_{W(X)^n}$ is constructed in \cite{Baryshnikov}, but for technical reasons related to allowing for domain restrictions, we introduced a distinction between a label $U_{\alpha\beta}^{\vec{\sigma}}$ and the set $s_D(U_{\alpha\beta}^{\vec{\sigma}})$.}.

\subsection{Social Welfare Functions as Chromatic Simplicial Maps}

Given a IIA-SWF $F$ on a domain $D$, we want to represent $F$ with simplicial map of the form $f\colon N_D \to N_{W(X)}$ that is chromatic w.r.t. the labels involving the alternatives. To be more precise, whenever we say a simplicial map of the form $f\colon N_D\to N_{W(X)}$ is \emph{chromatic} we mean that $f(U_{\alpha\beta}^{\vec{\sigma}})=U_{\alpha\beta}^{\sigma}$. Informally, $f$ preserves the $\alpha\beta$'s labels. %In Appendix \ref{sec:appendix}, we show that there is a bijection between the set of all IIA-SWFs defined on $D$ and the set of all chromatic simplicial maps of the form $f\colon N_D\to N_{W(X)}$. 

We want to define combinatorial topology versions of the unanimity and dictatorship conditions of the classical framework. But to do so, we first introduce some notation. If $G$ is a coalition, let $\vec{\sigma}^G$ denote the element of $\{+,-\}^n$ such that $\sigma^{G}_i=+$ iff $i\in G$. For instance, if $N=\{1,2,3\}$ and $G=\{1,3\}$, then $\vec{\sigma}^G$ denotes $(+,-,+)$. In particular, $\vec{\sigma}^N$ denotes the element of $\{+,-\}^n$ such that $\sigma^{N}_i=+$ for all $i\in N$. Analogously, $\vec{\sigma}^{\varnothing}$ denotes the element $\{+,-\}^n$ such that $\sigma^{\varnothing}_i=-$ for all $i\in N$. The following definitions are straightforward generalizations of the corresponding definitions for the unrestricted domain that appear in \cite{RajsbaumR2022preprint-new}. 

\begin{definition}\label{def:simplMap-unanimity-dict}
    Let $f\colon N_D\to N_{W(X)}$ be a chromatic simplicial map. We say that $f$ satisfies \emph{unanimity} if, for all $\alpha,\beta \in X$, we have that if $U_{\alpha\beta}^{\vec{\sigma}^N}$ is a vertex of $N_D$, then $f(U_{\alpha\beta}^{\vec{\sigma}^N})=U_{\alpha\beta}^+$. We say that $f$ is \emph{dictatorial} if there is a voter $i\in N$ such that: for all $\alpha,\beta \in X$, if $U_{\alpha\beta}^{\vec{\sigma}}$ is a vertex of $N_D$, then $f(U_{\alpha\beta}^{\vec{\sigma}})=U_{\alpha\beta}^{\vec{\sigma}_i}$. Such a voter is called a \emph{dictator} for f.
\end{definition}

The following result, whose proof is in Appendix \ref{sec:appendix}, can be interpreted as saying that finding possibility and impossibility results in the combinatorial topology framework is equivalent to finding them in the classical framework. 

\begin{theorem}\label{thrm:equivalence}
A domain $D$ is Arrow-inconsistent iff any chromatic simplicial map of the form $f\colon N_D \to N_{W(X)}$ satisfying unanimity is dictatorial.
\end{theorem}

\section{A Generalization of Arrow's Theorem}\label{sec:generalization}

In this section, we will expose the main contribution of this paper (Theorem \ref{thrm:arrow-generalization}). It is a generalization of Arrow's theorem to a class of domains, denoted $\mathcal{D}^{\text{PT}}\cap\mathcal{D}^{\text{DT}}$, satisfying a property called \emph{polarization and diversity over triples}. In Subsection \ref{subsec:pol-div-triples}, we define this class, which will be obtained by taking the intersection of two other classes that we will also define in this subsection, namely the class of \emph{polarization over triples}, $\mathcal{D}^{\text{PT}}$, and the class of \emph{diversity over triples}, $\mathcal{D}^{\text{DT}}$. In Subsection \ref{subsec:imp-proof}, we prove that $\mathcal{D}^{\text{PT}}\cap\mathcal{D}^{\text{DT}}$ in fact generalizes Arrow's theorem. 

\subsection{Polarization and Diversity over Triples}\label{subsec:pol-div-triples}

We first want to introduce the class $\mathcal{D}^{\text{PT}}$. To do so, we start by defining the notion of polarized profiles. Such profiles are explicitly used in a proof by \cite[Lemma 7 on p. 527]{DASGUPTAS1999Otso}, although not with this name.

\begin{definition}
A profile $\vec{P}$ on $Y\subseteq X$ is \emph{polarized} if there exist $P,P'\in W(Y)$, and a non-empty coalition $G$ distinct from $N$, such that $P_i=P$ for all $i\in G$ and $P_j=P'$ for all $j\in G^c$. We denote such a $\vec{P}$ as $(G\colon P, G^c\colon P')$.  
\end{definition}

For example, if $n=5$, $X=\{x,y,z\}$ and $G=\{1,4\}$, the profile $(xyz, yzx, \allowbreak yzx, xyz, yzx)$ is a polarized profile and can be denoted as $(G\colon xyz,G^c\colon yzx)$, the idea being to communicate that every voter in $G$ has $xyz$ as their ranking and every voter outside $G$ has $yzx$ as their ranking.

\begin{remark}\label{rmk:symmetric_polarized}
Of course, given a polarized profile $(G\colon P, G^c\colon P')$ w.r.t. a coalition $G$, we have that $(G\colon P', G^c\colon P)$ is also a polarized profile w.r.t. $G$.
\end{remark}

Certain polarized profiles over triples of alternatives are relevant to our results. For $n=2$, these profiles are called critical profiles by \cite{RajsbaumR2022preprint-new}, but we will call them \emph{strongly polarized} profiles in our more general setting. 

\begin{definition}\label{def:polarized-profiles}
    Let $Y\subseteq X$, such that $|Y|=3$, and $\vec{P}=(G\colon P, G^c\colon P')$ a polarized profile on $Y$. The profile $\vec{P}$ is \emph{strongly polarized} if $P$ and $P'$ differ on how they rank two different pairs of alternatives and coincide on how they rank the remaining pair of alternatives.
\end{definition}

\begin{remark}\label{rmk:12stronglyPolarized}
    For a given coalition $G$ and set $Y\subseteq X$, such that $|Y|=3$, there are exactly $12$ strongly polarized profiles on $Y$.  
\end{remark}

Now we define two sets of strongly polarized profiles that are going to be the basis to construct the class of domains $\mathcal{D}^{\text{PT}}$. These sets appeared in \cite[Lemma 2 on p. 87]{FISHBURNP.C1997Sadw} for the case of $3$ alternatives and $n\in \{2,3\}$ voters.  

\begin{definition}\label{def:B1}
Let $G$ be a non-empty coalition distinct from $N$ and $\{\alpha,\beta,\gamma\}\subseteq X$, $\alpha\neq \beta\neq \gamma\neq \alpha$. Let $D_1(G,\{\alpha,\beta,\gamma\})$ denote the set of preferences
\begin{align*}
   &\{(G\colon \beta\gamma\alpha, G^c\colon \alpha\beta\gamma), (G\colon \beta\alpha\gamma, G^c\colon \alpha\gamma\beta), (G\colon \alpha\beta\gamma, G^c\colon \gamma\alpha\beta),\\
    &(G\colon \alpha\gamma\beta G^c\colon \gamma\beta\alpha)
    ,(G\colon \gamma\alpha\beta, G^c\colon \beta\gamma\alpha),
    (G\colon \gamma\beta\alpha, G^c\colon \beta\alpha\gamma)\};
\end{align*}
and $D_2(G,\{\alpha,\beta,\gamma\})$ denote the set of preferences 
\begin{align*}
    &\{(G\colon \alpha\beta\gamma, G^c\colon \beta\gamma\alpha),(G\colon \alpha\gamma\beta, G^c\colon \beta\alpha\gamma),(G\colon \gamma\alpha\beta, G^c\colon \alpha\beta\gamma),\\
    &(G\colon \gamma\beta\alpha G^c\colon \alpha\gamma\beta),(G\colon \beta\gamma\alpha, G^c\colon \gamma\alpha\beta),(G\colon \beta\alpha\gamma, G^c\colon \gamma\beta\alpha)\}.
\end{align*}  
\end{definition}

\begin{figure}[ht]
\centering
\resizebox{1\textwidth}{!}{
    \begin{subfigure}{0.49\textwidth}
    \centering
    \tikzmath{\a = 1; \b =1.5; }
    \pgfdeclarelayer{background}
    \pgfsetlayers{background,main}

    \contourlength{0.1pt}
    \contournumber{10}

    \begin{tikzpicture}[scale=1, every node/.style={scale=1}]

    \begin{scope}[shift={(0,0)}
            , scale=1.1, every node/.style={scale=0.9}, rotate=71.56505118
                    ]
%\begin{comment}    
    \node (A) at (\a,0) {};
    \filldraw[black, opacity = 0.1] (A) circle (1.5pt) node[right, opacity = 0] {$U_{\gamma\alpha}^{(+,-)}$};
    \node (B) at (2*\a,0) {};
    \filldraw[black, opacity = 0.1] (B) circle (1.5pt) node[right, opacity = 0] {$U_{\alpha\beta}^{(-,+)}$};
    \node (C) at (3*\a,0) {};
    \filldraw[black, opacity = 0.1] (3*\a,0) circle (1.5pt) node[right, opacity = 0] {$U_{\beta\gamma}^{(+,-)}$};
    \node (D) at (4*\a,0) {};
    \filldraw[black, opacity = 0.1] (D) circle (1.5pt) node[right, opacity = 0] {$U_{\gamma\alpha}^{(-,+)}$};
    \node (E) at (5*\a,0) {};
    \filldraw[black, opacity = 0.1] (E) circle (1.5pt) node[right, opacity = 0] {$U_{\alpha\beta}^{(+,-)}$};
    \node (F) at (6*\a,0) {};
    \filldraw[black, opacity = 0.1] (F) circle (1.5pt) node[right, opacity = 0] {$U_{\beta\gamma}^{(-,+)}$};
    \node (G) at (7*\a,0) {};
    \filldraw[black, opacity = 0.1] (G) circle (1.5pt) node[right, opacity = 0] {$U_{\gamma\alpha}^{(+,-)}$};
%\end{comment}

    \node (A1) at (0.5*\a,\b) {};
    \filldraw[black] (A1) circle (1.5pt) node[below right, yshift=0.05cm] {\tiny $U_{\alpha\beta}^{\vec{\sigma}^N}$};
    \node (B1) at (1.5*\a,\b) {};
    \filldraw[black] (B1) circle (1.5pt) node[below right, rotate=-35, yshift=0.05cm] {\tiny $U_{\beta\gamma}^{\vec{\sigma}^{\varnothing}}$};
    \node (C1) at (2.5*\a,\b) {};
    \filldraw[black] (C1) circle (1.5pt) node[below right, rotate=-35, yshift=0.05cm] {\tiny $U_{\gamma\alpha}^{\vec{\sigma}^N}$};
    \node (D1) at (3.5*\a,\b) {};
    \filldraw[black] (D1) circle (1.5pt) node[below right, rotate=-35, yshift=0.05cm] {\tiny $U_{\alpha\beta}^{\vec{\sigma}^{\varnothing}}$};
    \node (E1) at (4.5*\a,\b) {};
    \filldraw[black] (E1) circle (1.5pt) node[below right, rotate=-35, yshift=0.05cm] {\tiny $U_{\beta\gamma}^{\vec{\sigma}^N}$};
    \node (F1) at (5.5*\a,\b) {};
    \filldraw[black] (F1) circle (1.5pt) node[below right, rotate=-35, yshift=0.05cm] {\tiny $U_{\gamma\alpha}^{\vec{\sigma}^{\varnothing}}$};
    \node (G1) at (6.5*\a,\b) {};
    \filldraw[black] (G1) circle (1.5pt) node[below right, rotate=-35, yshift=0.05cm] {\tiny $U_{\alpha\beta}^{\vec{\sigma}^N}$};
    
    \node (A2) at (0,2*\b) {};
    \filldraw[black] (A2) circle (1.5pt) node[left] {$U_{\beta\gamma}^{\vec{\sigma}^G}$};
    \node (B2) at (1*\a,2*\b) {};
    \filldraw[black] (B2) circle (1.5pt) node[left] {$U_{\gamma\alpha}^{\vec{\sigma}^{G^c}}$};
    \node (C2) at (2*\a,2*\b) {};
    \filldraw[black] (C2) circle (1.5pt) node[left] {$U_{\alpha\beta}^{\vec{\sigma}^G}$};
    \node (D2) at (3*\a,2*\b) {};
    \filldraw[black] (D2) circle (1.5pt) node[left] {$U_{\beta\gamma}^{\vec{\sigma}^{G^c}}$};
    \node (E2) at (4*\a,2*\b) {};
    \filldraw[black] (E2) circle (1.5pt) node[left] {$U_{\gamma\alpha}^{\vec{\sigma}^G}$};
    \node (F2) at (5*\a,2*\b) {};
    \filldraw[black] (F2) circle (1.5pt) node[left] {$U_{\alpha\beta}^{\vec{\sigma}^{G^c}}$};
    \node (G2) at (6*\a,2*\b) {};
    \filldraw[black] (G2) circle (1.5pt) node[left] {$U_{\beta\gamma}^{\vec{\sigma}^G}$};

%\begin{comment}    
    \draw[thick, opacity = 0.1] (A) -- (B1) -- node[below, opacity = 0] {\scalebox{.5}{$G\colon \gamma\beta\alpha$, $G^c\colon \alpha\gamma\beta$}} node[above, opacity = 0] {\scalebox{.2}{$G\colon \gamma\beta\alpha$, $\gamma\alpha\beta$}} (B) -- (A);
    \draw[thick, opacity = 0.1] (B) -- (C1) -- node[below, opacity = 0] {\tiny $\beta\gamma\alpha$, $\gamma\alpha\beta$} node[above, opacity = 0] {\tiny $\beta\gamma\alpha$, $\gamma\beta\alpha$} (C) -- (B);
    \draw[thick, opacity = 0.1] (C) -- (D1) -- node[below, opacity = 0] {\tiny $\beta\alpha\gamma$, $\gamma\beta\alpha$} node[above, opacity = 0] {\tiny $\beta\alpha\gamma$, $\beta\gamma\alpha$ }(D) -- (C);
    \draw[thick, opacity = 0.1] (D) -- (E1) -- node[below, opacity = 0] {\tiny $\alpha\beta\gamma$, $\beta\gamma\alpha$} node[above, opacity = 0] {\tiny $\alpha\beta\gamma$, $\beta\alpha\gamma$} (E) -- (D);
    \draw[thick, opacity = 0.1] (E) -- (F1) -- node[below, opacity = 0] {\tiny $\alpha\gamma\beta$, $\beta\alpha\gamma$} node[above, opacity = 0] {\tiny $\alpha\gamma\beta$, $\alpha\beta\gamma$} (F) -- (E);
    \draw[thick, opacity = 0.1] (F) -- (G1) -- node[below, opacity = 0] {\tiny $\gamma\alpha\beta$, $\alpha\beta\gamma$} (G) -- (F);
    
    \draw[thick, opacity = 0.1] (A1) -- node[above, opacity = 0] {\tiny $\gamma\alpha\beta$, $\alpha\gamma\beta$} (A) -- (B1) -- (A1);
    \draw[thick, opacity = 0.1] (B1) -- (B) -- (C1) -- (B1);
    \draw[thick, opacity = 0.1] (C1) -- (C) -- (D1) -- (C1);
    \draw[thick, opacity = 0.1] (D1) -- (D) -- (E1) -- (D1);
    \draw[thick, opacity = 0.1] (E1) -- (E) -- (F1) -- (E1);
    \draw[thick, opacity = 0.1] (F1) -- (F) -- (G1) -- (F1);
%\end{comment}    
    
    \draw[thick, draw= none] (A1) -- (B1) -- node[below, opacity = 0] {\tiny $\alpha\gamma\beta$, $\gamma\alpha\beta$} node[above,xshift=-0.15cm] {\scalebox{.5}{$G\colon \alpha\gamma\beta, G^c\colon \gamma\beta\alpha$}} (B2) -- (A1);
    \draw[thick, draw= none] (B1) -- (C1) -- node[below, opacity=0] {\tiny $\gamma\alpha\beta$, $\gamma\beta\alpha$} node[above, xshift=-0.15cm] { \scalebox{.5}{$G\colon \gamma\alpha\beta,G^c\colon \beta\gamma\alpha$}} (C2) -- (B1);
    \draw[thick, draw= none] (C1) -- (D1) -- node[below, opacity=0] {\tiny $\gamma\beta\alpha$, $\beta\gamma\alpha$} node[above, xshift=-0.15cm] {\scalebox{.5}{$G\colon \gamma\beta\alpha, G^c\colon \beta\alpha\gamma$}} (D2) -- (C1);
    \draw[thick, draw= none] (D1) -- (E1) -- node[below, opacity = 0] {\tiny $\beta\gamma\alpha$, $\beta\alpha\gamma$} node[above, xshift=-0.15cm] {\scalebox{.5}{$G\colon \beta\gamma\alpha, G^c\colon \alpha\beta\gamma$}}(E2) -- (D1);

    \draw[thick, draw= none] (E1) -- (F1) -- node[below, opacity=0] {\tiny $\beta\alpha\gamma$, $\alpha\beta\gamma$} node[above, xshift=-0.15cm] {\scalebox{.5}{$G\colon \beta\alpha\gamma, G^c\colon \alpha\gamma\beta$} } (F2) -- (E1);
    \draw[thick, draw= none] (F1) -- (G1) -- node[below, opacity=0] {\tiny $\alpha\beta\gamma$, $\alpha\gamma\beta$} (G2) -- (F1);
    
    \draw[thick, draw= none] (A2) -- node[above, xshift=-0.15cm] {\scalebox{0.5}{ $G\colon \alpha\beta\gamma G^c\colon \gamma\alpha\beta$}} (A1) -- (B2) -- (A2);
    \draw[thick, draw= none] (B2) -- (B1) -- (C2) -- (B2);
    \draw[thick, draw= none] (C2) -- (C1) -- (D2) -- (C2);
    \draw[thick, draw= none] (D2) -- (D1) -- (E2) -- (D2);
    \draw[thick, draw= none] (E2) -- (E1) -- (F2) -- (E2);
    \draw[thick, draw= none] (F2) -- (F1) -- (G2) -- (F2);
    
    \draw[thick] (A1) -- (B2) -- (B1) -- (C2) -- (C1) -- (D2) -- (D1) -- (E2) -- (E1) -- (F2) -- (F1) -- (G2);

\begin{pgfonlayer}{background}

    \draw[fill=red!10, fill opacity = 0, draw= none] (0.5*\a,\b) -- (1.5*\a,\b) -- (1*\a,2*\b) -- (0.5*\a,\b);
    \draw[fill=red!10, fill opacity = 0, draw= none] (1.5*\a,\b) -- (2.5*\a,\b) -- (2*\a,2*\b) -- (1.5*\a,\b);
    \draw[fill=red!10, fill opacity = 0, draw= none] (2.5*\a,\b) -- (3.5*\a,\b) -- (3*\a,2*\b) -- (2.5*\a,\b);
    \draw[fill=red!10, fill opacity = 0, draw= none] (3.5*\a,\b) -- (4.5*\a,\b) -- (4*\a,2*\b) -- (3.5*\a,\b);
    \draw[fill=red!10, fill opacity = 0, draw= none] (4.5*\a,\b) -- (5.5*\a,\b) -- (5*\a,2*\b) -- (4.5*\a,\b);
    \draw[fill=red!10, fill opacity = 0, draw= none] (5.5*\a,\b) -- (6.5*\a,\b) -- (6*\a,2*\b) -- (5.5*\a,\b);
    
    \draw[fill=red!10, fill opacity = 0.7, draw= none] (0,2*\b) -- (0.5*\a,\b) -- (1*\a,2*\b) -- (0,2*\b);
    \draw[fill=red!10, fill opacity = 0.7, draw= none] (1*\a,2*\b) -- (1.5*\a,\b) -- (2*\a,2*\b) -- (1*\a,2*\b);
    \draw[fill=red!10, fill opacity = 0.7, draw= none] (2*\a,2*\b) -- (2.5*\a,\b) -- (3*\a,2*\b) -- (2*\a,2*\b);
    \draw[fill=red!10, fill opacity = 0.7, draw= none] (3*\a,2*\b) -- (3.5*\a,\b) -- (4*\a,2*\b) -- (3*\a,2*\b);
    \draw[fill=red!10, fill opacity = 0.7, draw= none] (4*\a,2*\b) -- (4.5*\a,\b) -- (5*\a,2*\b) -- (4*\a,2*\b);
    \draw[fill=red!10, fill opacity = 0.7, draw= none] (5*\a,2*\b) -- (5.5*\a,\b) -- (6*\a,2*\b) -- (5*\a,2*\b);
\end{pgfonlayer}
    
    \draw[thick, opacity=0.1] (A1) -- node[left] {} (B1) -- node[left] {} (C1) -- node[left] {} (D1) -- node[left] {} (E1) -- node[left] {} (F1) -- node[left] {} (G1);
%   \draw[thick, blue] (A) -- (B) -- (C) -- (D) -- (E) -- (F) -- (G);
    \draw[thick, blue, dotted] (A2) -- node[left] {} (B2) -- node[left] {} (C2) -- node[left] {} (D2) -- node[left] {} (E2) -- node[left] {} (F2) -- node[left] {} (G2);
    
    \draw[thick, red, dashdotted] (A1) -- (A2);
    \draw[thick, red, dashdotted] (G1) -- (G2);
    \end{scope}
    \end{tikzpicture}
    \end{subfigure}
    \begin{subfigure}{0.49\textwidth}
    \centering
    \tikzmath{\a = 1; \b =1.5; }
    \pgfdeclarelayer{background}
    \pgfsetlayers{background,main}

    \contourlength{0.1pt}
    \contournumber{10}

    \begin{tikzpicture}[scale=1, every node/.style={scale=1}]

    \begin{scope}[shift={(0,0)}
            , scale=1.1, every node/.style={scale=0.9}, rotate=71.56505118
                    ]
%\begin{comment}    
    \node (A) at (\a,0) {};
    \filldraw[black] (A) circle (1.5pt) node[right] {$U_{\gamma\alpha}^{\vec{\sigma}^G}$};
    \node (B) at (2*\a,0) {};
    \filldraw[black] (B) circle (1.5pt) node[right] {$U_{\alpha\beta}^{\vec{\sigma}^{G^c}}$};
    \node (C) at (3*\a,0) {};
    \filldraw[black] (3*\a,0) circle (1.5pt) node[right] {$U_{\beta\gamma}^{\vec{\sigma}^G}$};
    \node (D) at (4*\a,0) {};
    \filldraw[black] (D) circle (1.5pt) node[right] {$U_{\gamma\alpha}^{\vec{\sigma}^{G^c}}$};
    \node (E) at (5*\a,0) {};
    \filldraw[black] (E) circle (1.5pt) node[right] {$U_{\alpha\beta}^{\vec{\sigma}^G}$};
    \node (F) at (6*\a,0) {};
    \filldraw[black] (F) circle (1.5pt) node[right] {$U_{\beta\gamma}^{\vec{\sigma}^{G^c}}$};
    \node (G) at (7*\a,0) {};
    \filldraw[black] (G) circle (1.5pt) node[right] {$U_{\gamma\alpha}^{\vec{\sigma}^G}$};
%\end{comment}

    \node (A1) at (0.5*\a,\b) {};
    \filldraw[black] (A1) circle (1.5pt) node[below left, yshift=0.05cm] {\tiny $U_{\alpha\beta}^{\vec{\sigma}^N}$};
    \node (B1) at (1.5*\a,\b) {};
    \filldraw[black] (B1) circle (1.5pt) node[below left, yshift=0.05cm] {\tiny $U_{\beta\gamma}^{\vec{\sigma}^{\varnothing}}$};
    \node (C1) at (2.5*\a,\b) {};
    \filldraw[black] (C1) circle (1.5pt) node[below left,  yshift=0.05cm] {\tiny $U_{\gamma\alpha}^{\vec{\sigma}^N}$};
    \node (D1) at (3.5*\a,\b) {};
    \filldraw[black] (D1) circle (1.5pt) node[below left,  yshift=0.05cm] {\tiny $U_{\alpha\beta}^{\vec{\sigma}^{\varnothing}}$};
    \node (E1) at (4.5*\a,\b) {};
    \filldraw[black] (E1) circle (1.5pt) node[below left,  yshift=0.05cm] {\tiny $U_{\beta\gamma}^{\vec{\sigma}^N}$};
    \node (F1) at (5.5*\a,\b) {};
    \filldraw[black] (F1) circle (1.5pt) node[below left, yshift=0.05cm] {\tiny $U_{\gamma\alpha}^{\vec{\sigma}^{\varnothing}}$};
    \node (G1) at (6.5*\a,\b) {};
    \filldraw[black] (G1) circle (1.5pt) node[below left, yshift=0.05cm] {\tiny $U_{\alpha\beta}^{\vec{\sigma}^N}$};
    
    \node (A2) at (0,2*\b) {};
    \filldraw[black, opacity=0.1] (A2) circle (1.5pt);
    \node (B2) at (1*\a,2*\b) {};
    \filldraw[black, opacity=0.1] (B2) circle (1.5pt);
    \node (C2) at (2*\a,2*\b) {};
    \filldraw[black, opacity=0.1] (C2) circle (1.5pt);
    \node (D2) at (3*\a,2*\b) {};
    \filldraw[black, opacity=0.1] (D2) circle (1.5pt);
    \node (E2) at (4*\a,2*\b) {};
    \filldraw[black, opacity=0.1] (E2) circle (1.5pt);
    \node (F2) at (5*\a,2*\b) {};
    \filldraw[black, opacity=0.1] (F2) circle (1.5pt);
    \node (G2) at (6*\a,2*\b) {};
    \filldraw[black, opacity=0.1] (G2) circle (1.5pt);

\draw[thick, opacity=0.1] (A2) -- (A1);
\draw[thick, opacity=0.1] (G2) -- (G1);

%\begin{comment}    
    \draw[thick, draw=none] (A) -- (B1) -- node[below, xshift=0.13cm] {\scalebox{.5}{$G\colon \gamma\beta\alpha, G^c\colon \alpha\gamma\beta$}} node[above, opacity=0] {\scalebox{.5}{$G\colon \gamma\beta\alpha, G^c\colon \gamma\alpha\beta$}} (B) -- (A);
    \draw[thick, draw=none] (B) -- (C1) -- node[below, xshift=0.13cm] {\scalebox{.5}{$G\colon \beta\gamma\alpha, G^c\colon \gamma\alpha\beta$}} node[above, opacity=0] {\scalebox{.5}{$G\colon \beta\gamma\alpha, G^c\colon \gamma\beta\alpha$}} (C) -- (B);
    \draw[thick, draw=none] (C) -- (D1) -- node[below, xshift=0.13cm] {\scalebox{.5}{$G\colon \beta\alpha\gamma, G^c\colon \gamma\beta\alpha$}} node[above, opacity = 0] {\tiny $\beta\alpha\gamma$, $\beta\gamma\alpha$ }(D) -- (C);
    \draw[thick, draw=none] (D) -- (E1) -- node[below, xshift=0.13cm] {\scalebox{.5}{$G\colon \alpha\beta\gamma, G^c\colon \beta\gamma\alpha$}} node[above, opacity = 0] {\tiny $\alpha\beta\gamma$, $\beta\alpha\gamma$} (E) -- (D);
    \draw[thick, draw=none] (E) -- (F1) -- node[below, xshift=0.13cm] {\scalebox{.5}{$G\colon \alpha\gamma\beta, G^c\colon \beta\alpha\gamma$}} node[above, opacity = 0] {\tiny $\alpha\gamma\beta$, $\alpha\beta\gamma$} (F) -- (E);
    \draw[thick, draw=none] (F) -- (G1) -- node[below, xshift=0.13cm] {\scalebox{.5}{$G\colon \gamma\alpha\beta, G^c\colon \alpha\beta\gamma$}} (G) -- (F);
    
    \draw[thick, opacity = 0.1] (A1) --  (A) -- (B1) -- (A1);
    \draw[thick, opacity = 0.1] (B1) -- (B) -- (C1) -- (B1);
    \draw[thick, opacity = 0.1] (C1) -- (C) -- (D1) -- (C1);
    \draw[thick, opacity = 0.1] (D1) -- (D) -- (E1) -- (D1);
    \draw[thick, opacity = 0.1] (E1) -- (E) -- (F1) -- (E1);
    \draw[thick, opacity = 0.1] (F1) -- (F) -- (G1) -- (F1);
%\end{comment}    
    
    \draw[thick, draw= none, opacity=0] (A1) -- (B1) -- node[below, opacity = 0] {\tiny $\alpha\gamma\beta$, $\gamma\alpha\beta$} node[above,xshift=-0.15cm, opacity = 0] {\scalebox{.5}{$G\colon \alpha\gamma\beta, G^c\colon \gamma\beta\alpha$}} (B2) -- (A1);
    \draw[thick, draw= none] (B1) -- (C1) -- node[below, opacity=0] {\tiny $\gamma\alpha\beta$, $\gamma\beta\alpha$} node[above, xshift=-0.15cm, opacity = 0] { \scalebox{.5}{$G\colon \gamma\alpha\beta,G^c\colon \beta\gamma\alpha$}} (C2) -- (B1);
    \draw[thick, draw= none] (C1) -- (D1) -- node[below, opacity=0] {\tiny $\gamma\beta\alpha$, $\beta\gamma\alpha$} node[above, xshift=-0.15cm, opacity = 0] {\scalebox{.5}{$G\colon \gamma\beta\alpha, G^c\colon \beta\alpha\gamma$}} (D2) -- (C1);
    \draw[thick, draw= none] (D1) -- (E1) -- node[below, opacity = 0] {\tiny $\beta\gamma\alpha$, $\beta\alpha\gamma$} node[above, xshift=-0.15cm, opacity = 0] {\scalebox{.5}{$G\colon \beta\gamma\alpha, G^c\colon \alpha\beta\gamma$}}(E2) -- (D1);

    \draw[thick, draw= none] (E1) -- (F1) -- node[below, opacity=0] {\tiny $\beta\alpha\gamma$, $\alpha\beta\gamma$} node[above, xshift=-0.15cm, opacity = 0] {\scalebox{.5}{$G\colon \beta\alpha\gamma, G^c\colon \alpha\gamma\beta$} } (F2) -- (E1);
    \draw[thick, draw= none] (F1) -- (G1) -- node[below, opacity=0] {\tiny $\alpha\beta\gamma$, $\alpha\gamma\beta$} (G2) -- (F1);
    
    \draw[thick, draw= none, opacity=0.1] (A2) -- node[above, xshift=-0.15cm, opacity=0] {\scalebox{0.5}{ $G\colon \alpha\beta\gamma G^c\colon \gamma\alpha\beta$}} (A1) -- (B2) -- (A2);
    \draw[thick, , draw= none, opacity=0.1] (B2) -- (B1) -- (C2) -- (B2);
    \draw[thick, draw= none, opacity = 0.1] (C2) -- (C1) -- (D2) -- (C2);
    \draw[thick, draw= none, opacity = 0.1] (D2) -- (D1) -- (E2) -- (D2);
    \draw[thick, draw= none, opacity = 0.1] (E2) -- (E1) -- (F2) -- (E2);
    \draw[thick, draw= none, opacity = 0.1] (F2) -- (F1) -- (G2) -- (F2);
    
    \draw[thick, opacity=0.1] (A1) -- (B2) -- (B1) -- (C2) -- (C1) -- (D2) -- (D1) -- (E2) -- (E1) -- (F2) -- (F1) -- (G2);

\begin{pgfonlayer}{background}

    \draw[fill=red!10, fill opacity = 0, draw= none] (0.5*\a,\b) -- (1.5*\a,\b) -- (1*\a,2*\b) -- (0.5*\a,\b);
    \draw[fill=red!10, fill opacity = 0, draw= none] (1.5*\a,\b) -- (2.5*\a,\b) -- (2*\a,2*\b) -- (1.5*\a,\b);
    \draw[fill=red!10, fill opacity = 0, draw= none] (2.5*\a,\b) -- (3.5*\a,\b) -- (3*\a,2*\b) -- (2.5*\a,\b);
    \draw[fill=red!10, fill opacity = 0, draw= none] (3.5*\a,\b) -- (4.5*\a,\b) -- (4*\a,2*\b) -- (3.5*\a,\b);
    \draw[fill=red!10, fill opacity = 0, draw= none] (4.5*\a,\b) -- (5.5*\a,\b) -- (5*\a,2*\b) -- (4.5*\a,\b);
    \draw[fill=red!10, fill opacity = 0, draw= none] (5.5*\a,\b) -- (6.5*\a,\b) -- (6*\a,2*\b) -- (5.5*\a,\b);
    
    \draw[fill=red!10, fill opacity = 0, draw= none] (0,2*\b) -- (0.5*\a,\b) -- (1*\a,2*\b) -- (0,2*\b);
    \draw[fill=red!10, fill opacity = 0, draw= none] (1*\a,2*\b) -- (1.5*\a,\b) -- (2*\a,2*\b) -- (1*\a,2*\b);
    \draw[fill=red!10, fill opacity = 0, draw= none] (2*\a,2*\b) -- (2.5*\a,\b) -- (3*\a,2*\b) -- (2*\a,2*\b);
    \draw[fill=red!10, fill opacity = 0, draw= none] (3*\a,2*\b) -- (3.5*\a,\b) -- (4*\a,2*\b) -- (3*\a,2*\b);
    \draw[fill=red!10, fill opacity = 0, draw= none] (4*\a,2*\b) -- (4.5*\a,\b) -- (5*\a,2*\b) -- (4*\a,2*\b);
    \draw[fill=red!10, fill opacity = 0, draw= none] (5*\a,2*\b) -- (5.5*\a,\b) -- (6*\a,2*\b) -- (5*\a,2*\b);

 \draw[fill=red!10, fill opacity = 0.7, draw= none] (6.5*\a,\b) -- (7*\a,0) -- (6*\a,0) -- (6.5*\a,\b);
    \draw[fill=red!10, fill opacity = 0.7, draw= none] (5.5*\a, \b) -- (6*\a,0) -- (5*\a,0) -- (5.5*\a,\b);
    \draw[fill=red!10, fill opacity = 0.7, draw= none] (4.5*\a, \b) -- (5*\a,0) -- (4*\a, 0) -- (4.5*\a,\b);
    \draw[fill=red!10, fill opacity = 0.7, draw= none] (3.5*\a, \b) -- (4*\a,0) -- (3*\a,0) -- (3.5*\a,\b);
    \draw[fill=red!10, fill opacity = 0.7, draw= none] (2.5*\a, \b) -- (3*\a,0) -- (2*\a,0) -- (2.5*\a,\b);
    \draw[fill=red!10, fill opacity = 0.7, draw= none] (1.5*\a, \b) -- (2*\a, 0) -- (1*\a,0) -- (1.5*\a,\b);

\end{pgfonlayer}

    \draw[thick, opacity=0.1] (A1) -- node[left] {} (B1) -- node[left] {} (C1) -- node[left] {} (D1) -- node[left] {} (E1) -- node[left] {} (F1) -- node[left] {} (G1);  
    \draw[thick, blue, dotted] (A) -- (B) -- (C) -- (D) -- (E) -- (F) -- (G);
    \draw[thick, opacity= 0.1] (A2) -- node[left] {} (B2) -- node[left] {} (C2) -- node[left] {} (D2) -- node[left] {} (E2) -- node[left] {} (F2) -- node[left] {} (G2);

    \draw[thick] (A) -- (B1) -- (B) -- (C1) -- (C) -- (D1) -- (D) -- (E1) -- (E) -- (F1) -- (F) -- (G1);
    
    \draw[thick, red, dashdotted] (A) -- (A1);
    \draw[thick, red, dashdotted] (G) -- (G1);

    \end{scope}
    \end{tikzpicture}
    \end{subfigure}}
\caption{On the left, the simplicial complex $B_1(G,\{\alpha,\beta,\gamma\})$ with vertices identified according to the patterns of the edges with dot-dash patterns. The dotted and solid patterns in the other edges of $B_1(G,\{\alpha,\beta,\gamma\})$ are there just for comparison with Figure \ref{fig:NWn}. On the right, the simplicial complex $B_2(G,\{\alpha,\beta,\gamma\})$ with patterns indicating analogous information to the one for the drawing on the left.}
\label{fig:B1-B2}
\end{figure}
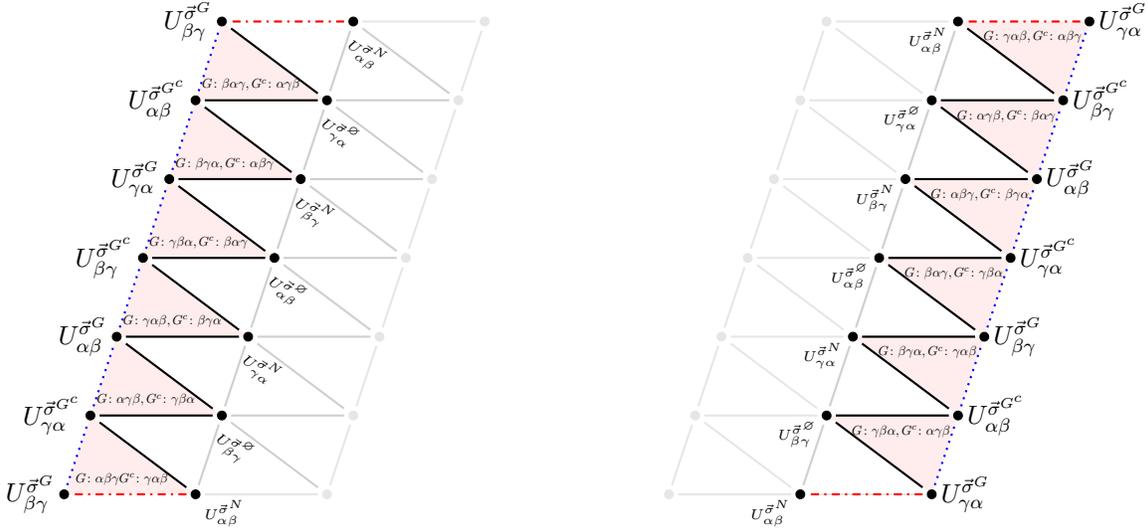

Let us comment on $D_1(G,\{\alpha, \beta, \gamma\})$ (resp. $D_2(G,\{\alpha, \beta, \gamma\})$) . It is easy to check that each of the six profiles in $D_1(G,\{\alpha, \beta, \gamma\})$ (resp. $D_2(G,\{\alpha, \beta, \gamma\})$) is in fact strongly polarized. Also, observe that for every strict total order $P$ on $\{\alpha,\beta,\gamma\}$, there exists a unique profile in $D_1(G,\{\alpha, \beta, \gamma\})$ (resp. $D_2(G,\{\alpha, \beta, \gamma\})$) such that every voter in $G$ has $P$ as her preference. Denoting the simplicial complex associated with $D_1(G,\{\alpha, \beta, \gamma\})$ (resp. $D_2(G,\{\alpha, \beta, \gamma\})$)  as $N_{D_1(G, \{\alpha,\beta,\gamma\})}$ (resp. $N_{D_2(G, \{\alpha,\beta,\gamma\})}$) is quite cumbersome, so let us denote it as $B_1(G, \{\alpha, \beta, \gamma\})$ (resp. $B_2(G, \{\alpha, \beta, \gamma\})$ . This simplicial complex is depicted on the left of Figure \ref{fig:B1-B2} (resp. on the right of this figure) drawn in such a way that is evident that it is isomorphic to a subcomplex of the torus in the base case in Figure \ref{fig:NWn}.

Observe that we can take each of the  profiles in $D_1(G,\{\alpha, \beta, \gamma\})$ and apply Remark \ref{rmk:symmetric_polarized} to obtain $D_2(G,\{\alpha,\beta,\gamma\}$ and \textit{viceversa}. Of course, $D_1(G,\{\alpha, \beta, \gamma\})\cap D_2(G,\{\alpha,\beta,\gamma\}) = \varnothing$, hence by Remark \ref{rmk:12stronglyPolarized}, we have that $D_1(G,\{\alpha, \beta, \gamma\})\cup D_2(G,\{\alpha,\beta,\gamma\})$ consists of the total $12$ strongly polarized profiles w.r.t. $G$ and $\{\alpha,\beta,\gamma\}$.

We are now ready to define the class $\mathcal{D}^{\text{PT}}$ of domains. 

\begin{definition}
The class of preference domains of \emph{polarization over triples}, $\mathcal{D}^{\text{PT}}$, is defined as follows: $D\in \mathcal{D}^{\text{PT}}$ iff for every coalition $G$ that is non-empty and distinct from $N$, and every triple $\{\alpha,\beta,\gamma\}\subseteq X$, $\alpha\neq \beta \neq \gamma\neq \alpha$, we have that $B_1(G,\{\alpha,\beta,\gamma\})$ is a subcomplex of $N_D$ or $B_2(G,\{\alpha,\beta,\gamma\})$ is a subcomplex of $N_D$ (or equivalently: $D_1(G, \{\alpha,\beta,\gamma\})$ is a subset of $D|_{\{\alpha,\beta,\gamma\}}$ or $D_2(G, \{\alpha,\beta,\gamma\})$ is a subset of $D|_{\{\alpha,\beta,\gamma\}}$).
\end{definition}

Now we define the classes of diversity over triples to complete the definition of the polarization and diversity over triples' class.

\begin{definition}\label{def:diversity-over-triples}
The class of preference domains of \emph{diversity over triples}, $\mathcal{D}^{\text{DT}}$, is defined as follows: $D\in\mathcal{D}^{\text{DT}}$ iff for every two coalitions $G$ and $G'$ such that $G\not\subseteq G'$ and $G'\not\subseteq G$, there exists three alternatives $\alpha,\beta,\gamma\in X$ such that there is a profile $\vec{P}\in D$ such that:
\begin{itemize}
    \vspace{-3mm}
    \begin{multicols}{2}
        \item if $i\in G\setminus {G'}$, then $\gamma P_i\alpha P_i\beta$;
        \item if $i\in G\cap G'$, then $\alpha P_i \beta P_i \gamma$; 
        \item if $i\in G'\setminus G$, then $\beta P_i\gamma P_i\alpha$; and
        \item if $i\in N\setminus(G\cup G')$, then $\gamma P_i\beta P_i\alpha$.
    \end{multicols}
    \vspace{-3mm}
\end{itemize}
We refer to $\mathcal{D}^{\text{PT}}\cap\mathcal{D}^{\text{DT}}$ as the class of \emph{polarization and diversity over triples}.
\end{definition}

We have already commented about the polarization part of $\mathcal{D}^{\text{PT}}\cap\mathcal{D}^{\text{DT}}$, so let us say a few words about $\mathcal{D}^{\text{DT}}$. Notice that if $G\cap G'$ is non-empty, then a profile $\vec{P}$ witnessing the condition established in Definition \ref{def:diversity-over-triples} (w.r.t. $G$ and $G'$) is such that there a voter whose preference in $\vec{P}$ restricted to $\{\alpha, \beta,\gamma \}$ is $\gamma\alpha\beta$; another voter with $\alpha\beta\gamma$; and a third voter $\beta\gamma\alpha$. In other words, for any distinct alternatives $a, b, c\in \{\alpha,\beta,\gamma\}$, there exists a voter that, in $\vec{P}$, ranks $a$ on top of $b$ and $c$; another voter that ranks $a$ in the middle of $b$ and $c$; an yet a third voter that ranks $a$ below $b$ and $c$. Therefore such a profile $\vec{P}$ is not value-restricted. 

If $n\geq 3$, there exists coalitions $G$ and $G'$ satisfying $G\not\subseteq G'$ and $G'\not\subseteq G$ such that $G\cap G'\neq \varnothing$. For example, $G=\{1,2\}$ and $G'=\{2,3\}$. Therefore, for $n\geq 3$, if $D\in \mathcal{D}^{\text{DT}}$, then $D$ has at least a profile that is not value-restricted. 

As another comment on $\mathcal{D^{\text{DT}}}$, when $G$ and $G'$ are (almost-)decisive w.r.t a given unanimous IIA-SWF, it is common for a profile with the structure specified in Definition \ref{def:diversity-over-triples} to be invoked as part of some proofs of Arrow's theorem or proofs of generalizations of it, e.g. \cite{CAMPBELL200235,KIRMAN1972267}. 

Clearly, for the general case of $|X|\ge 3$ alternatives and $n\ge 2$ voters, the unrestricted domain belongs to $\mathcal{D}^{\text{PT}}\cap\mathcal{D}^{\text{DT}}$. For the case of three alternatives and two voters, the super-Arrovian domain \say{$D^*$} in the proof in \cite[Lemma 2 on p. 87]{FISHBURNP.C1997Sadw} is clearly a member of $\mathcal{D}^{\text{PT}}\cap\mathcal{D}^{\text{DT}}$. For the case of three alternatives and three voters, it is easy to see that the super-Arrovian domain that appears in the proof in \cite[Lemma 3 on pp. 88--89]{FISHBURNP.C1997Sadw} is a subdomain of some domains in $\mathcal{D}^{\text{PT}}\cap\mathcal{D}^{\text{DT}}$. 

Since our proof in the next subsection will employ the combinatorial topology framework, we introduce a combinatorial topology version of $\mathcal{D}^{\text{DT}}$, which is equivalent to the one in Definition \ref{def:diversity-over-triples}.

\begin{definition}\label{def:diversity-over-triples-comb-top}
 A domain $D\in\mathcal{D}^{\text{DT}}$ iff for every two coalitions $G$ and $G'$ such that $G\not\subseteq G'$ and $G'\not\subseteq G$, there exists three alternatives $\alpha,\beta,\gamma\in X$ such that $\{U_{\alpha\beta}^{\vec{\sigma}^G}, U_{\beta\gamma}^{\vec{\sigma}^{G'}}, U_{\gamma\alpha}^{\vec{\sigma}^{(G\cap G')^c}}\}$ is a $2$-simplex of $N_D$. 
\end{definition}

\subsection{Proving the Impossibility through Ultrafilters and Combinatorial Topology}\label{subsec:imp-proof}

Our objective in this section is to prove that $\mathcal{D}^{\text{PT}}\cap \mathcal{D}^{\text{DT}}$ is a class of super-Arrovian domains. We will refer the reader to Appendix \ref{sec:appendix} for those proofs or parts of a proof that are more technical and only keep the interesting proofs or proof sketches in the main text. To prove that any $D\in \mathcal{D}^{\text{PT}}\cap \mathcal{D}^{\text{DT}}$ is Arrow-inconsistent we use an ultrafilter approach with combinatorial topology objects. Then we will prove that the property of belonging to $ \mathcal{D}^{\text{PT}}\cap \mathcal{D}^{\text{DT}}$ is closed upward under inclusion (and hence, any $D\in \mathcal{D}^{\text{PT}}\cap \mathcal{D}^{\text{DT}}$ is super-Arrovian). 

Given any unanimous chromatic $f\colon N_D \to N_{W(X)}$, where $D\in \mathcal{D}^{\text{PT}}\cap \mathcal{D}^{\text{DT}}$, the goal is to prove that the set of all almost-decisive coalitions w.r.t. $f$ is an ultrafilter w.r.t. $N$. Then we can easily show that $f$ is dictatorial using Lemma \ref{lem:ultra-dict} below, which is the combinatorial topology version of the step in the classical ultrafilter approach in which one deduces dictatorship. But to do all of this, we start by defining the combinatorial topology version of an almost-decisive coalition.

\begin{definition}\label{almost-dec}
Let $f\colon N_D\to N_W$ be a chromatic simplicial map, $Y\subseteq X$, and $G$ a coalition. If $ab$ is a ordered pair of distinct alternatives $a$ and $b$ in $X$, we say that $G$ is \emph{almost-decisive over $ab$ w.r.t. $f$} if $f(U_{ab}^{\vec{\sigma}^G})=U_{ab}^+$ whenever $U_{ab}^{\vec{\sigma}^G}$ is a vertex of $N_D$. We say that $G$ is \emph{almost-decisive over $Y$ w.r.t. $f$} if for all $a,b\in Y$ such that $U_{ab}^{\vec{\sigma}^G}$ is a vertex of $N_D$, we have that $f(U_{ab}^{\vec{\sigma}^G})=U_{ab}^+$. If $G$ is almost-decisive over $X$ w.r.t. $f$, we just say that it is \emph{almost-decisive w.r.t. $f$}.
\end{definition}

In words, if $G$ is almost-decisive then when everyone in $G$ agrees on ranking $a$ over $b$ and everyone not in $G$ agrees on ranking $b$ over $a$, then society ranks $a$ over $b$. Now we present a useful lemma that follows easily from this definition (see Appendix \ref{sec:appendix} for a proof).

\begin{lemma}\label{lem:almost-dec-compl}
Let $G$ be an almost-decisive coalition over $Y\subseteq X$ and $\beta, \alpha\in X$, where $\alpha\neq \beta$. If $U_{\alpha\beta}^{\vec{\sigma}^{G^c}}$ is a vertex of $N_D$, then $f(U_{\alpha\beta}^{\vec{\sigma}^{G^c}})=U_{\alpha\beta}^-$.
\end{lemma}

%\begin{proof}
%Observe that $U_{\beta\alpha}^{\vec{\sigma}^G}=U_{\alpha\beta}^{\vec{\sigma}^{G^c}}$, $U_{\beta\alpha}^+=U_{\alpha\beta}^-$ and $f(U_{\beta\alpha}^{\vec{\sigma}^G})=U_{\beta\alpha}^+$. Taking these observations together yields the desired result.
%\end{proof}

The following lemma is the one we mentioned at the second paragraph of this section (the proof is in Appendix \ref{sec:appendix}).

\begin{lemma}\label{lem:ultra-dict}
Let $f\colon N_D \to N_{W(X)}$ be a chromatic simplicial map and $\mathcal{G}$ the set of all almost-decisive coalitions w.r.t. $f$. If $\mathcal{G}$ is an ultrafilter of the set of all voters $N$, then $f$ is dictatorial. 
\end{lemma}

 So now we want to show that the set of all almost-decisive coalitions w.r.t. unanimous chromatic $f$ on $N_D$, with $D\in \mathcal{D}^{\text{PT}}\cap\mathcal{D}^{\text{DT}}$, satisfies the three properties of Definition \ref{def:ultrafilter} of an ultrafilter. We will see that $D\in \mathcal{D}^{\text{PT}}$ is enough for properties $1$ and $3$ to hold. Showing this for property $1$ is easy (we do it in Appendix \ref{sec:appendix}), but for property $3$ we will have to introduce some definitions and results related to domains $N_D$ having $B_i(G,Y)$ as a subcomplex, with $i=1$ or $i=2$, for a fixed coalition $G$ and triple of alternatives $Y$. We proceed to do so.

Definition \ref{def:boundary-interior} below is relevant to prove some subsequent lemmas. Lemmas \ref{lem:not-boundary-1}, \ref{lem:not-boundary-2} and \ref{lem:dec-G-Gc-Y} formalize and generalize an (heuristic) geometric argument made by \cite{RajsbaumR2022preprint-new,RajsbaumArmajacPODC}. We will explain more about this once we present the three lemmas and the geometric intuition of Lemmas \ref{lem:not-boundary-2} and \ref{lem:dec-G-Gc-Y}. Furthermore, in the context of only $3$ alternatives and only $n\in \{2,3\}$ voters, the proof of these lemmas taken together is very similar to a proof and part of another proof carried out in \cite[Lemma 2 and Lemma 3 on pp. 87--88]{FISHBURNP.C1997Sadw} to show that certain domains are super-Arrovian. However, the proofs in \cite{FISHBURNP.C1997Sadw} use the classical approach instead of the combinatorial topology approach and decisive coalitions instead of almost-decisive coalitions. 

\begin{definition}\label{def:boundary-interior}
An edge of $N_{W(X)}$ is said to be \emph{determined by transitivity} (\emph{DbT} edge, for short) if it is of the form $\{U_{\alpha\beta}^+, U_{\beta\gamma}^+\}$ for some $\alpha,\beta,\gamma\in X$. An edge of $N_{W(X)}$ is a \emph{non-DbT} edge if it is not a DbT edge. 
\end{definition}

To motivate our definition, notice that a DbT edge $\{U_{\alpha\beta}^+, U_{\beta\gamma}^+\}$ represents the strict total orders on $X$ ranking $\alpha$ over $\beta$ and $\beta$ over $\gamma$. Then $\{U_{\alpha\beta}^+, U_{\beta\gamma}^+\}$ is a face of exactly one $2$-simplex of $N_{W(X)}$ among the $2$-simplices of $N_{W(X)}$ that only involve alternatives in $\{\alpha,\beta,\gamma\}$, namely, it is a face of the $2$-simplex $\{U_{\alpha\beta}^+, U_{\beta\gamma}^+, U_{\alpha\gamma}^+\}$ (in contrast, notice that $\{U_{\alpha\beta}^+, U_{\beta\gamma}^+, U_{\gamma\alpha}^+\}$ is not a $2$-simplex since it represents the intransitive ranking $\alpha\beta\gamma\alpha$)\footnote{When $|X|=3$, it can be shown that an edge of $N_{W(X)}$ is a DbT edge iff it is a $1$-simplex in the boundary of $N_{W(X)}$. The \emph{boundary} of a pure simplicial complex $K$ is the simplicial complex induced by the $(\text{dim}(K)-1)$-simplices that each is the face of a unique facet of $N_{W(X)}$.}.

Notice that an edge of $N_{W(X)}$ of the form $\{U_{\alpha\beta}^-, U_{\gamma\alpha}^-\}$ is a DbT edge since it can be rewritten as $\{U_{\beta\alpha}^+, U_{\alpha\gamma}^+\}$. Of course, DbT edges live in the $1$-skeleton of $N_{W(X)}$. Fix three different alternatives, $x,y,z\in X$. In Figure \ref{fig:N_Wxyz}, it is easy to identify the six DbT edges and the six non-DbT edges involving alternatives in $\{x,y,z\}$.

\begin{lemma}\label{lem:not-boundary-1}
    Let $G$ be a non-empty coalition distinct from $N$; $Y\subseteq X$, such that $|Y|=3$; and $f\colon N_D \to N_{W(X)}$ a unanimous chromatic simplicial map. If $B_1(G, Y)$ (resp. $B_2(G, Y)$) is a subcomplex of $N_D$, then
    \begin{enumerate}
        \item Any edge of the form $\{U_{ac}^{\vec{\sigma}^G}, U_{ba}^{\vec{\sigma}^{G^c}}\}$, for some $a,b,c\in Y$, cannot be mapped to $\{U_{ac}^-, U_{ba}^-\}$ (resp. $\{U_{ac}^+, U_{ba}^+\}$). 
        \item Any edge of the form $\{U_{ac}^{\vec{\sigma}^{G^c}}, U_{ba}^{\vec{\sigma}^G}\}$, for some $a, b, c \in Y$, cannot be mapped to $\{U_{ac}^+, U_{ba}^+\}$ (resp. $\{U_{ac}^-, U_{ba}^-\}$).   
    \end{enumerate} 
\end{lemma}

See Appendix \ref{sec:appendix} for the proof.

\begin{lemma}\label{lem:not-boundary-2}
    Let $G$ be a non-empty coalition distinct from $N$; $Y\subseteq X$, such that $|Y|=3$; and $f\colon N_D \to N_{W(X)}$ a unanimous chromatic simplicial map. If $B_i(G, Y)$ is a subcomplex of $N_D$ for some $i\in \{1,2\}$, then any edge of the form $\{U_{\beta\gamma}^{\vec{\sigma}^G}, U_{\alpha\beta}^{\vec{\sigma}^{G^c}}\}$ or $\{U_{\beta\gamma}^{\vec{\sigma}^{G^c}}, U_{\alpha\beta}^{\vec{\sigma}^G}\}$, for some $\alpha,\beta,\gamma\in Y$, is mapped by $f$ to an non-DbT edge; in particular, to $\{U_{\beta\gamma}^+,U_{\alpha\beta}^-\}$ or $\{U_{\beta\gamma}^-,U_{\alpha\beta}^+\}$.
\end{lemma}

The proof of Lemma \ref{lem:not-boundary-2} is in Appendix \ref{sec:appendix}. Here we present a proof sketch that we hope will convince the reader that the lemma holds. 

\begin{figure}[ht!]
        \centering
        \includegraphics[scale=0.6]{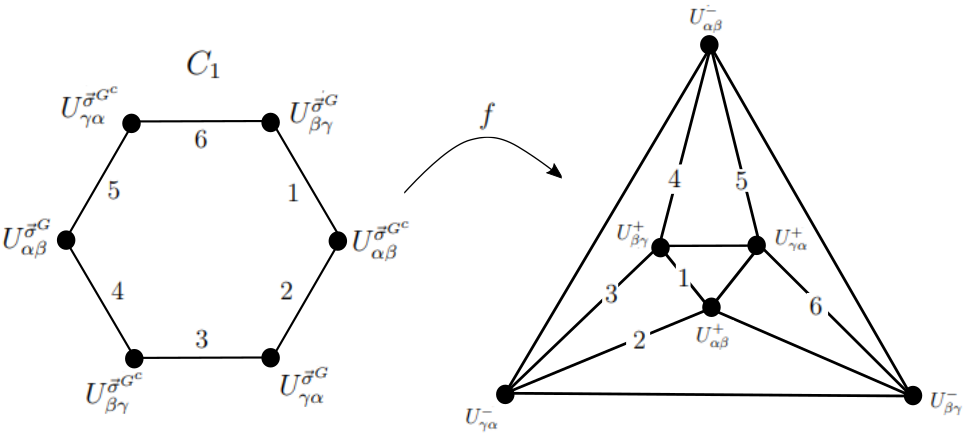}
        \caption{Geometric intuition behind the proof of Lemma
        \ref{lem:not-boundary-2}}
        \label{fig:mapping-CI-contradiction}
\end{figure}

W.l.o.g. assume $B_1(G,Y)$ is a subcomplex of $N_D$. Let $\alpha,\beta,\gamma\in Y$ such that $\alpha\neq\beta\neq\gamma\neq\alpha$. Then we can represent $B_1(G,Y)$ as it is depicted on the left of Figure \ref{fig:B1-B2}. Let $C_1$ denote the dotted-line cycle of $B_1(G,\{\alpha,\beta,\gamma\})$ in this figure \footnote{Formally, $C_1$ is subcomplex of $B_1(G,\{\alpha,\beta,\gamma\})$ of dimension $1$}. With the goal of reaching a contradiction, we assume that the edge $\{U_{\beta\gamma}^{\vec{\sigma}^G},U_{\alpha\beta}^{\vec{\sigma}^{G^c}}\}\in C_1$ is mapped by $f$ to the DbT edge $\{U_{\beta\gamma}^+,U_{\alpha\beta}^+\}$. This is indicated in Figure \ref{fig:mapping-CI-contradiction} by labelling edge $\{U_{\beta\gamma}^{\vec{\sigma}^G},U_{\alpha\beta}^{\vec{\sigma}^{G^c}}\}$ in $C_1$ and labelling with the same number the edge of $N_{W(X)}$ to which $\{U_{\beta\gamma}^{\vec{\sigma}^G},U_{\alpha\beta}^{\vec{\sigma}^{G^c}}\}$ is mapped under $f$. By chromaticity of $f$ we know that every edge in $C_1$ is mapped to some edge in the part of $skel^1(N_{W(X)})$ only involving alternatives in $\{\alpha,\beta,\gamma\}$. This is why only this subcomplex of $N_{W(X)}$ is depicted in Figure \ref{fig:mapping-CI-contradiction}. Applying, succesively, chromaticity of $f$ and the relevant part of Lemma \ref{lem:not-boundary-1} implies that $C_1$ has to be mapped over $skel^1(N_{W(X)})$ as indicated by the numbers that act as labels. As it can be seen, $f(U_{\beta\gamma}^{\vec{\sigma}^G})=U_{\beta\gamma}^+$ and $f(U_{\beta\gamma}^{\vec{\sigma}^G})=U_{\beta\gamma}^-$, a contradiction.  

\begin{lemma}\label{lem:dec-G-Gc-Y}
Let $G$ be a coalition; $Y=\{\alpha,\beta,\gamma\}\subseteq X$, such that $|Y|=3$; and $f\colon N_D \to N_{W(X)}$ a unanimous chromatic simplicial map. If $B_i(G,Y)$ is a subcomplex of $N_D$ for some $i\in \{1,2\}$ whenever $G$ is non-empty and distinct from $N$, then (either) $G$ or $G^c$ is almost-decisive over $Y$.
\end{lemma}

\begin{figure}[ht!]
        \centering
        \includegraphics[scale=0.6]{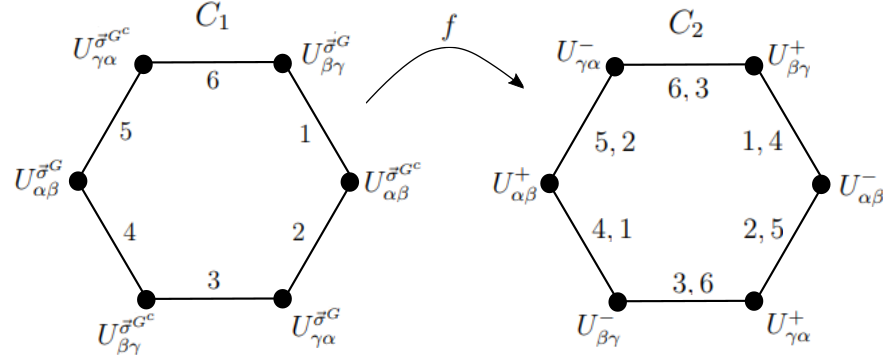}
        \caption{Geometric intuition behind the proof of Lemma \ref{lem:dec-G-Gc-Y}.}
        \label{fig:dec-G-Gc-Y}
\end{figure}

Again, we just present a geometric proof sketch of why this result holds, for a formal proof see Appendix \ref{sec:appendix}. Let $C_2$ denote the cycle that consists of the non-DbT edges that are represented in Figure \ref{fig:N_Wxyz}. Cycle $C_2$ is represented on the right of Figure \ref{fig:dec-G-Gc-Y}, along with cycle $C_1$ represented on the left of this figure. By Lemma \ref{lem:not-boundary-2}, $C_1$ has to be mapped over $C_2$ under $f$. So we can start by asking where could edge $\{U_{\beta\gamma}^{\vec{\sigma}^G}, U_{\alpha\beta}^{\vec{\sigma}^{G^c}}\}$ by mapped under $f$. There are two options: $\{U_{\beta\gamma}^+, U_{\alpha\beta}^-\}$ or $\{U_{\beta\gamma}^-, U_{\alpha\beta}^+\}$. So let us see both cases. 

Case 1 (resp. 2): $f(\{U_{\beta\gamma}^{\vec{\sigma}^G}, U_{\alpha\beta}^{\vec{\sigma}^{G^c}}\})=\{U_{\beta\gamma}^+, U_{\alpha\beta}^-\}$ (resp. $f(\{U_{\beta\gamma}^{\vec{\sigma}^G}, U_{\alpha\beta}^{\vec{\sigma}^{G^c}}\})=\{U_{\beta\gamma}^-, U_{\alpha\beta}^+\}$) . In this case, by chromaticity, $C_1$ has to be mapped as follows: if $e$ is an edge of $C_2$ and $x$ is the number that acts as a label for $e$, then $f$ maps $e$ to the edge in $C_1$ that has $x$ as the first (resp. second) number appearing in its label. For example, edge $\{U_{\beta\gamma}^{\vec{\sigma}^{G^c}}, U_{\gamma\alpha}^{\vec{\sigma}^G}\}$, with label $3$, gets mapped to $\{U_{\beta\gamma}^-, U_{\gamma\alpha}^+\}$ (resp. $\{U_{\beta\gamma}^+, U_{\gamma\alpha}^-\}$), with label $3,6$ (resp. $6,3$). Looking at how the vertices are mapped, we can see that $G$ (resp. $G^c$) is almost-decisive over $Y$. 

For the case of only two voters and three alternatives,  \cite{RajsbaumR2022preprint-new,RajsbaumArmajacPODC} say that the cycle that we call $C_1$ has to be mapped over the cycle $C_2$ due to the unanimity edges in dashed-lines in Figure \ref{fig:NWn}, but they do not go into the details of why. We formalized this via Lemmas \ref{lem:not-boundary-1} and \ref{lem:not-boundary-2}. Furthermore, we generalized their argument because we do not need the unanimity edges, only the unanimity vertices. Moreover, we showed that this same argument can be applied when there are $n\ge 2$ voters and $|X|\ge 3$ alternatives if we have the structure provided by the $B_i(\cdot,\cdot)$'s and focus on the relevant part of the $2$-skeleton of $N_D$. Finally, \cite{RajsbaumR2022preprint-new,RajsbaumArmajacPODC}, like us, say that $C_1$ can be mapped over $C_2$ in two ways, one of which makes voter $1$ the dictator and the other makes voter the dictator. In our case, since we are dealing $n\ge 2$ voters and $|X|\ge 3$ alternatives, we can only conclude that $G$ or $G^c$ is almost-decisive over the triple of alternatives $Y$.

The following lemma says that, given a domain in $\mathcal{D}^{\text{PT}}$, the almost-decisiveness of a coalition over an ordered pair of alternatives spreads to all ordered pairs of alternatives. This sort of \say{contagion} result has been used in other ultrafilter proofs. For instance, for the case of the unrestricted domain, \cite{KIRMAN1972267} shows that this contagion of almost-decisiveness occurs. As another example, \cite{CAMPBELL200235} has a contagion lemma for the case of domains satisfying the chain property.

\begin{lemma}\label{lem:contagioLemma}
Let $f\colon N_D\to N_W$ be a unanimous chromatic simplicial map, where $D\in \mathcal{D}^{\text{PT}}$, let $G$ be a non-empty coalition distinct from $N$, and $\alpha$ and $\beta$ two different alternatives in $X$ such that $f(U_{\alpha\beta}^{\vec{\sigma}^G})=U_{\alpha\beta}^+$, then $G$ is almost-decisive. 
\end{lemma}

\begin{proof}
Suppose there are two different alternatives $\alpha$ and $\beta$ in $X$ such that $f(U_{\alpha\beta}^{\vec{\sigma}^G})=U_{\alpha\beta}^+$. Let $\gamma,\delta\in X\setminus\{\alpha,\beta\}$ with $\gamma\neq \delta$. To show: $(1)$ $f(U_{\gamma\delta}^{\vec{\sigma}^G})=U_{\gamma\delta}^+$, $(2)$ $f(U_{\alpha\gamma}^{\vec{\sigma}^G})=U_{\alpha\gamma}^+$, $(3)$ $f(U_{\gamma\alpha}^{\vec{\sigma}^G})=U_{\gamma\alpha}^+$, $(4)$ $f(U_{\beta\gamma}^{\vec{\sigma}^G})=U_{\beta\gamma}^+$, $(5)$ $f(U_{\gamma\beta}^{\vec{\sigma}^G})=U_{\gamma\beta}^+$, $(6)$ $f(U_{\beta\alpha}^{\vec{\sigma}^G})=U_{\beta\alpha}^+$.
We focus on proving 1 and we will prove 2-6 along the way. 

Let $Y_1=\{\alpha,\beta,\gamma\}$.  Since $f(U_{\alpha\beta}^{\vec{\sigma}^G})=U_{\alpha\beta}^+$, by Lemma \ref{lem:almost-dec-compl}, $G^c$ cannot be almost-decisive over $Y_1$. Then by Lemma \ref{lem:dec-G-Gc-Y}, $G$ is almost-decisive over $Y_1$. Therefore, 2-6 hold. In particular, $f(U_{\alpha\gamma}^{\vec{\sigma}^G})=U_{\alpha\gamma}^+$.

Let $Y_2=\{\alpha,\gamma,\delta\}$. Since $f(U_{\alpha\gamma}^{\vec{\sigma}^G})=U_{\alpha\gamma}^+$, by Lemma \ref{lem:almost-dec-compl}, $G^c$ cannot be almost-decisive over $Y_2$. Then by Lemma \ref{lem:dec-G-Gc-Y}, $G$ is almost-decisive over $Y_2$. Then 
$f(U_{\gamma\delta}^{\vec{\sigma}^G})=U_{\gamma\delta}^+$, i.e., 1 holds. 
\end{proof}

\begin{figure}
    \centering
    \tikzmath{\a = 1; \b =1.5; }
\pgfdeclarelayer{background}
\pgfsetlayers{background,main}

\begin{tikzpicture}[scale=1, every node/.style={scale=1}]

\begin{scope}[shift={(1.7,3.8)}
        , scale=1, every node/.style={scale=1.2}
                ]

        \tikzmath{\x = 3; \y =1.5; \op = 0.3; }
    
    \node (A) at (-\x,0) {};

    \node (A1) at (0,0) {};

    \node (B) at (\x,0) {};

    \node (B1) at (0,\y) {};

    \draw [thick, blue, line width=0.6mm] plot [smooth, tension=0.7] coordinates {(A) (-0.6*\x,0.8*\y) (B1)};
    \draw [thick, blue, line width=0.6mm] plot [smooth, tension=0.7] coordinates {(A) (-0.5*\x,-0.2*\y) (A1)};
    \draw [thick, blue, line width=0.6mm] plot [smooth, tension=0.7] coordinates {(A1) (-0.1*\x,0.5*\y) (B1)};
    \draw [thick, blue, line width=0.6mm] plot [smooth, tension=0.7] coordinates {(B) (0.6*\x,0.8*\y) (B1)};
    \draw [thick, blue, line width=0.6mm] plot [smooth, tension=0.7] coordinates {(B) (0.5*\x,-0.2*\y) (A1)};
    \draw [thick, blue, line width=0.6mm] plot [smooth, tension=0.7] coordinates {(A1) (0.1*\x,0.5*\y) (B1)};

    \node (A) at (-\x,0) {};
    \filldraw[black] (A) circle (1.5pt) node[left] {$U_{\alpha\beta}^{\vec{\sigma}^G}$};

    \node (A1) at (0,0) {};
    \filldraw[black] (A1) circle (1.5pt) node[below] {$U_{\gamma\alpha}^{\vec{\sigma}^G}$};

    \node (B) at (\x,0) {};
    \filldraw[black] (B) circle (1.5pt) node[right] {$U_{\gamma\delta}^{\vec{\sigma}^G}$};

    \node (B1) at (0,\y) {};
    \filldraw[black] (B1) circle (1.5pt) node[above] {$U_{\alpha\gamma}^{\vec{\sigma}^G}$};

\end{scope}

\end{tikzpicture}
    \caption{Geometric intuition behind the proof of Lemma \ref{lem:contagioLemma}.}
    \label{fig:connection}
\end{figure}
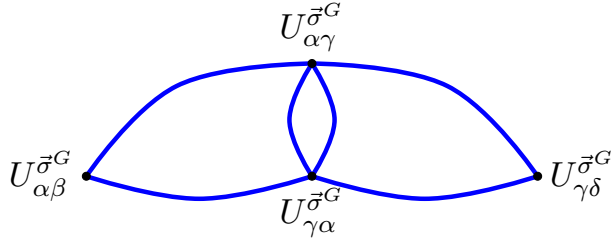

Now we give the geometric intuition of this proof. Observe that $Y_1$ and $Y_2$ are triples that share exactly two alternatives, i.e., $\alpha$ and $\gamma$. Let $i\in \{1,2\}$. Since, $D\in \mathcal{D}^{\text{PT}}$, there exists $j\in \{1,2\}$ such that $B_j(G, Y_i)$ is a subcomplex of $N_D$. Let $B(G, Y_i)=B_j(G, Y_i)$. It is easy to see that $B(G, Y_1)$ and $B(G, Y_2)$ share exactly two non-unanimous vertices, namely $U_{\alpha\gamma}^{\vec{\sigma}^G}$ and $U_{\gamma\alpha}^{\vec{\sigma}^G}$, as depicted in Figure \ref{fig:connection}. So intuitively, $f(U_{\alpha\beta}^{\vec{\sigma}^G})=U_{\alpha\beta}^+$ is spreading almost-decisiviness along the cycles depicted in Figure \ref{fig:connection} until it reaches all the target vertices.

%\begin{figure}[ht!]
%        \centering
%        \includegraphics[scale=0.6]{dos_ciclos.png}
%        \caption{Geometric intuition behind the proof of Lemma \ref{lem:contagioLemma}.}
%        \label{fig:connection}
%\end{figure}
The subsequent theorem states that membership to $\mathcal{D}^\text{PT}$ provides a sufficient condition for the set of all almost-decisive coalitions (w.r.t. a given $f$) to satisfy the third property of ultrafilters. 

\begin{lemma}\label{lem:prop3ultra}
Let $f\colon N_D \to N_W$ be a unanimous chromatic simplicial map, where $D\in \mathcal{D}^{\text{PT}}$ and let $G$ be a coalition. We have that $G$ or $G^c$ is almost-decisive. 
\end{lemma}

This lemma follows easily from Lemma \ref{lem:contagioLemma} (see Appendix \ref{sec:appendix} for details). 

In the following lemma we show that $\mathcal{D}^{\text{PT}}\cap\mathcal{D}^{\text{DT}}$ is sufficient to induce the second property of ultrafilters. The proof of this lemma is essentially a combinatorial topology version of the steps in the proof in \cite[Lemma B on pp. 274-275]{KIRMAN1972267} that demonstrate the intersection property of ultrafilters in the classical Arrovian framework. We derive our proof assuming domains in $\mathcal{D}^{\text{PT}}\cap\mathcal{D}^{\text{DT}}$, instead of the unrestricted domain, which is assumed in \cite{KIRMAN1972267}.

\begin{lemma}\label{lem:prop2ultra}
Let $f\colon N_D\to N_{W(X)}$ be a unanimous chromatic simplicial map such that $D\in \mathcal{D}^{\text{PT}}\cap\mathcal{D}^{\text{DT}}$. If $G$ and $G'$ are two almost-decisive coalitions (w.r.t. $f$), then the coalition $G\cap G'$ is almost-decisive.
\end{lemma} 

\begin{proof}
Suppose $G$ and $G'$ are two almost-decisive coalitions (w.r.t $f$). If $G$ or $G'$ is empty. Then $G\cap G'$ is empty and we are done. Suppose then $G$ and $G'$ are non-empty. 

If $G\subseteq G'$ or $G'\subseteq G$ then $G\cap G'$ is $G$ or $G'$ and we are done. Suppose then $G\not\subseteq G'$ and $G'\not\subseteq G$. We proceed by contradiction, suppose $G\cap G'$ is not almost-decisive. Then, since $D\in \mathcal{D}^{\text{PT}}$, by Lemma \ref{lem:prop3ultra}, $(G\cap G')^c$ is almost-decisive. Since $D\in \mathcal{D}^{\text{DT}}$, $G\not\subseteq G'$ and $G\not\subseteq G'$, there exists alternatives $\alpha,\beta,\gamma\in X$ such that $\{U_{\alpha\beta}^{\vec{\sigma}^G}, U_{\beta\gamma}^{\vec{\sigma}^{G'}}, U_{\gamma\alpha}^{\vec{\sigma}^{(G\cap G')^c}}\}$ is a $2$-simplex of $N_D$, denote it $T$. 

Since, $G$, $G'$ and $(G\cap G')^c$ are almost-decisive, $T$ is mapped to $\{U_{\alpha\beta}^+, U_{\beta\gamma}^+, U_{\gamma\alpha}^+\}$, but this is not a simplex of $N_{W(X)}$ (since it corresponds to the intransitive ranking $\alpha\beta\gamma\alpha$), a contradiction. 
\end{proof}

Now we present our generalized version of Arrow's theorem: 

\begin{theorem}\label{thrm:arrow-generalization}
If $D\in \mathcal{D}^{\text{PT}}\cap \mathcal{D}^{\text{DT}}$, then $D$ is Arrow-inconsistent.
\end{theorem}

The proof follows from combining lemmas \ref{lem:ultra-dict}, \ref{lem:prop3ultra}, \ref{lem:prop2ultra}, and an additional lemma in Appendix \ref{sec:appendix} (see this appendix for a detailed proof). 

We finalize this section via a proposition that says that the property of belonging to $\mathcal{D}^{\text{PT}}\cap\mathcal{D}^{\text{DT}}$ is closed upward under inclusion (the proof is in Appendix \ref{sec:appendix}). 

\begin{proposition}\label{prop:closed-under-up-incl}
Let $D$ and $D'$ are domains such that $D\in \mathcal{D}^{\text{PT}}\cap\mathcal{D}^{\text{DT}}$ and $D\subseteq D'$. We have that $D'\in \mathcal{D}^{\text{PT}}\cap\mathcal{D}^{\text{DT}}$. In particular, every domain in $\mathcal{D}^{\text{PT}}\cap\mathcal{D}^{\text{DT}}$ is super-Arrovian.  
\end{proposition}

%\begin{proposition}\label{prop:superArrovian}
%    If $D\in \mathcal{D}^{\text{PT}}\cap\mathcal{D}^{\text{DT}}$, then $D$ is super-Arrovian.
%\end{proposition}

%\begin{proof}
%Follows from combining Theorem \ref{thrm:equivProps} and Proposition \ref{prop:closed-under-up-incl}.
%\end{proof}

\section{Conclusions}

In this work, we proved a generalization of Arrow's theorem (Theorem \ref{thrm:arrow-generalization}) through combinatorial topology. In contrast with the domain restrictions of the impossibility results of \cite{CAMPBELL200235,KalaiMullerSatterthwaite}, the one of our generalization does not require the unrestricted domain at the level of some triple of alternatives. 

%Regarding the intuition of our results, we can say that diversity (involving non-value-restricted preferences) and confronted opinions in some profiles (strong polarization) lead to impossibility.
Regarding the technical aspects, one of the main drawbacks of using Baryshnikov's construction is that the dimension of $N_{W^(X)^n}$ increases with $|X|$. As we said in the introduction, \cite{Baryshnikov} says that the $2$-skeleton of $N_{W(X)^n}$ allows him to prove a generalization of Arrow's theorem to domains with the free triple property. In this paper, we have shown that even when you have domain restrictions at the level of triples, the $2$-skeleton is still useful in deriving impossibility results in arbitrary (finite) dimensions. %for arbitrary (finite) number of alternatives and voters. 
Notice that this is equivalent to using restricted profiles in the classical framework, which is common in social choice. However, we believe that the $2$-skeleton introduces geometric understanding, which can inspire results, as it did for the base case characterization in the context of a broad class of domains in \cite{RajsbaumR2022preprint-new}.

%We conclude by mentioning some possibilities for future research:
%\begin{itemize}
%    \item It would be interesting to further compare the domains in $\mathcal{D}^{\text{DT}}\cap \mathcal{D}^{\text{PT}}$ with other domains and domain restrictions in the social choice literature.
%    \item \textcolor{red}{\cite{BaryRoot}} provide an algebraic topology proof of the famous Gibbard-Satterthwaite impossibility theorem \cite{Gibbard,SATTERTHWAITE1975187}. It would be interesting to come up with a combinatorial topology proof. 
%    \item One might want to ask if domains in  $\mathcal{D}^{\text{DT}}\cap \mathcal{D}^{\text{PT}}$ escape Gibbard-Satterthwaite.
%    \item Can we use the $2$-skeleton of $N_D$ together with the characterization for the base case for a broad class of domains in \cite{RajsbaumR2022preprint-new} to come up with characterizations for the general case of any (finite) number of voters and alternatives?
%\end{itemize}

In future work, one may want to analyze in more detail the relation between the domains in $\mathcal{D}^{\text{DT}}\cap \mathcal{D}^{\text{PT}}$ and other domains in social choice. 
Moreover, we plan to provide a combinatorial topology proof for the Gibbard-Satterthwaite theorem and study its relation with the homological approach \cite{baryshnikov2024topological}.
Last, we hope that the $2$-skeleton strategy of this paper can be combined with the characterization in  \cite{RajsbaumR2022preprint-new}, which precisely works at the level of two voters and three alternatives, to derive characterizations in arbitrary (finite) dimensions.

\section*{Acknowledgments}
This paper (including the appendix) is adapted from parts of Isaac Lara's master's thesis \cite{Lara2023thesis} at \textit{Universidad Nacional Autónoma de México} (UNAM), which was done with the financial support of the \textit{Consejo Nacional de Humanidades, Ciencia y Tecnología} (CONAHCYT). Isaac Lara thanks CONAHCYT for this financial support. 

Isaac Lara also thanks the \textit{Programa de Apoyo a los Estudiantes de Posgrado} 2023 of
UNAM for providing him funding to pursue a research stay at the \textit{Institut de Recherche
en Informatique Fondamentale} (IRIF) in Paris, France, from March 15, 2023, to April
13, 2023. This stay contributed positively to the development of Isaac Lara's thesis.

Thanks to Armando Castañeda, César Hernández-Cruz, Natalia Jonard Pérez, Edwin Muñoz-Rodriguez and Ber Lorke for reading the thesis from which this paper is adapted and for providing very valuable comments.

\printbibliography[title=References]

\appendix
\section{Appendix}\label{sec:appendix}

In Subsection \ref{subsec:N_D-scratch}, given a domain $D$, we formally construct the chromatic simplicial complex $N_D$. In Subsection \ref{subsec:DY-to-SY}, we show that for all $Y\subseteq X$, such that $|Y|\ge 2$, there exists a bijection between the set of all subpreferences of $D$ on $Y$, denoted $D|_Y$, and the set of all $(\binom{|Y|}{2}-1)$-simplices that only involve alternatives in $Y$, denoted $S(Y)$. In Subsection \ref{subsec:WX-to-NWX}, we prove that there is a bijection between $W(X)$ to the set of facets of $N_{W(X)}$. This, together with any other result in Subsection \ref{subsec:WX-to-NWX}, were already proven by \cite{Baryshnikov}, but we present them in this appendix for it to be a self-contained reference for the equivalence between the classical and the combinatorial topology frameworks. In Subsection \ref{subsec:FD-to-MD}, we show that, for any given $D$, there is a bijection between the set of all IIA-SWFs on $D$, denoted $\mathcal{F}_D$,  and the set of all chromatic simplicial maps of the form $f\colon N_D \to N_{W(X)}$, denoted $\mathcal{M}_D$. All of these bijections allow us to refer interchangeably to subprofiles and their corresponding simplices, to strict total orders and their corresponding simplices, as well as to SWFs in $\mathcal{F}_D$ and their corresponding simplicial maps in $\mathcal{M}_D$. Finally, we provide the proofs missing in the main text, as well as some lemmas used for that purpose.

\subsection{Constructing $N_D$ from Scratch}\label{subsec:N_D-scratch}
Let $Y\subseteq X$ such that $|Y|\ge 2$. Let $L'$ be the following set:
    \begin{align*}
        \bigcup_{\substack{
        \vec\sigma\in \{+,-\}^n \\
        \alpha, \beta \in X, \alpha \neq \beta 
}}\{U_{\alpha\beta}^{\vec{\sigma}}\}.
\end{align*}

Let $D\subseteq W(X)^n$ such that $D\neq \varnothing$. If $U_{\alpha\beta}^{\vec{\sigma}}\in L'$, we define 
\begin{align*}
    s_D(U_{\alpha\beta}^{\vec{\sigma}})=\{\vec{P}\in D \colon \text{for all }\: i\in N, \: \alpha P_i \beta \: \text{iff} \: \sigma_i=+\}.
\end{align*}

Let $\sim_{s_D}$ be a binary relation on $L'$ defined as follows: $U_{\alpha\beta}^{\vec{\sigma}}\sim_{s_D}U_{\gamma\delta}^{\vec{\sigma}'}$ iff $s_D(U_{\alpha\beta}^{\vec{\sigma}})=s_D(U_{\gamma\delta}^{\vec{\sigma}'})$. Clearly, $\sim_{s_D}$ is an equivalence relation. It is not hard to see that if $U_{\alpha\beta}^{\vec{\sigma}}\in L'$ the equivalence class $[U_{\alpha\beta}^{\vec{\sigma}}]$ induced by $\sim_{s_D}$ is $\{U_{\alpha\beta}^{\vec{\sigma}}, U_{\beta\alpha}^{-\vec{\sigma}}\}$. Abusing notation, we drop the brackets from $[U_{\alpha\beta}^{\vec{\sigma}}]$ and just write $U_{\alpha\beta}^{\vec{\sigma}}$ or $U_{\beta\alpha}^{-\vec{\sigma}}$ to refer to this equivalence class. Hence, if $L$ is the partition of $L'$ induced by $\sim_{s_D}$, we simply write $L$ as

\begin{align*}
        \bigcup_{\substack{
        \vec\sigma\in \{+,-\}^n \\
        \alpha, \beta \in X, \alpha \neq \beta 
}}\{U_{\alpha\beta}^{\vec{\sigma}}\}.
\end{align*}

Therefore, it makes sense to write $U_{\alpha\beta}^{\vec{\sigma}}= U_{\beta\alpha}^{-\vec{\sigma}}$. We also abuse notation in defining $s_D(U_{\alpha\beta}^{\vec{\sigma}})$ as 
\begin{align*}
\{\vec{P}\in D \colon \text{for all }\: i\in N, \: \alpha P_i \beta \: \text{iff} \: \sigma_i=+\}.
\end{align*}
where $U_{\alpha\beta}^{\vec{\sigma}}$ is interpreted as an equivalence class in $L$ (instead of an element of $L'$).

Let $N_D$ is a simplicial complex defined as follows:
\begin{itemize}
    \item Its set of vertices, denoted $V(N_D)$ is 
    \begin{align*}
      \{u\in L\colon s_D(u) \neq \varnothing \}.
    \end{align*}
    \item a non-empty subset $S\subseteq V(N_D)$, where $S=\{v_1,\dots,v_k\}$, is a $(k-1)$-simplex of $N_D$ iff
\begin{align*}
\bigcap_{i=1}^k s_D(v_i) \neq \varnothing
\end{align*}
\end{itemize}

%Before showing that $N_D$ is in fact a chromatic simplicial complex, we introduce a function from which we obtain a coloring. Let $\chi\colon L \to (X^2)^n$ defined as $\chi (U_{\alpha\beta}^{\vec{\sigma}}) = (x_1,\dots,x_n)$ such that for all $i\in N$, we have $x_i=\alpha\beta$ if $\sigma_i=+$ and $x_i=\beta\alpha$ if $\sigma_i=-$-. Clearly, $\chi$ is injective. 

\begin{proposition}\label{prop:chromaticity-of-ND}
    If $D$ is a domain, the simplicial complex $N_D$ together with a labeling $\chi\colon V(N_D) \to \{Y\subseteq X\colon |Y|=2\}$ defined as
    \begin{align*}
    \chi(U_{\alpha\beta}^{\vec{\sigma}})=\{\alpha,\beta\}
    \end{align*}
    is a chromatic simplicial complex.
\end{proposition}

\begin{proof}
It is easy to show that $N_D$ is a simplicial complex, so we only have to prove that the $\chi$ labeling is a coloring, i.e., we have to show that if $t$ is a simplex of $N_D$, for all $U_{\alpha\beta}^{\vec{\sigma}}, U_{\gamma\delta}^{\vec{\sigma}'}\in t$ such that $U_{\alpha\beta}^{\vec{\sigma}}\neq  U_{\gamma\delta}^{\vec{\sigma}'}$, we have that $\chi(U_{\alpha\beta}^{\vec{\sigma}})\neq \chi(U_{\gamma\delta}^{\vec{\sigma}'})$. 

    By definition of $\chi$, notice that is suffices to show that
    ($\alpha\neq\gamma$ and $\alpha\neq\delta$) or ($\beta\neq\gamma$ and $\beta\neq\delta$).

    Since $t$ is a simplex of $N_D$ there exists $\vec{P}\in D$ such that $\vec{P}\in s_D(U_{\alpha\beta}^{\vec{\sigma}})\cap s_D(U_{\gamma\delta}^{\vec{\sigma}'})$.
    
    Suppose $\alpha = \gamma$ or $\alpha = \delta$. To show: $\beta\neq \gamma$ and $\beta \neq \delta$. 
    We proceed by contradiction supposing that $\beta=\gamma$ or $\beta=\delta$. We proceed by cases.

    Case $1$: $\beta= \gamma$. Then $\alpha=\delta$. This leads to $U_{\gamma\delta}^{\vec{\sigma}'}=U_{\beta\alpha}^{\vec{\sigma}'}=U_{\alpha\beta}^{-\vec{\sigma}'}$. We proceed by subcases.

    Subcase $1.1$: $\vec{\sigma}= -\vec{\sigma}'$. Then $U_{\alpha\beta}^{\vec{\sigma}}=U_{\gamma\delta}^{\vec{\sigma}'}$, a contradiction.

    Subcase $1.2$: $\vec{\sigma}\neq -\vec{\sigma}'$. Then since $\vec{P}\in s_D(U_{\alpha\beta}^{\vec{\sigma}})\cap s_D(U_{\gamma\delta}^{\vec{\sigma}'})=s_D(U_{\alpha\beta}^{\vec{\sigma}})\cap s_D(U_{\alpha\beta}^{-\vec{\sigma}'})$, there is a voter $i\in N$ such that $\alpha P_i \beta$ and $\beta P_i \alpha$, a contradiction to the asymmetry of $P_i$.

    Case $2$: $\beta=\delta$. Then $\alpha=\gamma$. This leads to $U_{\gamma\delta}^{\vec{\sigma}'}=U_{\alpha\beta}^{\vec{\sigma}'}$. We proceed by subcases.

    Subcase $2.1$: $\vec{\sigma}= \vec{\sigma}'$. Then $U_{\alpha\beta}^{\vec{\sigma}}=U_{\gamma\delta}^{\vec{\sigma}'}$, a contradiction.

    Subcase $2.2$: $\vec{\sigma} \neq \vec{\sigma}'$. Then since $\vec{P}\in s_D(U_{\alpha\beta}^{\vec{\sigma}})\cap s_D(U_{\gamma\delta}^{\vec{\sigma}'})=s_D(U_{\alpha\beta}^{\vec{\sigma}})\cap s_D(U_{\alpha\beta}^{\vec{\sigma}'})$, there is a voter $i\in N$ such that $\alpha P_i \beta$ and $\beta P_i \alpha$, a contradiction to the asymmetry of $P_i$.

    \end{proof}
\subsection{A Bijection from $D|_Y$ to $S(Y)$}\label{subsec:DY-to-SY}

Let $S(Y)$ be the set of all $(\binom{|Y|}{2}-1)$-simplex of $N_D$ that only involve alternatives of $Y$, i.e., if $U_{\alpha\beta}^{\vec{\sigma}}\in S(Y)$, then $\alpha,\beta\in Y$. 

Our objective is to define a bijection between $D|_Y$ and $S(Y)$ to talk about subprofiles of $D$ and simplices in an interchangeably way. Consider then a  function $g_Y\colon D|_Y \to S(Y)$ defined as follows: 
\begin{align*}
    \text{for all $\vec{P}\in D|_Y$,}\: g_Y(\vec{P})=\{U_{xy}^{\vec{\sigma}}\in L\colon x,y\in Y;x\neq y;\text{for all}\:i,\:\vec{\sigma}_i=+\:\text{iff}\:xy\in P_i\}.
\end{align*}

\begin{proposition}
    For all $\vec{P}\in D|_Y$, we have that $g_Y(\vec{P})$ is indeed in $S(Y)$.
\end{proposition}

\begin{proof}
Let $\vec{P}\in D|_Y$. By construction,
\begin{align*}
    \vec{P}\in \bigcap_{g_Y(\vec{P})}s_D(v).
\end{align*}
Therefore, $g_Y(\vec{P})$ is a simplex of $N_D$. Also by construction, it only involves alternatives in $Y$, i.e., if $U_{\alpha\beta}^{\vec{\sigma}}\in g_Y(\vec{P})$, then $\alpha,\beta\in Y$. We still have to prove that $g_Y(\vec{P})$ is a $(\binom{|Y|}{2}-1)$-simplex, i.e., that it has cardinality $\binom{|Y|}{2}$. 

Observe that $\vec{P}$ is an $n$-tuple of strict total orders on $Y$, which in particular are asymmetric and total. Then for all $x,y\in Y$, $x\neq y$ there exists a unique $\vec{\sigma}\in \{+,-\}^n$ such that for all $i\in N$,  $\vec{\sigma}_i=+$ iff $xy\in  P_i $. Since there are $\binom{|Y|}{2}$ different pairs of alternatives in $Y$, we have that $g_Y(\vec{P})$ is of cardinality $\binom{|Y|}{2}$. Therefore, $g_Y(\vec{P})\in S(Y)$.
\end{proof}

We will show that the function that we define next is the inverse function of $g_Y$. Let $h_Y\colon S(Y)\to D|_Y$ defined as follows:
\begin{align*}
    &\text{for all $\{v_1,v_2,\dots,v_{\binom{|Y|}{2}}\}\in S(Y)$}, \:  h_Y(\{v_1,v_2,\dots,v_{\binom{|Y|}{2}}\})=\vec{P}|_Y, \\
    &\text{for all} \: \vec{P}\in \bigcap_{i=1}^{\binom{|Y|}{2}}s_D(v_i).
\end{align*}

\begin{proposition}
$h$ is well-defined. 
\end{proposition}

\begin{proof}
Firstly, notice that $\bigcap_{i=1}^{\binom{|Y|}{2}}s_D(v_i)$ is non-empty since $\{v_1,v_2,\dots,v_{\binom{|Y|}{2}}\}\in S(Y)$. 

Finally, notice that if $\vec{P}, \vec{P}'\in \bigcap_{i=1}^{\binom{|Y|}{2}}s_D(v_i)$, then $\vec{P}|_Y=\vec{P}'|_Y$.
\end{proof}

\begin{proposition}
$g_Y$ is a bijection with inverse function $h_Y$.    
\end{proposition}

\begin{proof}
To show: $h_Y(g_Y(x))=x$ for all $x\in D|_Y$ and $g_Y(h_Y(x))=x$ for all $x\in S(Y)$. 

Let $\vec{P}\in D|_Y$. We have that
\begin{align*}
    h_Y(g_Y(\vec{P}))=\vec{P}'|_Y\: \text{for all} \: \vec{P}'\in \bigcap_{u\in g_Y(\vec{P})}s_D(u).
\end{align*}

To show: $\vec{P}=\vec{P}'|_Y$. Let $i\in N$. Since $P_i$ and $P'_i|_Y$ are strict total orders on $Y$, it suffices to show that for all $x,y\in Y$, if $xy\in P_i$, then $xy\in P'_i|_Y$. Let $xy\in P_i$. Since $g_Y(\vec{P})\in S(Y)$ and $N_D$ is chromatic, there exists $U_{xy}^{\vec{\sigma}}$ such that for all $j\in N$, $\vec{\sigma}_j=+$ iff $xy\in P_j$. Hence, the fact that $xy\in P_i$ implies $\vec{\sigma}_i=+$. On the other hand, notice that $\vec{P}'\in s_D(U_{xy}^{\vec{\sigma}})$. Hence, $xy\in \vec{P}'_i$. Finally we get $xy\in P'_i|_Y$ due to the fact that $x,y\in Y$. 

Let $\{v_1,v_2,\dots,v_{\binom{|Y|}{2}}\}\in S(Y)$. We have that
\begin{align*}
     g_Y(h_Y(\{v_1,v_2,\dots,v_{\binom{|Y|}{2}}\}))&=\{U_{xy}^{\vec{\sigma}}\in L\colon x,y\in Y;x\neq y;\text{for all}\:i,\:\sigma_i=+\\
    &\text{iff}\:xy\in P_i|_Y\}, \:\text{for all}\:\vec{P}\in\bigcap_{i=1}^{\binom{|Y|}{2}}s_D(v_i). 
\end{align*}

To show: $\{v_1,v_2,\dots,v_{\binom{|Y|}{2}}\}=g(h(\{v_1,v_2,\dots,v_{\binom{|Y|}{2}}\}))$. 

Notice that both $\{v_1,v_2,\dots,v_{\binom{|Y|}{2}}\}$ and $g_Y(h_Y(\{v_1,v_2,\dots,v_{\binom{|Y|}{2}}\}))$ belong to $S(Y)$ and remember that $N_D$ is chromatic. Therefore, for all distinct $x$ and $y$ in $Y$, there exists a unique $\vec{\sigma}\in \{+,-\}^n$ and 
a unique $\vec{\sigma}'\in \{+,-\}^n$ such that $U_{xy}^{\vec{\sigma}}\in \{v_1,v_2,\dots,v_{\binom{|Y|}{2}}\}$ and $U_{xy}^{\vec{\sigma}'}\in g_Y(h_Y(\{v_1,v_2,\dots,v_{\binom{|Y|}{2}}\}))$. Clearly, if we show that $\vec{\sigma}=\vec{\sigma}'$ we are done. Fix $\vec{P}\in \bigcap_{i=1}^{\binom{|Y|}{2}}s_D(v_i)$ and observe the following three things:
\begin{enumerate}
    \item Since $U_{xy}^{\vec{\sigma}'}\in g_Y(h_Y(\{v_1,v_2,\dots,v_{\binom{|Y|}{2}}\}))$, we have that, for all $i\in N$, $\vec{\sigma}'_i=+$ iff $xy\in P_i|Y$.
    \item Since $\vec{P}\in \bigcap_{i=1}^{\binom{Y}{2}}s_D(v_i)$ and $U_{xy}^{\vec{\sigma}}\in \{v_1,v_2,\dots,v_{\binom{|Y|}{2}}\}$, we have $\vec{P}\in s_D(U_{xy}^{\vec{\sigma}})$. Then, for all $i\in N$, $\vec{\sigma}_i=+$ iff $xy \in P_i$. 
    \item Since $x,y\in Y$, for all $i\in N$, $xy\in P_i$ iff $xy\in P_i|_Y$.
\end{enumerate}

Taking these three facts together, we obtain, for all $i\in N$, $\vec{\sigma}_i=+$ iff $\vec{\sigma}'_i=+$. Therefore, $\vec{\sigma}=\vec{\sigma}'$. 
\end{proof}

\subsection{A Bijection from $W(X)$ to the facets of $N_{W(X)}$}\label{subsec:WX-to-NWX}
All the results in this section were already proven (or implicitly implied in an obvious way) in \cite{Baryshnikov}, but we will be more explicit about certain details and omit others. We as well use some different terminology than \cite{Baryshnikov}.

\begin{proposition}\label{prop:chromaticity-of-NWX}
    The simplicial complex $N_{W(X)}$ together with a labeling \\ $\chi\colon V(N_{W(X)}) \to \{Y\subseteq X\colon |Y|=2\}$ defined as
    \begin{align*}
    \chi(U_{\alpha\beta}^{\sigma})=\{\alpha,\beta\}
    \end{align*}
    is a chromatic simplicial complex.
\end{proposition}

The proof of this proposition can be easily adapted from the proof of Proposition \ref{prop:chromaticity-of-ND}, hence it is omitted.

\begin{corollary}\label{cor:color}
If $t$ is a simplex of $N_{W(X)}$, then $\text{dim}(t)\le \binom{|X|}{2}-1$.
\end{corollary}

Let $A(N_W(X))$ be the set of simplices of maximum dimension among the simplices of $N_{W(X)}$ (we will later show that $N_{W(X)}$ is pure, so we will see that $A(N_W(X))$ coincide with the set of facets of $N_{W(X)}$). We introduce a function between $W(X)$ and the set of all facets of $A(N_{W(X)})$.
Let $\bar{g}\colon W(X) \to A(N_{W(X)})$ defined as follows:
\begin{align*}
    \bar{g}(P)=\{U_{xy}^+\colon xy\in P\}.
\end{align*}

\begin{proposition}
    $\bar{g}$ is well-defined and $\bar{g}(P)$ is a $(\binom{|X|}{2}-1)$-simplex for all $\vec{P}\in W(X)$
\end{proposition}

\begin{proof}
    Let $P\in W(X)$. By totality and asymmetry of $P$, we have $|P|=\binom{|X|}{2}-1$. Then $|g(P)|=\binom{|X|}{2}$. Then by construction, $P$ belongs to the intersection of the elements of $g(P)$. So $g(P)$ is a $(\binom{|X|}{2}-1)$-simplex. Finally, by Corollary \ref{cor:color}, $g(P)$ is of maximum dimension among the simplices of $N_{W(X)}$.
\end{proof}

Consider a function $\bar{h}\colon A(N_{W(X)})\to W(X)$ defined as:
\begin{align*}
    \bar{h}(\{v_1,\dots,v_{\binom{|X|}{2}}\})=P \: \text{such that} \: P\in \bigcap_{i=1}^{\binom{|X|}{2}} v_i.
\end{align*}
It is not hard to see that $\bar{h}$ is well-defined (observe that $\bigcap_{i=1}^{\binom{|X|}{2}} v_i$ has to have a unique element).

We want to show the following:
\begin{proposition}
$\bar{g}$ is a bijection with inverse function $\bar{h}$.
\end{proposition}

\begin{proof}
Let $P\in W(X)$. We have that $\bar{h}(\bar{g}(P)))=\bar{h}(\{U_{xy}^+\colon xy\in P\})=P'$ such that $P'\in \bigcap_{u\in \bar{g}(P)} u$, then, it is easy to see that that $P=P'$. 

Let $\{v_1,\dots,v_{\binom{|X|}{2}}\}\in A(N_{W(X)})$. We have that 
\begin{align*}
\bar{g}(\bar{h}(\{v_1,\dots,v_{\binom{|X|}{2}}\}))=\{U_{xy}^+\colon xy\in P\}
\end{align*}
such that $P\in \bigcap_{i=1}^{\binom{|X|}{2}} v_i$. Invoking chromaticity of $N_{W(X)}$, it is not hard to see that the desired result holds.
\end{proof}

So we have shown that there is a bijection between the strict total orders on $X$ and the set of simplices of maximum dimension of $N_{W(X)}$. Finally, if we show that $N_{W(X)}$ is pure, $A(N_{W(X)})$ is also the set of facets of $N_{W(X)}$, and so $g$ could then be thought as a bijection between the strict total orders on $X$ and the set of facets of $N_{W(X)}$. So let us prove it.

\begin{proposition}
    The simplicial complex $N_{W(X)}$ is pure.
\end{proposition}

\begin{proof}
Let $t$ be a facet of $N_{W(X)}$. Since this facet was chosen in an arbitrary way, to prove that the simplicial complex $N_{W(X)}$ is pure, it suffices to show that $\text{dim}(t)=\binom{|X|}{2}-1$. 

Observe that we can write $t$ as $\{u_1,\dots,u_{\text{dim}(t)+1}\}$. Also, since $t$ is a simplex of $N_{W(X)}$, there exists a strict total order $P$ on $X$ in $\cap_{i=1}^{\text{dim}(t)+1}u_i$. Clearly, $P$ is of cardinality $\binom{|X|}{2}$.   

Since we have seen that simplices of maximum dimension of $N_{W(X)}$ are of dimension $\binom{|X|}{2}-1$, we have that $\text{dim}(t)\le \binom{|X|}{2}-1$. Hence, if we show that $\text{dim}(t)<\binom{|X|}{2}-1$ leads to contradiction we are done. Let us do it. Suppose that $\text{dim}(t)<\binom{|X|}{2}-1$. Since $|P|=\binom{|X|}{2}$, the hypothesis $\text{dim}(t)<\binom{|X|}{2}-1$ implies the existence of a pair of alternatives $x,y\in X$ such that $xy\in P$ and $U_{xy}^+\not\in t$. Observe that $P\in (\cap_{i=1}^{\text{dim}(t)+1}u_i)\cap U_{xy}^+$. But then $t'=t\cup\{U_{xy}^+\}$ is a simplex of $N_{W(X)}$ with $t$ as a face. But then $t$ is not a facet of $N_{W(X)}$, a contradiction.
\end{proof}

\subsection{A Bijection from $\mathcal{F}_D$ to $\mathcal{M}_D$}\label{subsec:FD-to-MD}

The following construction is a straightforward generalization of the bijection from $\mathcal{F}_{N(W)^n}$ and to $\mathcal{M}_{N(W)^n}$ established by \cite{Baryshnikov}. 

\begin{definition}
    Let $\mathcal{B}\colon \mathcal{F}_D\to \mathcal{M}_D$ such that $\mathcal{B}(F)$ is the chromatic simplicial map defined as follows: $\mathcal{B}(F)$ assigns any vertex $U_{\alpha\beta}^{\vec{\sigma}}$ of $N_D$ to the vertex $U_{\alpha\beta}^{\sigma}$ of $N_{W(X)}$, where $\sigma=+$ iff we have the following: $\alpha F(\vec{P}) \beta$, for every $\vec{P}\in s_D(U_{\alpha\beta}^{\vec{\sigma}})$. 
\end{definition}

\begin{proposition}
If $F\in \mathcal{F}_D$, then $\mathcal{B}(F)$ is well-defined.
\end{proposition}

The proof is an easy generalization of the proof that appears in \cite{RajsbaumArmajacPODC} for the case of the unrestricted domain.

\begin{proof}
Let $U_{\alpha\beta}^{\vec{\sigma}}$ be a vertex of $N_D$. Let $\vec{P},\vec{P}'\in s_D(U_{\alpha\beta}^{\vec{\sigma}})$. To show: $\alpha F(\vec{P}) \beta$ iff $\alpha F(\vec{P}') \beta$. Since $\vec{P},\vec{P}'\in s(U_{\alpha\beta}^{\vec{\sigma}})$, for all $i\in N$, $\alpha P_i \beta$ iff $\sigma_i=+$ and $\alpha P'_i \beta$ iff $\sigma_i=+$. Hence, for all $i\in N$, $\alpha P_i \beta$ iff $\alpha P'_i \beta$. Then, since $F$ satisfies IIA, we get $\alpha F(\vec{P}) \beta$ iff $\alpha F(\vec{P}') \beta$.
\end{proof}

It is not hard to show that $\mathcal{B}(F)$ is in fact a chromatic simplicial map.

The next step is to prove that it is a bijection, whose inverse function, 
\begin{align*}
    \mathcal{B}^{-1}\colon M_D \to F_D,
\end{align*}
is defined as follows: for every $f\in \mathcal{M}_D$, we have that $\mathcal{B}^{-1}(f)\colon D \to W(X)$ is a function defined as $(\mathcal{B}^{-1}(f))(\vec{P})=\bar{h}(f(g_X(\vec{P})))$, where the functions $g_X\colon D \to S(X) $ and $\bar{h}\colon A(N_{W(X)}) \to W(X)$ are defined in Sections \ref{subsec:DY-to-SY} and \ref{subsec:WX-to-NWX}, respectively. Notice that $g_X(p)$ lives in $S(X)$, and since $f$ is rigid and $N_D$ is of dimension $\binom{|X|}{2}-1$ (these things are easy to check), $f(g_X(p))$ is a facet of $N_{W(X)}$, i.e., it belongs to $A(N_{W(X)})$. 

Before showing that $\mathcal{B}$ is a bijection, we first show the following:

\begin{proposition}\label{prop:Binverse-IIA}
    Let $f\in \mathcal{M}_D$. The SWF $\mathcal{B}^{-1}(f)$ indeed satisfies IIA.
\end{proposition}

\begin{proof}
Let $\alpha,\beta\in X$, $\alpha\neq \beta$, and $\vec{P}',\vec{P}''\in D$ such that for all $i\in N$, $\alpha\beta\in P'_i$ iff $\alpha\beta\in P''_i$. 
To show: $\alpha\beta\in (\mathcal{B}^{-1}(f))(\vec{P}')$ iff $\alpha\beta\in (\mathcal{B}^{-1}(f))(\vec{P}'')$.

By definition, 
\begin{align*}
    (\mathcal{B}^{-1}(f))(\vec{P}')&=\bar{h}(f(g_X(\vec{P'})))=P' \:\text{such that} \: P'\in \bigcap_{v\in f(g_X(\vec{P}'))}v,
\end{align*}
where $f(g_X(\vec{P}'))=\{f(U_{xy}^{\vec{\sigma}})\colon x,y\in X; x\neq y; \text{for all}\:i,\vec{\sigma}_i=+\:\text{iff}\:xy\in P'_i\}$. Also,
\begin{align*}
    (\mathcal{B}^{-1}(f))(\vec{P}'')&=\bar{h}(f(g_X(\vec{P''})))=P'' \:\text{such that} \: P''\in \bigcap_{v\in f(g_X(\vec{P}''))}v,
\end{align*}
where $f(g_X(\vec{P}''))=\{f(U_{xy}^{\vec{\sigma}})\colon x,y\in X; x\neq y; \text{for all}\:i,\vec{\sigma}_i=+\:\text{iff}\:xy\in P''_i\}$. 

By chromaticity of $N_{W(X)}$ and $f$, there exist $U_{\alpha\beta}^{\vec{\sigma}'}\in g_X(\vec{P}')$ and $U_{\alpha\beta}^{\vec{\sigma}''}\in g_X(\vec{P}'')$ such that (for all $i\in N$, $\vec{\sigma}'_i=+$ iff $\alpha\beta\in P'_i$) and (for all $i\in N$, $\vec{\sigma}''_i=+$ iff $\alpha\beta\in P''_i$). But by hypothesis, for all $i\in N$, $\alpha\beta\in P'_i$ iff $\alpha\beta\in P''_i$. But then, for all $i\in N$, $\vec{\sigma}'_i=+$ iff $\vec{\sigma}''_i=+$. Therefore,  $U_{\alpha\beta}^{\vec{\sigma}'}=U_{\alpha\beta}^{\vec{\sigma}''}$. Hence, $f(U_{\alpha\beta}^{\vec{\sigma}'})=f(U_{\alpha\beta}^{\vec{\sigma}''})$. Therefore, $\alpha\beta\in (\mathcal{B}^{-1}(f))(\vec{P}')$ iff $\alpha\beta\in (\mathcal{B}^{-1}(f))(\vec{P}'')$.
\end{proof}

\begin{proposition}\label{prop:bijectionB}
$\mathcal{B}$ is a bijection with inverse function $\mathcal{B}^{-1}$.
\end{proposition}

\begin{proof}
First, we have to show that for all $f\in M_D$, it holds that $\mathcal{B}(\mathcal{B}^{-1}(f))\colon N_D \to N_{W(X)}$ is such that $\mathcal{B}(\mathcal{B}^{-1}(f))=f$.

Let $U_{\alpha\beta}^{\vec{\sigma}}\in V(N_D)$. To show: 
$\mathcal{B}(\mathcal{B}^{-1}(f))(U_{\alpha\beta}^{\vec{\sigma}})=f(U_{\alpha\beta}^{\vec{\sigma}})$.

By definition of $\mathcal{B}$, we have that $\mathcal{B}(\mathcal{B}^{-1}(f))(U_{\alpha\beta}^{\vec{\sigma}})=U_{\alpha\beta}^\sigma$ such that $\sigma=+$ iff $\alpha\beta\in (\mathcal{B}^{-1}(f))(\vec{P})$ for all $\vec{P}\in s_D(U_{\alpha\beta}^{\vec{\sigma}})$. 

This is well-defined since $\mathcal{B}^{-1}(f)$ satisfies IIA by Proposition \ref{prop:Binverse-IIA}. So we can fix $\vec{P}\in s_D(U_{\alpha\beta}^{\vec{\sigma}})$ and we have that 
\begin{align*}
    (\mathcal{B}^{-1}(f))(U_{\alpha\beta}^{\vec{\sigma}})=U_{\alpha\beta}^{\sigma}\:\text{such that}\:(\sigma=+\:\text{iff}\:\alpha\beta\in (\mathcal{B}^{-1}(f))(\vec{P})).
\end{align*}
Then, 
\begin{align}\label{cond:cond1}
(\mathcal{B}^{-1}(f))(U_{\alpha\beta}^{\vec{\sigma}})=&U_{\alpha\beta}^{\sigma}\:\text{such that}\:(\sigma=+\:\text{iff}\:\alpha\beta\in P'\:
\text{such that} \: P'\in \bigcap_{v\in f(g_X(\vec{P}))}v),
\end{align}
where $f(g_X(\vec{P}))=\{f(U_{xy}^{\vec{\sigma}})\colon x,y\in X; X\neq y; \text{for all}\:i,\vec{\sigma}_i=+\:\text{iff}\:xy\in P_i\}$
On the other hand, if $f(U_{\alpha\beta}^{\vec{\sigma}})=U_{\alpha\beta}^{\sigma'}$, since $\vec{P}\in s_D(U_{\alpha\beta}^{\vec{\sigma}})$, we have the following:
\begin{align}\label{cond:cond2}
    f(U_{\alpha\beta}^{\vec{\sigma}})=U_{\alpha\beta}^{\sigma'}\in f(g_X(\vec{P})).
\end{align}
 We proceed by contradiction supposing $\sigma\neq \sigma'$. We now proceed by cases.

 Case 1: $\sigma=+$ and $\sigma'=-$. Since $\sigma=+$, by expression (\ref{cond:cond1}), $\alpha\beta\in P'$. Also notice that by expression (\ref{cond:cond2}), $U_{\alpha\beta}^-\in f(g_X(\vec{P}))$, then, by definition of $(\mathcal{B}^{-1}(f))(\vec{P})$, we have $\beta\alpha \in P'$, a contradiction to the asymmetry of $P'$.

 Case 2: $\sigma=-$ and $\sigma'=+$. This case leads to a contradiction in an analogous way. 

The second part of this proof consists on proving that 
\begin{align*}
\mathcal{B}^{-1}(\mathcal{B}(F))\colon D \to W(X)
\end{align*}
is such that $\mathcal{B}^{-1}(\mathcal{B}(F))=F$, for all $F\in \mathcal{F}_D$.

Let $\vec{P}\in D$. To show: $(\mathcal{B}^{-1}(\mathcal{B}(F)))(\vec{P})=F(\vec{P})$. 

By definition of our functions,
\begin{align*}
(\mathcal{B}^{-1}(\mathcal{B}(F)))(\vec{P})&=\bar{h}((\mathcal{B}(F))(g_X(\vec{P})))=P \: \text{such that} \: P\in \bigcap_{v\in (\mathcal{B(F)})(g_X(\vec{P}))}v,
\end{align*}
where $(\mathcal{B(F)})(g_X(\vec{P}))=\{(\mathcal{B}(F))(U_{xy}^{\vec{\sigma}}\colon\text{for all}\:i,\vec{\sigma}_i=+\:\text{iff}\:xy\in P_i\}$.

So, we want to prove that $P=F(\vec{P})$. Let $\alpha\beta\in X$, $\alpha\neq \beta$. To show: $\alpha\beta\in P$ iff $\alpha\beta\in F(\vec{P})$.

Since $(\mathcal{B}(F))(g_Y(\vec{P}))$ is a facet of the chromatic simplicial complex $N_{W(X)}$ and since $\mathcal{B}(F)$ is chromatic, there exists a unique $\vec{\sigma}'\in \{+,-\}^n$ such that $U_{\alpha\beta}^{\vec{\sigma}}\in g_Y(\vec{P})$. Then,
\begin{align}\label{cond:cond3}
    \vec{P}\in s_D(U_{\alpha\beta}^{\vec{\sigma}}).
\end{align}
On the other hand, by definition of $\mathcal{B}$,
\begin{align}\label{cond:cond4}
(\mathcal{B}(F))(U_{\alpha\beta}^{\vec{\sigma}'})=U_{\alpha\beta}^{\sigma}\:
\text{such that} \: (\sigma=+ \: \text{iff} \: \alpha\beta\in F(\vec{P}') \: \text{for all}\: \vec{P'}\in s_D(U_{\alpha\beta}^{\vec{\sigma}'})).
\end{align}
Combining expressions (\ref{cond:cond3}) and (\ref{cond:cond4}), we get that
\begin{align}\label{cond:cond5}
(\mathcal{B}(F))(U_{\alpha\beta}^{\vec{\sigma}'})=U_{\alpha\beta}^{\sigma}\:\text{such that} \: (\sigma=+ \: \text{iff} \: \alpha\beta\in F(\vec{P})).
\end{align}

First we show that $P\subseteq F(\vec{P})$. Suppose $\alpha\beta\in P$. Then since $P\in U_{\alpha\beta}^{\sigma}$, we have that $\sigma=+$ (otherwise, we would contradict the asymmetry of $P$). But then, by \ref{cond:cond5}, $\alpha\beta\in F(\vec{P})$. 

Finally, we show that $F(\vec{P})\subseteq P$. Suppose $\alpha\beta\in F(\vec{P})$. Then by \ref{cond:cond5}, we have $\sigma=+$, so $(\mathcal{B}(F))(U_{\alpha\beta}^{\vec{\sigma}'})=U_{\alpha\beta}^{+}$. Then $P\in U_{\alpha\beta}^{+}$. But then $\alpha\beta\in P$. 
\end{proof}

\subsection{Proof of Theorem \ref{thrm:equivalence}}

We will use the following lemma to prove Theorem \ref{thrm:equivalence}. An analogous proposition and proof for the particular case of the unrestricted domain can be consulted in \cite{RajsbaumR2022preprint-new}. 

\begin{lemma}\label{lem:equivalence}
Let $F\in F_D$ and $\mathcal{B}$ the function defined in Subsection \ref{subsec:FD-to-MD}. The following hold:
\begin{enumerate}
    \item $\mathcal{B}(F)$ is unanimous iff $F$ is unanimous.
    \item  $\mathcal{B}(F)$ is dictatorial iff $F$ is dictatorial.
\end{enumerate}
\end{lemma}

\begin{proof}
Let us prove $(1)$. By definition of $\mathcal{B}$, for every vertex of $N_D$ of the form $U_{\alpha\beta}^{\vec{\sigma}^N}$, we have that
$(\mathcal{B}(F))(U_{\alpha\beta}^{\vec{\sigma}^N})=U_{\alpha\beta}^+$ iff $\alpha F(\vec{P})\beta$ for all $\vec{P}\in s_D(U_{\alpha\beta}^{\vec{\sigma}^N})$. 

We start with the $\Rightarrow$ direction. Suppose $\mathcal{B}(F)$ is unanimous. Then \\ $(\mathcal{B}(F))(U_{\alpha\beta}^{\vec{\sigma}^N})=U_{\alpha\beta}^+$. Hence, $\alpha F(\vec{P})\beta$ for all $\vec{P}\in s_D(U_{\alpha\beta}^{\vec{\sigma}^N})$. Hence, $F$ is unanimous.

Now we prove the $\Leftarrow$ direction. Suppose $F$ is unanimous. Then for all $\vec{P}\in s_D(U_{\alpha\beta}^{\vec{\sigma}^N})$, we have that $\alpha F(\vec{P})\beta$, then $(\mathcal{B}(F))(U_{\alpha\beta}^{\vec{\sigma}^N})=U_{\alpha\beta}^+$. Therefore, $\mathcal{B}(F)$ is unanimous.

 We now want to prove $(2)$. Remember that for every vertex $U_{\alpha\beta}^{\vec{\sigma}}$ of $N_D$, we have that
$(\mathcal{B}(F))(U_{\alpha\beta}^{\vec{\sigma}})=U_{\alpha\beta}^+$ iff $\alpha F(\vec{P})\beta$ for all $\vec{P}\in s_D(U_{\alpha\beta}^{\vec{\sigma}})$. 

We begin with the $\Rightarrow$ direction. Suppose $\mathcal{B}(F)$ is dictatorial. Let $i\in N$ be a dictator. If $\vec{P}\in D$, we want to show that $\alpha P_i \beta$ implies $\alpha F(\vec{P}) \beta$. Suppose $\vec{P}\in D$. Let $\vec{\sigma}_{\vec{P}}\in \{+,-\}^n$ be such that for all $j\in N$, $(\vec{\sigma}_{\vec{P}})_j=+$ iff $\alpha P_j \beta$. Then $\vec{P}\in s_D(U_{\alpha\beta}^{\vec{\sigma}_{\vec{P}}})$. On the other hand, since $\mathcal{B}(F)$ is dictatorial, $(\mathcal{B}(F))(U_{\alpha\beta}^{\vec{\sigma}_{\vec{P}}})=U_{\alpha\beta}^+$. Hence, since $\vec{P}\in s_D(U_{\alpha\beta}^{\vec{\sigma}_{\vec{P}}})$, it holds that $\alpha F(\vec{P})\beta$. 

We now show that $\Leftarrow$ direction. Suppose $F$ is dictatorial. Let $i\in N$ be a dictator. Let $U_{\alpha\beta}^{\vec{\sigma}}$ be a vertex of $N_D$. To show: $f(U_{\alpha\beta}^{\vec{\sigma}})=U_{\alpha\beta}^{\vec{\sigma}_i}$. For every $\vec{P}\in s_D(U_{\alpha\beta}^{\vec{\sigma}})$, we have that $F$ being dictatorial implies that $\alpha F(\vec{P}) \beta$ iff $\vec{\sigma}_i=+$. But then $f(U_{\alpha\beta}^{\vec{\sigma}})=U_{\alpha\beta}^{\vec{\sigma}_i}$.
\end{proof}

Here is the proof of Theorem \ref{thrm:equivalence}:

\begin{proof}
We start with the $\Rightarrow$ direction. Suppose $D$ is Arrow-inconsistent. Let $f\colon N_D\to N_W(X)$ be a chromatic simplicial map, i.e., $f\in \mathcal{M}_D$, satisfying unanimity. To show: $f$ is dictatorial. 

Recall that in Proposition \ref{prop:bijectionB}, we proved that the function $\mathcal{B}$ (defined in Subsection \ref{subsec:FD-to-MD}) is a bijection with inverse $\mathcal{B}^{-1}$. By definition of $\mathcal{B}^{-1}$, we have that $\mathcal{B}^{-1}(f)\in \mathcal{F}_D$. Also, since $\mathcal{B}^{-1}$ is the inverse function of the bijection $\mathcal{B}$, we have $\mathcal{B}(\mathcal{B}^{-1}(f))=f$. Therefore, by part $1$ of Lemma \ref{lem:equivalence}, $f$ being unanimous implies that $\mathcal{B}^{-1}(f)$ is unanimous. Then, since $D$ is Arrow-inconsistent, $\mathcal{B}^{-1}(f)$ is dictatorial. Then, by part $2$ of Lemma \ref{lem:equivalence}, $f$ is dictatorial.

Let us prove the $\Leftarrow$ direction. Suppose that for all $f\in \mathcal{M}_D$, $f$ unanimous implies that $f$ is dictatorial. We want to show that $D$ is Arrow-inconsistent. Let $F\in \mathcal{F}_D$ satisfying unanimity. To show: $F$ is dictatorial. 

Observe that $\mathcal{B}(F)\in \mathcal{M}_D$. Since $F$ is unanimous, by part $1$ of Lemma \ref{lem:equivalence} we have that $\mathcal{B}(F)$ is unanimous. Then applying our hypothesis, $\mathcal{B}(F)$ is dictatorial. Then by part $2$ of Lemma \ref{lem:equivalence}, $F$ is dictatorial. 
\end{proof}

\subsection{Proof of Lemma \ref{lem:almost-dec-compl}}

\begin{proof}
Observe that $U_{\beta\alpha}^{\vec{\sigma}^G}=U_{\alpha\beta}^{\vec{\sigma}^{G^c}}$, $U_{\beta\alpha}^+=U_{\alpha\beta}^-$ and $f(U_{\beta\alpha}^{\vec{\sigma}^G})=U_{\beta\alpha}^+$. Taking these observations together yields the desired result.
\end{proof}

\subsection{Proof of Lemma \ref{lem:ultra-dict}}

\begin{proof}
Suppose $\mathcal{G}$ is an ultrafilter of $N$. Since $N$ is finite, by Theorem \ref{dictator_in_ultra} there exists a voter, call it $d$, such that $\mathcal{G}=\{B\subseteq N: d\in B\}$.

Let $U_{\alpha\beta}^{\vec{\sigma}}$ be a vertex of $N_D$. Since $U_{\alpha\beta}^{\vec{\sigma}}$ is an arbitrary vertex of $N_D$, by Definition \ref{def:simplMap-unanimity-dict} we have that $d$ is a dictator for $f$ if $f(U_{\alpha\beta}^{\vec{\sigma}})=U_{\alpha\beta}^{\vec{\sigma}_d}$. Clearly, there exists a coalition $G$ of $N$ such that $\vec{\sigma}=\vec{\sigma}^G$. Therefore, it suffices to show that $f(U_{\alpha\beta}^{\vec{\sigma}^G})=U_{\alpha\beta}^{\vec{\sigma}^G_d}$.

By property $3$ of the definition of an ultrafilter, $G$ or $G^c$ is an element of $\mathcal{G}$, i.e., one of them is almost-decisive. We proceed by checking the two possible cases. 

Case $1$: $G\in \mathcal{G}$. Then by definition of almost-decisiveness $f(U_{\alpha\beta}^{\vec{\sigma}^G})=U_{\alpha\beta}^+$. Also, since $G\in \mathcal{G}$, voter $d$ is in $G$, so $\vec{\sigma}^G_d=+$. Therefore, we have $f(U_{\alpha\beta}^{\vec{\sigma}^G})=U_{\alpha\beta}^{\vec{\sigma}^G_d}$.

Case $2$: $G^c\in \mathcal{G}$. Then by Lemma \ref{lem:almost-dec-compl}, we have  
$f(U_{\alpha\beta}^{\vec{\sigma}^G})=U_{\alpha\beta}^-$. Also, since $G\in \mathcal{G}$, voter $d$ is in $G^c$, we have that $\vec{\sigma}^G_d=-$. Therefore, we have $f(U_{\alpha\beta}^{\vec{\sigma}^G})=U_{\alpha\beta}^{\vec{\sigma}^G_d}$.

Therefore, $d$ is a dictator for $f$, so $f$ is dictatorial.
\end{proof}

\subsection{Proof of Lemma \ref{lem:not-boundary-1}}

We will use the subsequent lemma to prove Lemma \ref{lem:not-boundary-1}.

\begin{lemma}\label{lem:critical-triangle}
Let $f\colon N_D\to N_{W(X)}$ be a unanimous chromatic simplicial map. If $\{U_{bc}^{\vec{\sigma}}, U_{ca}^{\vec{\sigma}'}, U_{ab}^{\vec{\sigma}^N}\}$ is a triangle of $N_D$, then the edge  $\{U_{bc}^{\vec{\sigma}}, U_{ca}^{\vec{\sigma}'}\}$ cannot by mapped by $f$ to $\{U_{bc}^+, U_{ca}^+\}$.  
\end{lemma}

\begin{proof}
 Suppose $\{U_{bc}^{\vec{\sigma}}, U_{ca}^{\vec{\sigma}'}, U_{ab}^{\vec{\sigma}^N}\}$ is a triangle of $N_D$ and denote it $T$. We proceed by contradiction: suppose $\{U_{bc}^{\vec{\sigma}}, U_{ca}^{\vec{\sigma}'}\}$ is mapped by $f$ to $\{U_{bc}^+, U_{ca}^+\}$. By unanimity, $f(U_{ab}^{\vec{\sigma}^N})=U_{ab}^+$. Then $T$ is mapped by $f$ to $\{U_{ab}^+, U_{bc}^+, U_{ca}^+\}$, which is not a simplex (since it corresponds to the intransitive ranking $abca$). 
\end{proof}

Here is the proof of Lemma \ref{lem:not-boundary-1}.

\begin{proof}
Suppose $B_1(G, Y)$ (resp. $B_2(G, Y)$) is a subcomplex of $N_D$. Then the triangles $\{U_{ac}^{\vec{\sigma}^G}, U_{ba}^{\vec{\sigma}^{G^c}}, U_{cb}^{\vec{\sigma}^{\varnothing}}\}$ and $\{U_{ac}^{\vec{\sigma}^{G^c}}, U_{ba}^{\vec{\sigma}^G}, U_{cb}^{\vec{\sigma}^N}\}$ (resp. $\{U_{ac}^{\vec{\sigma}^G}, U_{ba}^{\vec{\sigma}^{G^c}}, U_{cb}^{\vec{\sigma}^N}\}$ and $\{U_{ac}^{\vec{\sigma}^{G^c}}, U_{ba}^{\vec{\sigma}^G}, U_{cb}^{\vec{\sigma}^{\varnothing}}\}$) are triangles of $N_D$. The desired results follow by applying Lemma \ref{lem:critical-triangle}.
\end{proof}

\subsection{Proof of Lemma \ref{lem:not-boundary-2}}

We start with a remark.

\begin{remark}\label{rmk:edges-of-Bi}
The simplicial complex $B_i(G, \{\alpha, \beta, \gamma\})$, for all $i\in \{1,2\}$, contains all the edges of the form $\{U_{ab}^{\vec{\sigma}^G}, U_{ca}^{\vec{\sigma}^{G^c}}\}$ for all $a,b,c\in \{\alpha,\beta,\gamma\}$ (see Figure \ref{fig:B1-B2}).
\end{remark}

To provide additional details, if a profile $\vec{P}$ in a domain $D$, i.e., a facet of $N_D$, has an edge of the form $\{U_{ab}^{\vec{\sigma}^G}, U_{ca}^{\vec{\sigma}^{G^c}}\}$ as a face, that means that in that profile any voter in $G$ disagrees with any voter in $G^c$ on at least two pairs of alternatives: $\{a,b\}$ and $\{a,c\}$. Notice, in Figure \ref{fig:B1-B2}, that every profile of $B_i(G,\{\alpha,\beta,\gamma\})$, for both $i=1,2$, has an edge of this form. Moreover, every profile of $B_i(G,\{\alpha,\beta,\gamma\})$ has a unanimity vertex, i.e., a vertex for which everyone at $N$ agrees on the pair in question. Therefore, as we already said before, every triangle in $B_i(G,\{\alpha,\beta,\gamma\})$ represents a profile in which $G$ and $G^c$ disagree on two pairs of alternatives and agree on the remaining pair. 

We now proceed wíth the proof of Lemma \ref{lem:not-boundary-2}.

\begin{proof}
W.l.o.g. suppose $B_1(G,Y)$ is a subcomplex of $N_D$ (the other case is analogous). Let $\alpha,\beta,\gamma\in Y$, $\alpha\neq \beta\neq \gamma\neq \alpha$. We will only prove the case of an edge of the form $\{U_{\beta\gamma}^{\vec{\sigma}^G}, U_{\alpha\beta}^{\vec{\sigma}^{G^c}}\}$ since the other case is analogous. Denote $\{U_{\beta\gamma}^{\vec{\sigma}^G}, U_{\alpha\beta}^{\vec{\sigma}^{G^c}}\}$ by $e$. Since $f$ is a chromatic simplicial map, to get the desired result it suffices to show that $e$ cannot be mapped to a DbT edge. 

By part $1$ of Lemma \ref{lem:not-boundary-1}, the edge $e$ cannot be mapped to $\{U_{\beta \gamma}^-, U_{\alpha\beta}^-\}$ under $f$, so let us show that it cannot be mapped to the other DbT edge:  $\{U_{\beta \gamma}^+, U_{\alpha\beta}^+\}$. 

We proceed by contradiction: suppose $e$ is mapped by $f$ to $\{U_{\beta\gamma}^+, U_{\alpha\beta}^+\}$. Since $f$ is chromatic, 
\begin{align}\label{eq:f=a+}    f(U_{\beta\gamma}^{\vec{\sigma}^{G^c}})=U_{\beta\gamma}^+.
\end{align}
Observe the following three things:
\begin{itemize}
    \item By chromaticity of $f$, we have that $f(U_{\alpha\beta}^{\vec{\sigma}^{G^c}})=U_{\alpha\beta}^+$.
    \item By Remark \ref{rmk:edges-of-Bi}, we have that $\{U_{\alpha\beta}^{\vec{\sigma}^{G^c}}, U_{\gamma \alpha}^{\vec{\sigma}^G}\}$ is an edge in $B_1(G,Y)$.
    \item By part $2$ of Lemma \ref{lem:not-boundary-1}, we have that $\{ U_{\alpha\beta}^{\vec{\sigma}^{G^c}}, U_{\gamma \alpha}^{\vec{\sigma}^G}\}$ cannot be mapped to $\{U_{\alpha\beta}^+, U_{\gamma\alpha}^+\}$ under $f$.
\end{itemize}
Taking these three observations together as well as the chromaticy of $f$, we get that $\{ U_{\alpha\beta}^{\vec{\sigma}^{G^c}}, U_{\gamma \alpha}^{\vec{\sigma}^G}\}$ is mapped to $\{ U_{\alpha\beta}^+, U_{\gamma\alpha}^-\}$. Now observe the following three things: 
\begin{itemize}
    \item By chromaticity of $f$, we have that $f(U_{\gamma \alpha}^{\vec{\sigma}^G})=U_{\gamma \alpha}^-$.
    \item By Remark \ref{rmk:edges-of-Bi}, we have that $\{U_{\gamma \alpha}^{\vec{\sigma}^G}, U_{\beta \gamma}^{\vec{\sigma}^{G^c}}\}$ is an edge in $B_1(G,Y)$.
    \item By part $1$ of Lemma \ref{lem:not-boundary-1}, we have that $\{U_{\gamma \alpha}^{\vec{\sigma}^G}, U_{\beta \gamma}^{\vec{\sigma}^{G^c}}\}$ cannot be mapped to $\{ U_{\gamma \alpha}^-, U_{\beta \gamma}^-\}$ under $f$.
\end{itemize}
Taking these three observations together as well as the chromaticy of $f$, we get that $\{ U_{\gamma \alpha}^{\vec{\sigma}^G}, U_{\beta \gamma}^{\vec{\sigma}^{G^c}}\}$ is mapped to $\{ U_{\gamma \alpha}^-, U_{\beta \gamma}^+\}$. Now observe the following three things: 
\begin{itemize}
    \item By chromaticity of $f$, we have that $f(U_{\beta \gamma}^{\vec{\sigma}^{G^c}})=U_{\beta \gamma}^+$.
    \item By Remark \ref{rmk:edges-of-Bi}, we have that $\{U_{\beta \gamma}^{\vec{\sigma}^{G^c}},U_{\alpha \beta}^{\vec{\sigma}^G}\}$ is an edge in $B_1(G,Y)$.
    \item By part $2$ of Lemma \ref{lem:not-boundary-1}, we have that $\{U_{\beta \gamma}^{\vec{\sigma}^{G^c}},U_{\alpha\beta}^{\vec{\sigma}^G}\}$ cannot be mapped to $\{ U_{\beta \gamma}^+, U_{\alpha\beta}^+\}$ under $f$.
\end{itemize}
Taking these three observations together as well as the chromaticy of $f$, we get that $\{U_{\beta\gamma}^{\vec{\sigma}^{G^c}}, U_{\alpha\beta}^{\vec{\sigma}^G}\}$ is mapped to $\{ U_{\beta\gamma}^+, U_{\alpha\beta}^-\}$. Now observe the following three things: 
\begin{itemize}
    \item By chromaticity of $f$, we have that $f(U_{\alpha\beta}^{\vec{\sigma}^G})=U_{\alpha\beta}^-$.
    \item By Remark \ref{rmk:edges-of-Bi}, we have that $\{U_{\alpha\beta}^{\vec{\sigma}^G},U_{\gamma \alpha}^{\vec{\sigma}^{G^c}}\}$ is an edge in $B_1(G,Y)$.
    \item By part $1$ of Lemma \ref{lem:not-boundary-1}, we have that $\{U_{\alpha\beta}^{\vec{\sigma}^G},U_{\gamma \alpha}^{\vec{\sigma}^{G^c}}\}$ cannot be mapped to $\{ U_{\alpha\beta}^-, U_{\gamma \alpha}^-\}$ under $f$.    
\end{itemize}
Taking these three observations together as well as the chromaticy of $f$, we get that $\{U_{\alpha\beta}^{\vec{\sigma}^G}, U_{\gamma \alpha}^{\vec{\sigma}^{G^c}}\}$ is mapped to $\{ U_{\alpha\beta}^-, U_{\gamma \alpha}^+\}$. Now observe the following three things: 
\begin{itemize}
    \item By chromaticity of $f$, we have that $f(U_{\gamma \alpha}^{\vec{\sigma}^{G^c}})=U_{\gamma \alpha}^+$.
    \item By Remark \ref{rmk:edges-of-Bi}, we have that $\{U_{\gamma \alpha}^{\vec{\sigma}^{G^c}}, U_{\beta \gamma }^{\vec{\sigma}^G}\}$ is an edge in $B_1(G,Y)$.
    \item By part $2$ of Lemma \ref{lem:not-boundary-1}, we have that $\{U_{\gamma \alpha}^{\vec{\sigma}^{G^c}}, U_{\beta \gamma}^{\vec{\sigma}^G}\}$ cannot be mapped to $\{U_{\gamma \alpha}^+, U_{\beta \gamma}^+\}$ under $f$.    
\end{itemize}
Taking these three observations together as well as the chromaticy of $f$, we get that $\{U_{\gamma\alpha}^{\vec{\sigma}^{G^c}}, U_{\beta \gamma}^{\vec{\sigma}^G}\}$ is mapped to $\{U_{\gamma\alpha}^+, U_{\beta \gamma}^-\}$. Since $f$ is chromatic, we have that $f(U_{\beta \gamma}^{\vec{\sigma}^G})=U_{\beta \gamma}^-$, a contradiction to equation \ref{eq:f=a+}.
\end{proof}

\subsection{Proof of Lemma \ref{lem:dec-G-Gc-Y}}

\begin{proof}
Suppose $B_i(G,Y)$ is a subcomplex of $N_D$ for some $i\in \{1,2\}$ whenever $G$ is non-empty and distinct from $N$.

If $G=\varnothing$ or $G=N$, then $G$ or $G^c$ equals $N$, but then by unanimity of $f$, $G$ or $G^c$ is almost-decisive over $Y$. 

Suppose then $G\neq \varnothing$ and $G\neq N$. Hence, there is $i\in \{1,2\}$ such that $B_i(G,Y)$ is a subcomplex of $N_D$. Let $\alpha,\beta,\gamma\in Y$ with $\alpha\neq \beta\neq \gamma\neq \alpha$. We begin by asking: where could $f$ map edge $\{U_{\beta\gamma}^{\vec{\sigma}^G}, U_{\alpha\beta}^{\vec{\sigma}^{G^c}}\}$ of $B_i(G,Y)$? By Lemma \ref{lem:not-boundary-2}, there are only two options: $\{U_{\beta\gamma}^+,U_{\alpha\beta}^-\}$ or $\{U_{\beta\gamma}^-,U_{\alpha\beta}^+\}$. We proceed by cases. 

Case $1$: $f(\{U_{\beta\gamma}^{\vec{\sigma}^G}, U_{\alpha\beta}^{\vec{\sigma}^{G^c}}\})=\{U_{\beta\gamma}^+,U_{\alpha\beta}^-\}$. Then, by chromaticity of $f$, it holds that 
\begin{align}\label{eqs:v1-and-v2}
    f(U_{\beta\gamma}^{\vec{\sigma}^G})=U_{\beta\gamma}^+\:\text{and}\: f(U_{\alpha\beta}^{\vec{\sigma}^{G^c}})=U_{\alpha\beta}^-.
\end{align}
But then by Lemma \ref{lem:not-boundary-2}, we have that $f(\{U_{\alpha\beta}^{\vec{\sigma}^{G^c}}, U_{\gamma\alpha}^{\vec{\sigma}^G}\})=\{U_{\alpha\beta}^-, U_{\gamma\alpha}^+\}$. Then, by chromaticity of $f$, it holds that 
\begin{align}\label{eq:v3}
    f(U_{\gamma\alpha}^{\vec{\sigma}^G})=U_{\gamma\alpha}^+.
\end{align}
But then by Lemma \ref{lem:not-boundary-2}, we have that $f(\{U_{\gamma\alpha}^{\vec{\sigma}^G},U_{\beta\gamma}^{\vec{\sigma}^{G^c}}\})=\{U_{\gamma\alpha}^+, U_{\beta\gamma}^-\}$. Then, by chromaticity of $f$, it holds that 
\begin{align}\label{eq:v4}
    f(U_{\beta\gamma}^{\vec{\sigma}^{G^c}})=U_{\beta\gamma}^-.
\end{align}
But then by Lemma \ref{lem:not-boundary-2}, we have that $f(\{U_{\beta\gamma}^{\vec{\sigma}^{G^c}},U_{\alpha\beta}^{\vec{\sigma}^G}\})=\{ U_{\beta\gamma}^-, U_{\alpha\beta}^+\}$. Then, by chromaticity of $f$, it holds that 
\begin{align}\label{eq:v5}
    f(U_{\alpha\beta}^{\vec{\sigma}^G})=U_{\alpha\beta}^+.
\end{align}
But then by Lemma \ref{lem:not-boundary-2}, we have that $f(\{U_{\alpha\beta}^{\vec{\sigma}^G}, U_{\gamma\alpha}^{\vec{\sigma}^{G^c}}\})=\{ U_{\alpha\beta}^+, U_{\gamma\alpha}^-\}$. Then, by chromaticity of $f$, it holds that 
\begin{align}\label{eq:v6}
    f(U_{\gamma\alpha}^{\vec{\sigma}^{G^c}})=U_{\gamma\alpha}^-.
\end{align}

Taking \ref{eqs:v1-and-v2} to \ref{eq:v6}, we obtain that $G$ is almost-decisive over $Y$.  

Case $2$: $f(\{U_{\beta\gamma}^{\vec{\sigma}^G}, U_{\alpha\beta}^{\vec{\sigma}^{G^c}}\})=\{U_{\beta\gamma}^-,U_{\alpha\beta}^+\}$. Analogously to case $1$, successively applying chromaticity of $f$ and Lemma \ref{lem:not-boundary-2} we get: 
\begin{align}
&f(U_{\beta\gamma}^{\vec{\sigma}^G})=U_{\beta\gamma}^-,f(U_{\alpha\beta}^{\vec{\sigma}^{G^c}})=U_{\alpha\beta}^+,f(U_{\gamma\alpha}^{\vec{\sigma}^G})=U_{\gamma\alpha}^-,\\
&f(U_{\beta\gamma}^{\vec{\sigma}^{G^c}})=U_{\beta\gamma}^+,f(U_{\alpha\beta}^{\vec{\sigma}^G})=U_{\alpha\beta}^-, \: \text{and} \: f(U_{\gamma\alpha}^{\vec{\sigma}^{G^c}})=U_{\gamma\alpha}^+.
\end{align}
Therefore, for case $2$, it holds that $G^c$ is almost-decisive over $Y$.
\end{proof} 

\subsection{Proof of Lemma \ref{lem:prop3ultra}}

\begin{proof}
If $G=\varnothing$ or $G=N$, then $G$ or $G^c$ equals $N$, but then by unanimity of $f$, $G$ or $G^c$ is almost-decisive. Suppose then $G\neq \varnothing$ and $G\neq N$.

Let $\alpha, \beta \in X$. Then $f(U_{\alpha\beta}^{\vec{\sigma}^G})=U_{\alpha\beta}^+$ or $f(U_{\alpha\beta}^{\vec{\sigma}^G})=U_{\alpha\beta}^-$ (equivalently, $f(U_{\beta\alpha}^{\vec{\sigma}^{G^c}})=U_{\beta\alpha}^+$). In the first case, Lemma \ref{lem:contagioLemma} implies that $G$ is almost-decisive. In the second case, Lemma \ref{lem:contagioLemma} implies that $G^c$ is almost-decisive. 
\end{proof}

\subsection{Proof of Theorem \ref{thrm:arrow-generalization}}

We want to present a lemma that will be used as part of Theorem \ref{thrm:arrow-generalization}'s proof. But before, let us recall what do we mean by unanimity vertex (although it might be obvious). If $D$ is a domain, a vertex of $N_D$ is of \emph{unanimity} if it is of the form $U_{\alpha\beta}^{\vec{\sigma}^N}$ for distinct $\alpha,\beta\in X$. 

\begin{lemma}\label{lem:prop1ultra}
Let $f\colon N_D \to N_{W(x)}$ be a chromatic and unanimous simplicial map and $\mathcal{G}$ the set of all
almost-decisive coalitions (over $X$) w.r.t. $f$. We have that $\varnothing\not\in \mathcal{G}$ iff $N_D$ has an unanimity vertex.
\end{lemma}

\begin{proof}
We start with the $\Rightarrow$ direction. We proceed by contrapositive. Assume there is no unanimity vertex in $N_D$. But then $\varnothing$ satisfies the definition of almost-decisiveness by vacuity. 

Now, we prove the $\Leftarrow$ direction. Suppose there exists a unanimity vertex $U_{\alpha\beta}^{\vec{\sigma}^N}$ for some distinct $\alpha,\beta\in X$. Notice that, by unanimity of $f$, society is almost-decisive, i.e. $N\in \mathcal{G}$. Then by Lemma \ref{lem:almost-dec-compl}, $f(U_{\alpha\beta}^{\vec{\sigma}^\varnothing)}=U_{\alpha\beta}^-$ for all distinct $\alpha,\beta\in X$.  Hence, $\varnothing\not\in \mathcal{G}$. 
\end{proof}

Of course, in other Arrovian proofs using ultrafilters, e.g. \cite{CAMPBELL200235,KIRMAN1972267}, there is an argument of this sort to prove property $1$ of their definition.

We next present the proof of Theorem \ref{thrm:arrow-generalization}.

\begin{proof}
Let $D\in \mathcal{D}^{\text{PT}}\cap \mathcal{D}^{\text{DT}}$, $f\colon N_D\to N_{W(X)}$ be a unanimous chromatic simplicial map, and $\mathcal{G}$ the set of all almost-decisive coalitions w.r.t. $f$.

By Lemma \ref{lem:ultra-dict}, if we show that $\mathcal{G}$ is an ultrafilter w.r.t. $N$, we are done.

Since $D\in \mathcal{D}^{\text{PT}}$, we have that $N_D$ has a unanimity vertex. Therefore, by Lemma \ref{lem:prop1ultra}, $\varnothing\not\in \mathcal{G}$. Hence, property $1$ of ultrafilters hold.

Also, since $D\in \mathcal{D}^{\text{PT}}$, Lemma \ref{lem:prop3ultra} guarantees that property $3$ of ultrafilters holds.

Finally, having $D\in \mathcal{D}^{\text{PT}}\cap \mathcal{D}^{\text{DT}}$ guarantees, by Lemma \ref{lem:prop2ultra}, that property $2$ of ultrafilters hold.

Therefore, $\mathcal{G}$ is an ultrafilter w.r.t. $N$.
\end{proof}

\subsection{Proof of Proposition \ref{prop:closed-under-up-incl}}

\begin{proof}
Since $D\subseteq D'$, it is easy to see that $N_D$ is a subcomplex of $N_{D'}$, we denote this fact as $N_D\subseteq N_{D'}$. 

Firstly, let us see that $D'\in \mathcal{D}^{\text{PT}}$. To see this, let $G$ be a non-empty coalition distinct from $N$ and $Y\subseteq X$ such that $|Y|=3$. Since $D\in \mathcal{D}^{\text{PT}}$, there exists an $i\in \{1,2\}$  such that $B_i(G,Y)\subseteq N_D$. Since $N_D\subseteq N_{D'}$, we have that $B_i(G,Y)\subseteq N_{D'}$. Hence, $D'\in \mathcal{D}^{\text{PT}}$.

Secondly, let us prove that $D'\in \mathcal{D}^{\text{DT}}$. To see this, let $G$ and $G'$ coalitions such that $G\setminus G'$ and $G'\setminus G$ are non-empty. Since $D\in \mathcal{D}^{\text{DT}}$, there exists $\alpha,\beta,\gamma\in X$, all different from each other, such that $\{U_{\alpha\beta}^{\vec{\sigma}^G}, U_{\beta\gamma}^{\vec{\sigma}^{G'}}, U_{\gamma\alpha}^{\vec{\sigma}^{(G\cap G')^c}}\}$ is a $2$-simplex of $N_D$. Since $N_D\subseteq N_{D'}$, the set $\{U_{\alpha\beta}^{\vec{\sigma}^G}, U_{\beta\gamma}^{\vec{\sigma}^{G'}}, U_{\gamma\alpha}^{\vec{\sigma}^{(G\cap G')^c}}\}$ is a $2$-simplex of $N_D'$. Hence, $D'\in \mathcal{D}^{\text{DT}}$. Therefore, $D'\in \mathcal{D}^{\text{PT}}\cap\mathcal{D}^{\text{DT}}$.

To see that if $D\in \mathcal{D}^{\text{PT}}\cap\mathcal{D}^{\text{DT}}$, then $D$ is super-Arrovian, just combine what we just proved with Theorem \ref{thrm:arrow-generalization}.
\end{proof}

\end{document}